\documentclass[12pt]{iopart}

\usepackage{bm}
\usepackage{graphicx}
\usepackage{iopams}
\usepackage{setstack}
\usepackage{url}\usepackage[backend=bibtex8,sorting=none,doi=true, url=true, isbn=false]{biblatex}
\renewbibmacro{in:}{}
\usepackage{color}
\usepackage[dvipsnames]{xcolor}
\usepackage{outlines}
\usepackage{caption}
\usepackage{cancel}
\usepackage{cleveref}
\usepackage{listings} %
\bibliography{EverythingPlasmaBibthesis} %
\crefformat{appendix}{#2#1#3} %

\makeatletter
\DeclareFontFamily{U}{tipa}{}
\DeclareFontShape{U}{tipa}{m}{n}{<->tipa10}{}
\newcommand{\arc@char}{{\usefont{U}{tipa}{m}{n}\symbol{62}}}%

\newcommand{\arc}[1]{\mathpalette\arc@arc{#1}}

\newcommand{\arc@arc}[2]{%
  \sbox0{$\m@th#1#2$}%
  \vbox{
    \hbox{\resizebox{\wd0}{\height}{\arc@char}}
    \nointerlineskip
    \box0
  }%
}
\makeatother

\begin{document}

\title[Toroidal and slab ETG instability in JET-ILW pedestals]{Toroidal and slab ETG instability dominance in the linear spectrum of JET-ILW pedestals}

\author{Jason F. Parisi$^{1,2,3}$, Felix I. Parra$^{1}$, Colin M. Roach$^{2}$, Carine Giroud$^{2}$, William Dorland$^{4,1}$, David R. Hatch$^{5}$, Michael Barnes$^{1}$, Jon C. Hillesheim$^{2}$,  Nobuyuki Aiba$^{6}$, Justin Ball$^{7}$, Plamen G. Ivanov$^{1,2}$, and JET Contributors*}
\address{$^1$Rudolf Peierls Centre for Theoretical Physics, University of Oxford, Oxford OX1 3PU, UK}
\address{$^2$Culham Centre for Fusion Energy, Culham Science Centre, Abingdon OX14 3DB, UK}
\address{$^3$Merton College, Merton Street, Oxford, OX1 4JD, UK}
\address{$^4$Department of Physics, University of Maryland, College Park, Maryland 20742, USA}
\address{$^5$Institute for Fusion Studies, University of Texas at Austin, Austin, Texas, 78712, USA}
\address{$^6$National Institutes for Quantum and Radiological Science and Technology, Rokkasho, Aomori 039-3212, Japan}
\address{$^7$\'Ecole Polytechnique F\'ed\'erale de Lausanne (EPFL), Swiss Plasma Center (SPC), CH-1015 Lausanne, Switzerland}
\address{*See the author list of Joffrin E. et al. 2019 Nucl. Fusion 59, 112021}

\ead{jason.parisi@physics.ox.ac.uk}
\vspace{10pt}

\begin{abstract}
Local linear gyrokinetic simulations show that electron temperature gradient (ETG) instabilities are the fastest growing modes for $k_y \rho_i \gtrsim 0.1$ in the steep gradient region for a JET pedestal discharge (92174) where the electron temperature gradient is steeper than the ion temperature gradient. Here, $k_y$ is the wavenumber in the direction perpendicular to both the magnetic field and the radial direction, and $\rho_i$ is the ion gyroradius. At $k_y \rho_i \gtrsim 1$, the fastest growing mode is often a novel type of toroidal ETG instability. This toroidal ETG mode is driven at scales as large as $k_y \rho_i \sim (\rho_i/\rho_e) L_{Te} / R_0 \sim 1$ and at a sufficiently large radial wavenumber that electron finite Larmor radius effects become important; that is, $K_x \rho_e \sim 1$, where $K_x$ is the effective radial wavenumber. Here, $\rho_e$ is the electron gyroradius, $R_0$ is the major radius of the last closed flux surface, and $1/L_{Te}$ is an inverse length proportional to the logarithmic gradient of the equilibrium electron temperature. The fastest growing toroidal ETG modes are often driven far away from the outboard midplane. In this equilibrium, ion temperature gradient instability is subdominant at all scales and kinetic ballooning modes are shown to be suppressed by $\mathbf{ E} \times \mathbf{ B} $ shear. ETG modes are very resilient to $\mathbf{ E} \times \mathbf{ B}$ shear. Heuristic quasilinear arguments suggest that the novel toroidal ETG instability is important for transport.
\end{abstract}

\section{Introduction} \label{sec:introduction}

H-mode is currently the most favored high confinement operating regime in tokamaks. In H-mode, plasma confinement significantly improves once plasma heating exceeds a certain threshold \cite{Wagner1982,Ryter1996}. H-mode was first discovered in ASDEX \cite{Wagner1982}, and subsequently in most other tokamaks \cite{Burrell1989,Bush1990,Greenwald1997,Keilhacker1999}. The precursor to H-mode, L-mode \cite{Burrell1992}, has fairly constant equilibrium gradients across its radius, whereas H-mode is characterized by the presence of a pedestal with decreased turbulent particle and heat diffusivities, and therefore significantly increased equilibrium gradients. These increased gradients drive MHD instabilities, which set hard limits on the maximum achievable pressure gradient \cite{Keilhacker1984, Connor1998, Huysmans2005, Snyder2009}. Turbulent transport caused by microinstabilities driven unstable by equilibrium gradients that steepen during the inter-ELM (inter-edge-localized mode) period \cite{Hill1997} can constrain other pedestal dynamics such as MHD stability \cite{Rogers1998,Hatch2017}, scrape off layer and divertor physics \cite{Neuhauser2002}, and neoclassical transport \cite{Pusztai2016}, and hence studying H-mode inter-ELM  pedestal microstability is of great interest. 

To study the pedestal microinstabilities, we use gyrokinetics \cite{Taylor1968, Catto1978,Antonsen1980,Frieman1982,Parra2008,Abel2013} --- an asymptotic approach to solving the Fokker-Planck kinetic equation. Gyrokinetics is well-suited for studying anisotropic turbulence in highly magnetized plasmas. One of its main results, the gyrokinetic equation, is a nonlinear partial differential equation for the time evolution of the perturbed gyroaveraged distribution function. We will use the linearized gyrokinetic equation in conjunction with Maxwell's equations to study microinstabilities in JET pedestals. The local $\delta f$ gyrokinetic code GS2 \cite{Dorland2000} is used to simulate the pedestal plasmas presented in this article. 

We study the stability of a JET ITER-like wall (JET-ILW) inter-ELM magnetic equilibrium with different ion and electron temperature profiles. The ion and electron temperatures are obtained using impurity charge exchange emission and Thomson scattering, respectively. Since $\mathbf{ E} \times \mathbf{ B} $ shear is hypothesized to play a key role in pedestal formation \cite{Burrell1992,Hahm1995,Biglari1990}, we focus on the radial region near the maximum value of the equilibrium $\mathbf{ E} \times \mathbf{ B}$ shear. The region of maximum $\mathbf{ E} \times \mathbf{ B}$ shear is estimated by balancing the radial electric field with the pressure gradient.

Gyrokinetic studies of pedestals have been performed before. Local gyrokinetic analysis of MAST found the main instabilities at $k_{\perp} \rho_i \sim 1$ to be kinetic ballooning modes (KBMs) in the steep pressure gradient region and microtearing modes (MTMs) in the less steep pressure gradient region inside the pedestal top, throughout the inter-ELM recovery of the pedestal \cite{Dickinson2012}. A follow up study using DBS and cross-polarization scattering found that $k_{\perp} \rho_i \approx 3 - 4$ turbulence at the pedestal top in MAST was most consistent with the electron temperature gradient (ETG) instability \cite{Hillesheim2016}. Using the Gyrokinetic Toroidal Code \cite{Lin1998}, PIC simulations in the steep gradient region of DIII-D discharges found electrostatic electron-driven modes peaking at poloidal angle $\theta = \pm \pi/2$ \cite{Fulton2014}. More recently, in JET-ILW discharges where the ion temperature was not measured and was assumed to be equal to the measured electron temperature, nonlinear global gyrokinetic calculations were performed using the GENE code \cite{Jenko2000,Gorler2011}. These global simulations predict pedestal heat transport fluxes that are comparable with experiment, and suggest that pedestal fluxes will be increasingly dominated by ion temperature gradient (ITG) turbulence as the heating power increases \cite{Hatch2017}. Hatch et al. also proposed that impurity seeding reduces ion-scale and ETG instability transport via ion-dilution and increased collisionality \cite{Hatch2017}. In \cite{Hatch2016}, it was again demonstrated that the sum of neoclassical, MTM, and ETG turbulent transport was in good agreement with another JET-ILW pedestal measurement. Another recent work that used experimental ion temperature profiles found that ITG was suppressed in JET Carbon discharges, but not in JET-ILW cases, where ITG turbulence carried a substantial fraction of the total heat flux \cite{Hatch2019}. The difference between JET Carbon and JET-ILW was attributable to a decreased density gradient in JET-ILW discharges, which increased the growth rates of slab ITG and ETG instabilities.

In this work, we identify a novel type of toroidal ETG instability that appears in regions of steep equilibrium temperature gradients. These sub-ion Larmor scale modes have a radial wavenumber larger than its poloidal wavenumber, and have been observed (but not explained) in previous pedestal simulations \cite{Told2008,Jenko2009,Told2012,Fulton2014,Baumgaertel2011,Kotschenreuther2019}. The particularly large radial wavenumber means that the radial magnetic drift plays an important role in these toroidal ETG modes. We find that this toroidal ETG has a large critical gradient threshold, which occurs due to the pedestal's magnetic geometry and the radial magnetic drift. Moreover, because of the large equilibrium temperature gradients, we show theoretically and numerically that both toroidal and slab ETG modes are extended from perpendicular scales of $k_y \rho_e \sim 1$ in the core, to $k_y \rho_i \sim 1$ in the pedestal, where $k_y$ is the binormal wavenumber, defined in \Cref{sec:gyrokineticreviews}, and $\rho_s$ is the gyroradius for a species $s$. 

We primarily examine microinstability at a single radial location in the steep gradient region of JET-ILW shot 92714 \cite{Giroud2018}, a highly-fueled deuterium discharge with deuterated ethylene ($\mathrm{C}_2 \mathrm{D}_4$) injection. For this discharge, at all scales where instability occurs --- $0.005 \lesssim k_y \rho_i \lesssim 400$ --- we find that electron temperature gradient-driven modes are the fastest growing modes. For $k_y \rho_i \gtrsim 1$, the novel toroidal ETG mode is usually the fastest growing mode. We also show that the gradients of the measured ion temperature profiles are insufficiently steep to drive ITG instability. With the measured ion temperature profiles, the ion temperature gradient is close to the critical gradient needed for linear instability and hence subdominant. We also show that if ion temperature gradients are made sufficiently steep, toroidal and slab ITG modes become unstable at $k_y \rho_i \ll 1$, but \textcolor{black}{can be} suppressed by $\mathbf{ E} \times \mathbf{ B}$ shear. Our findings suggest that the toroidal and slab ITG mode are stable in many radial pedestal locations, even in the steep gradient region that we examine.

The layout of this paper is as follows: we first introduce gyrokinetics and the notation used throughout this paper in \Cref{sec:gyrokineticreviews}. We then present JET-ILW density, temperature, and rotation profiles from an inter-ELM pedestal in \Cref{sec:jetprofiles}. Here, we also give a broad overview of the growth rates and unusual mode structures for the fastest growing modes in this pedestal, including a discussion of electromagnetic effects and $\mathbf{ E} \times \mathbf{ B}$ shear. At a wide range of scales, we find an ETG mode with unusual character. This mode typically has a radial wavenumber that is significantly larger than the poloidal wavenumber, and is insensitive to finite $\beta$ effects and $\mathbf{ E} \times \mathbf{ B}$ shear. Motivated by the results of \Cref{sec:jetprofiles} and using the notation of \Cref{sec:gyrokineticreviews}, we then make analytic predictions about microinstability in steep gradient regions in \Cref{sec:linGKwithlargegrads}. This theoretical analysis explains the existence of the novel toroidal ETG modes that we see in \Cref{sec:jetprofiles}. We then examine ETG and ITG (or lack thereof) instability in linear gyrokinetic simulations in \Cref{sec:GKsims,sec:ITG}, respectively. The effect of $\mathbf{ E} \times \mathbf{ B} $ shear is discussed further in \Cref{sec:ExBeffects}. Finally, we conclude in \Cref{sec:discussion}. Experimentally-minded readers might wish to jump to \Cref{sec:jetprofiles,sec:GKsims}, while those more theoretically inclined and with a background in gyrokinetics might wish to begin at \Cref{sec:linGKwithlargegrads}.

\section{Gyrokinetics} \label{sec:gyrokineticreviews}

In this section, we introduce the system of gyrokinetic equations and notation used throughout this paper. This section can be skipped for readers well-acquainted with gyrokinetics, or who mainly wish to see gyrokinetic simulations results in \Cref{sec:jetprofiles,sec:GKsims,sec:ITG}. Gyrokinetics \cite{Taylor1968,Catto1978,Antonsen1980,Frieman1982,Parra2008,Abel2013} is used to investigate turbulence and transport using an asymptotic expansion in the ratio of  $ \rho_{*s} \equiv \rho_s / L_{Ts} \ll 1$. We express the gradients by the equilibrium length scales $L_{Q} \equiv -( \partial \ln Q / \partial  r )^{-1}$, where $Q$ can be the equilibrium density, temperature, or pressure, and the distance $r$ is half of the diameter of the flux surface at the midplane. Assuming $k_{\perp} \rho_i \sim 1$ and $\omega \ll \Omega_s$, gyrokinetics describes plasma behavior on spatial scales comparable to the ion gyroradius, and on timescales much longer than the gyro period. The quantity $k_{\perp}$ is the perpendicular turbulence wavenumber, $\omega $ is the frequency for turbulence quantities, $\Omega_s = Z_s e B / m_s c $ is the gyrofrequency, $Z_s$ is the charge number, $e$ is the proton charge, $B$ is the \textcolor{black}{leading order} magnetic field strength, $m_s$ is the species mass, and $c$ is the speed of light. The gyrokinetic ordering is $\rho_{*s} \sim  \omega / \Omega_s \sim  \nu_s / \Omega_s \sim  k_{\parallel} / k_{\perp} \ll 1$, where $\nu_s$ is a typical collision frequency for species $s$, and $k_{\parallel}$ is the turbulence parallel wavenumber. To obtain a rough estimate for the radial electric field (see \Cref{eq:pressgradientradEfield}), we will impose that the radial electric field is comparable to the pressure gradient, which implies a low flow ordering \cite{Hinton1976,Helander2002,Parra2015} for the electric field, $| \mathbf{E} | \sim T_{0e} / e L_{Te}$, that is, the equilibrium $\mathbf{ E} \times \mathbf{ B} $ drift is small compared with the thermal velocity $v_{ts} = \sqrt{2T_{0s}/m_s} $ by a factor of $\rho_{*s}$, where $T_{0s}$ is the leading order temperature.

We expand the magnetic field in $\rho_{*s} $, $\mathbf{B} + \mathbf{B}_1 + \mathbf{B}_2 + \ldots $, where $\mathbf{B}_n = \rho_{*s}^n \mathbf{B}  $ \textcolor{black}{(we reserve $\mathbf{B}$ for the leading order magnetic field, and do not explicitly use a symbol for the total magnetic field in this paper)}. The lowest order magnetic field is written as $\mathbf{B} = I(r) \nabla \zeta + \nabla \zeta \times \nabla \psi $, where $\zeta$ is the toroidal angle, $\psi$ is the poloidal flux divided by $2\pi$, and $I(r)$ is a flux function. For $n \geq 1$, we further split $\mathbf{B}_n $ into long-wavelength and turbulence components, $\mathbf{B}_n = \mathbf{B}_n^{lw} + \textcolor{black}{ \overline{\mathbf{B}}_n^{tb}} $. We reserve the overline notation for some turbulent quantities because later we will write their Fourier components without an overline, which will keep the notation tidier. Long wavelength quantities, $g ^{lw}$, spatially change on equilibrium length scales, $\nabla g ^{lw} \sim g ^{lw} / L_{Ts}$, and temporally change on slow time scales, $\partial g ^{lw}/ \partial t \sim g ^{lw} / \tau_E$, where $\tau_E$ is the energy confinement time and $t$ is the time variable. Turbulence quantities, $g ^{tb} $, spatially change on equilibrium length scales along the mean magnetic field, $ \hat{\mathbf{ b} } \cdot  \nabla g ^{tb} \sim g ^{tb}/ L_{Ts}$, but on gyroradius scales across the mean field, $\nabla_{\perp} g ^{tb} \sim g ^{tb} / \rho_s$, and temporally change on fast time scales, $\partial g ^{tb} / \partial t \sim \omega g ^{tb} $. Here, $\hat{\mathbf{ b} } = \mathbf{B} / B $, and $\nabla_{\perp}$ is a spatial derivative perpendicular to $\mathbf{B} $. We ignore the correction, $\mathbf{B}_1 ^{lw}  $, which is mainly due to the effect of the neoclassical pressure anisotropy on the magnetic field. One can show that the turbulent component of $\mathbf{B}_1 $ can be written as $\textcolor{black}{ \overline{ \mathbf{B}}_1 ^{tb}} = \nabla \overline{  A}_{\parallel 1 } ^{tb} \times \hat{\mathbf{ b} }  + \overline{  B}_{\parallel 1 } ^{tb} \hat{\mathbf{ b} }  $, where $\overline{B}_{\parallel 1}  ^{tb}  $ and $\overline{  A}_{\parallel 1} ^{tb} $ are the leading order parallel components of the turbulent magnetic field and magnetic vector potential, respectively.

\textcolor{black}{ We also expand the electric field $\mathbf{E}$} in $\rho_{*s} $, $\mathbf{E} = \mathbf{E}_0 + \mathbf{E}_1 + \ldots $, where $\mathbf{E}_n \sim \rho_{*s}^n T_{0s} / e L_{Ts} $. We split $\mathbf{E}_n $ into long wavelength and turbulent parts, $\mathbf{E}_n = \mathbf{E}_n ^{lw} +\textcolor{black}{ \overline{  \mathbf{E}}_n ^{tb}} $. To lowest order, $\mathbf{E}_0 $ is electrostatic; $\mathbf{E}_0 ^{lw} = - \nabla \phi_0 $, and $\textcolor{black}{ \overline{ \mathbf{E}}_0 ^{tb}} = - \nabla_{\perp} \overline{\phi} ^{tb}_1 $. Here, $\phi_0$ is the leading order electric potential and $\textcolor{black}{ \overline{ \phi} ^{tb} _1}$ is the leading order turbulent electric potential, where $\textcolor{black}{ \overline{ \phi }^{tb} _1} \sim \rho_{*s} \phi_0$. Since $\phi_0$ is a flux function, $\mathbf{E}_0 \cdot \hat{\mathbf{ b} } = 0 $. To leading order, the parallel components of the electric field are $E_{\parallel } ^{lw} = - \hat{\mathbf{ b} } \cdot \nabla \phi_1 ^{lw} $ and $\textcolor{black}{ \overline{ E}_{\parallel } ^{tb}} = - \hat{\mathbf{ b} } \cdot \nabla \overline{  \phi}_1 ^{tb} - (1/c) (\partial \overline{A}_{\parallel 1 } ^{tb} / \partial t )$. The electrostatic potential $\phi ^{lw} _1$ is mainly due to neoclassical physics.

We expand the distribution function in $\rho_{*s}$, $f_{s} = F_{Ms} + f_{1s} + \ldots$, where the lowest order distribution function, $F_{Ms}$, is a stationary Maxwellian,
\begin{equation}
F_{Ms} (r, v) = n_{0s} (r) \Big{(} \frac{m_s}{2 \pi T_{0s} (r)} \Big{)}^{3/2} \exp \Big{(} - \frac{m_s v^2 }{2 T_{0s} (r)} \Big{)},
\end{equation}
with particle speed $v$, and flux functions $n_{0s}$ and $T_{0s}$, where $n_{0s}$ is the leading order density. The Maxwellian is stationary because the mean flow is subsonic. Higher order corrections to the distribution function can be split into long-wavelength and turbulent quantities, $f_{ns} = f ^{lw}_{ns} + f ^{tb}_{ns}$, where neoclassical corrections would be included in $f ^{lw}_{ns} $.

To describe phase space, we will employ gyrokinetic variables. These are the guiding center, $\mathbf{R}_s$, the kinetic energy, $\mathcal{E} = v^2 /2$ , the magnetic moment, $\mu = v_{\perp}^2 / 2B$ where $\mathbf{v}_{\perp} = \mathbf{v} - \mathbf{v} \cdot \hat{\mathbf{ b} } \hat{\mathbf{ b} } $, and the gyrophase, $\varphi$, which is a particle's angular location during its gyromotion. The guiding center is given by $\mathbf{R}_s = \mathbf{r} - \bm{\rho}_s $, the gyroradius position is given by $\bm{\rho}_s = \hat{\mathbf{ b} } \times \mathbf{v} / \Omega_s $, and the quantity $\mathbf{r} $ is the particle position. The first order turbulent component of the distribution function can be written as
\begin{equation}
f^{tb}_{1s} (\mathbf{R}_s , \mathcal{E}, \mu, \varphi, t) =  \overline{{h}}_s \big{(} \mathbf{R}_s , \mathcal{E}, \mu, t \big{)} - \frac{Z_s e \overline{ \phi} ^{tb}_1 }{T_{0s}}  F_{Ms} (\mathbf{r} , \mathcal{E}, t).
\label{eq:hdefinition}
\end{equation}
Note that the function $\overline{ h}_s$ is independent of the gyrophase --- our task is to find an evolution equation for $\overline{ h}_s$. 

To find $\overline{ h}_s$, we substitute \Cref{eq:hdefinition} into the Fokker-Planck equation. Because only the variable $\varphi $ varies over a single gyroperiod, it is convenient to average the Fokker-Planck equation over the gyromotion using a gyroaverage, defined as $\langle \ldots \rangle = (1/2\pi) \int_0^{2\pi} \ldots d \varphi |_{\mathbf{R}_s , \mathcal{E}, \mu}$, evaluated at fixed $\mathbf{R}_s, \mathcal{E}$, and $\mu$. Gyroaveraging the Fokker-Planck equation and taking its turbulent component, we obtain the low flow electromagnetic gyrokinetic equation,
\begin{equation}
\fl
\eqalign{
\bigg{(} \frac{\partial }{\partial t} + \Omega_E \frac{\partial}{\partial \zeta} \bigg{)} \overline{h}_{s} + & (v_{\parallel } \hat{\mathbf{b} } + \mathbf{ v}_{Ms} + \langle \mathbf{ v}_{\chi} ^{tb} \rangle ) \cdot \nabla_{\mathbf{R_s} } \overline{ h}_{s} +  \sum_{s'} \bigg{\langle} C^{(l)}_{ss'} \bigg{\rangle} \\ & = \frac{Z_s e F_{Ms}}{T_{0s}} \bigg{(} \frac{\partial }{\partial t} + \Omega_E \frac{\partial}{\partial \zeta} \bigg{)}  \langle \overline{ \chi} ^{tb}_1 \rangle - \langle \mathbf{ v}_{\chi} ^{tb} \rangle \cdot \nabla_{\mathbf{R}_s } F_{Ms}, 
}
\label{eq:lowfloweq}
\end{equation}
where $\Omega_E (r) = - c \partial \phi_0 / \partial \psi$ is the $\mathbf{ E} \times \mathbf{ B} $ toroidal angular velocity, $C^{(l)}_{ss'}$ is a linearized Fokker-Planck collision operator, $\nabla_{\mathbf{R}_s } \equiv \partial / \partial \mathbf{R}_s $,
the magnetic drift is
\begin{equation}
\mathbf{v}_{Ms} = \frac{ \hat{\mathbf{ b} }}{\Omega_s} \times \left[ \left( v_{\parallel }^2 + \frac{v_{\perp}^2}{2} \right) \nabla \ln B + v_{\parallel}^2 \frac{4 \pi }{B^2} \frac{\partial p_0}{\partial r} \nabla r \right].
\label{eq:magneticdrift}
\end{equation}
Here, $p_0 = \sum_s p_{0s}$ is the total pressure and $p_{0s} = n_{0s} T_{0s}$ is the lowest order pressure. The parallel velocity is $v_{\parallel } = \mathbf{v} \cdot \hat{\mathbf{ b} }$, and the gyrokinetic drift is $ \mathbf{ v}_{\chi}^{tb} = (c/B) \hat{\mathbf{b} } \times \nabla \overline{  \chi} ^{tb}_1$. Here, $\overline{ \chi} ^{tb}_1 $ is the leading order turbulent gyrokinetic potential defined as
\begin{equation}
\overline{\chi} ^{tb}_1  = \overline{  \phi }^{tb}_1 - \frac{ v_{\parallel } \overline{A} ^{tb} _{\parallel 1}}{c} + \frac{m_s}{Z_s e} \int_0^{\mu} \overline{ B}_{\parallel 1} ^{tb} (\mathbf{R}_s  + \bm{\rho}_s (\mu') ) d\mu'.
\end{equation} 

In Equation (3), $\Omega_E(r)$ can be approximated around the radial location $r_c$ of interest by $\Omega_E(r_c) + (r - r_c) (\partial  \Omega_E/ \partial r)$ because the characteristic size of the eddies is small compared with $L_{Te}$. In the low flow ordering that we use, the term $(r - r_c) (\partial \Omega_E/ \partial r)$, which represents the $\mathbf{E} \times \mathbf{B}$ shear, should be neglected because it is of the same size as other terms that we have not kept. Even so, we perform some simulations with $\mathbf{E} \times \mathbf{B}$ shear. We will justify using this small term in \Cref{sec:ExBeffects}.

To close the system of equations, we need to find $ \overline{  \phi} ^{tb} _1 $, $\overline{A}_{\parallel 1} ^{tb} $, and $\overline{ B}_{\parallel 1} ^{tb} $ using $ \overline{ h}_s$. To find $\overline{ \phi} ^{tb} _1$, we use the first order turbulent quasineutrality condition,
\begin{equation}
\sum_s \frac{Z_s^2 e^2 \overline{  \phi} ^{tb} _1}{T_{0s}} n_{0s} = \sum_s Z_s e \int  \overline{  h}_s (\mathbf{r}  - \bm{\rho}_s, \mathcal{E}, \mu)  d^3 v.
\label{eq:turbQN}
\end{equation}
The parallel vector potential, $\overline{   A} ^{tb}_{\parallel 1}$, is found using the parallel component of Amp\`ere's law,
\begin{equation}
- \nabla^2_{\perp} \overline{A}_{\parallel 1} ^{tb}  = \frac{4 \pi e}{c} \sum_s Z_s \int v_{\parallel } \overline{ h}_s (\mathbf{r}  - \bm{\rho}_s, \mathcal{E}, \mu) d^3 v.
\end{equation} 
Finally, $\overline{ B}_{\parallel 1} ^{tb}$ is determined by perpendicular pressure balance,
\begin{equation}
\frac{B \overline{ B}_{\parallel 1} ^{tb} }{4 \pi} + \sum_s \int m_s B \int_0^{\mu} \overline{h}_s (\mathbf{r}  - \bm{\rho}_s (\mu'), \mathcal{E}, \mu) d \mu' d^3 v = 0.
\end{equation}
Note that the integral over $\mu'$ only affects the $\mu$ dependence of $\bm{\rho}_s$.

Throughout this paper, we will examine the stability properties of the gyrokinetic equation in the linear local limit. To understand how these linear instabilities then cause turbulent transport, one needs to keep the nonlinear term of \Cref{eq:lowfloweq}, which we will neglect in this work. The local limit, $k_{\perp} L_{Ts} \gg 1$, is useful for analytic treatment and numerically efficient simulations. If $k_{\perp} L_{Ts} \gg 1$, modes can be Fourier analyzed in the perpendicular domain. In JET shot 92174 at the radial location we examine, $L_{Te} \simeq 0.02 \mathrm{m}$, and thus the local approximation is good provided that $k_{\perp} \rho_i \gg \rho_i / L_{Te} \simeq 0.12$. \textcolor{black}{Note that throughout this work, the quantity $k_y \rho_i$ will be a deceiving measure of $k_{\perp} \rho_i$; the modes that we find typically have a very large radial wavenumber compared to $k_y \rho_i$. Hence, for these modes $k_{\perp} \rho_i \gg k_y \rho_i$.}

To describe the properties of the turbulent pieces, $\overline{ \phi }^{tb}_1, \overline{A}_{\parallel 1} ^{tb}, \overline{B}_{\parallel 1} ^{tb}$, and $\overline{ h}_s $, we use the flux coordinates $(x, y, \theta)$. The coordinate $x$ is a local flux surface label defined around the flux surface $r_c$ (note that it is different from the flux label $r$), $y$ is a field line label, and $\theta$ is a poloidal coordinate that labels the position along the magnetic field line. The coordinates $x$ and $y$ are given by
\begin{equation}
x = \frac{\textcolor{black}{q_c}}{r_c B_a}(\psi(r) - \psi(r_c)), \;\;\; y = \frac{1}{B_a} \frac{\partial \psi}{\partial r} \bigg{|}_{r_c} \alpha,
\end{equation}
where $B_a$ is the toroidal magnetic field strength evaluated at $r_c$ and $R_c$, $R_c$ is the distance from the axis of symmetry of the tokamak to the center of the flux surface $r_c$ at the midplane, \textcolor{black}{$q_c = q(r_c)$}, $\alpha = \zeta - q \theta + \nu (r, \theta)$, and $\nu (r, \theta)$ is a function $2 \pi$-periodic in $\theta$,
\begin{equation}
\fl \nu (r, \theta) = - I(r) \bigg{(} \int_0^{\theta} d \theta' \bigg{[} \frac{1}{R^2(\theta') \mathbf{B} (\theta') \cdot \nabla \theta' } - \frac{1}{2\pi} \oint \frac{d \theta''}{R^2(\theta'') \mathbf{B}(\theta'') \cdot \nabla \theta''} \bigg{]}  \bigg{)}.
\end{equation}
The safety factor, $q(r)$, is given by $2\pi q(r) = \oint I(r) d \theta / R^2 \mathbf{B} \cdot \nabla \theta $. We choose to define the poloidal angle $\theta$ as
\begin{equation}
\theta = 2\pi l /L_{\theta},
\label{eq:thetaequation}
\end{equation}
where $l$ is the arclength along the magnetic field, and $L_{\theta}$ is the distance along a field line for one complete poloidal turn.

Spatial anisotropy, $k_{\perp}/k_{\parallel} \gg 1$, implies that $\partial / \partial x \sim \partial / \partial y \gg (2\pi / L_{\theta}) \partial / \partial \theta$. In the linear local limit, we Fourier analyze $\overline{  \phi}^{tb}_1$ locally in the perpendicular plane and in time,
\begin{equation}
\overline{\phi} ^{tb}_1 (x, y, \theta, t) = \sum_{k_{x}, k_{y}, \omega } \phi ^{tb} _1 (k_{x}, k_{y}, \theta, \omega) \exp(ik_{x} x + i k_{y} y - i \omega t).
\label{eq:phifourier}
\end{equation}
The electromagnetic fluctuations $\overline{A}_{\parallel } ^{tb} $ and $\overline{B}_{\parallel } ^{tb} $ are Fourier analyzed in a similar way. It will also be useful to Fourier analyze $\overline{h}_s$,
\begin{equation}
\fl \overline{h}_s (X_s, Y_s, \theta, \mathcal{E} ,\mu, t) = \sum_{k_{x}, k_{y}, \omega} h_s (k_{x}, k_{y}, \theta, \mathcal{E}, \mu, \omega) \exp(ik_{x} X_s + i k_{y} Y_s - i \omega t),
\label{eq:hfourier}
\end{equation}
where $X_s = x - \bm{\rho}_s \cdot \nabla x $ and $Y_s = y - \bm{\rho}_s \cdot \nabla y$ are guiding center variables. In the next section we present the profiles for the JET shot that we are examining, as well as an overview of the gyrokinetic results. These gyrokinetic results will motivate the work for the rest of the paper.

\section{Pedestal Gyrokinetic Simulations of JET Shot 92174} \label{sec:jetprofiles}

In this section, we present the significant linear microstability features of a single JET-ILW inter-ELM pedestal discharge at a single radial location. This equilibrium exhibits properties such as temperature, magnetic geometry, injected neutral beam power, and fueling that are typical for JET-ILW inter-ELM H-mode pedestals: key experimental parameters for this discharge are $I_p = 1.4$ MA, $B_{T0} = 1.9$ T, $H_{98(y,2)} = 1.0$, $n_G = 0.7$, $P_{\mathrm{NBI}} = 17.4$ MW, $\beta_N = 2.5$, and $R_D = 0.9 \times 10^{22}$ electrons/s. Here, $I_p$ is the poloidal current, $B_{T0}$ is the toroidal magnetic field at $R = 2.96$m, $H_{98(y,2)}$ is the H factor relative to the IPB98${}_{(y,2)}$ scaling \cite{Doyle2007}, $n_G$ is the Greenwald density fraction \cite{Greenwald1988} defined as the line averaged density divided by the Greenwald density limit, $P_{\mathrm{NBI}}$ is the neutral beam injection power, $\beta_N$ is the normalized $\beta$ factor \cite{Troyon1984}, and $R_D$ is the deuterium electron flow rate.

In \Cref{subsec:JETprofiles}, we show the pedestal equilibrium temperature, density, and flow profiles, which will have significant implications for microstability. In \Cref{subsec:gyroresults}, we present an overview of linear results from gyrokinetic simulations, run both with and without finite $\beta$ effects. From these results, we justify an electrostatic study. Here, we find a range of modes, including an unusual toroidal ETG instability that is driven at a very wide range of perpendicular scales, and has a radial wavenumber that is typically much larger than its poloidal wavenumber. A significant portion of the paper will be devoted to understanding this mode. We show that this mode is largely unaffected by finite $\beta$ effects and $\mathbf{ E} \times \mathbf{ B}$ shear, and in subsequent sections, that it could play an important role in transport. Finally, in \Cref{subsec:linearfeatures}, we present the prominent features of the electrostatic growth rate spectrum.

\subsection{JET-ILW Profiles} \label{subsec:JETprofiles}

In this paper, we focus on simulation results from JET shot 92174. We run linear gyrokinetic simulations with a single deuterium ion species and no impurities, assuming that $n_{0e} = n_{0i}$ (note that experimentally $Z_{\mathrm{eff}} = 1.8$, where $Z_{\mathrm{eff}} = \sum_i \textcolor{black}{n_{0i}} Z_i^2 / \textcolor{black}{n_{0e}}$). The three other pedestals that we have analyzed (82550, 92167, 92168) give qualitatively similar results, which is notable, given that the nature of these discharges varies quite significantly. The experimental and simulation parameters and linear gyrokinetic growth rates for these additional three discharges are shown in \cref{app:dischargeparams}.

\begin{figure}[t]
\centering
    \includegraphics[width=\textwidth]{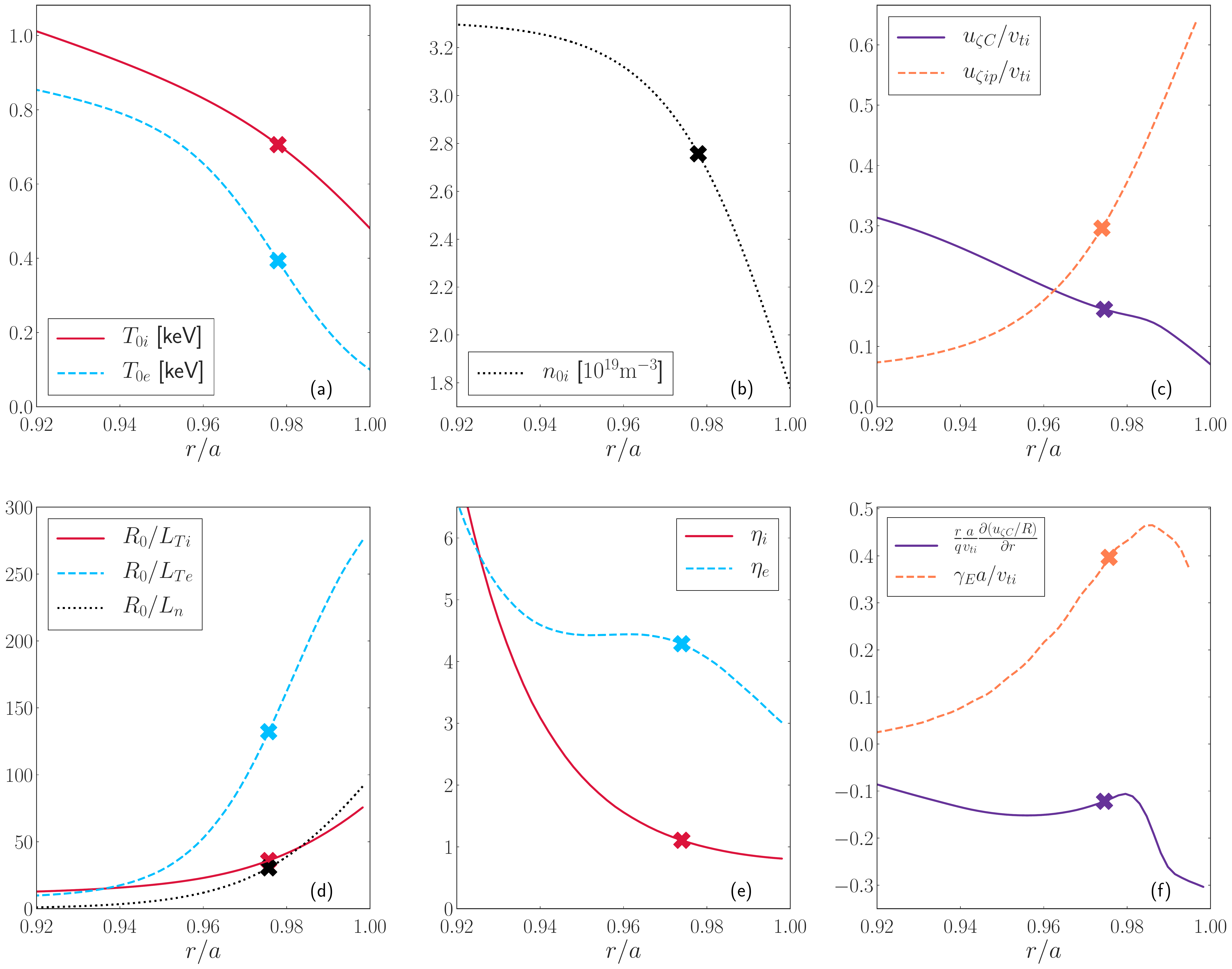}
 \caption{Pedestal profiles and their gradients for JET shot 92174. Crosses indicate simulation location of $r/a = 0.9743$. (a): Ion and electron temperatures profiles. (b): Density profiles. (c): Flow profiles \textcolor{black}{for $u_{\zeta C}$, the experimental value for the toroidal component of the ${}^{12}_6 C^+$ flow, and $u_{\zeta i p}$, the toroidal component of the ion diamagnetic flow, defined in \Cref{eq:uzeta}.} (d): Temperature and density gradients profiles. (e): $\eta_s$ profiles, where the parameter $\eta_s$ is defined as $\eta_s \equiv L_n/L_{Ts}$. (f): \textcolor{black}{Flow shear} profiles.}
\label{fig:1profiles}
\end{figure}

The temperature and density profiles for shot 92174 and associated gradients, are shown in \Cref{fig:1profiles}(a), (b), and (d) as functions of $r/a$. The distance $a$ is the value of $r$ at the last closed flux surface (LCFS). In \Cref{fig:1profiles}(c), we also show the toroidal velocity of ${}^{12}_6 C^+$, $u_{\zeta C}$, at the outboard midplane, normalized by the ion thermal speed $v_{ti} = \sqrt{2 T_{0i}/m_i}$. We assume that this velocity is a good proxy for the toroidal ion velocity, $u_{\zeta i} $. We normalize the gradient length scales using the major radius of the last closed flux surface, $R_0$, which is the radial distance to the center of the last closed flux surface at the midplane. The profiles in \Cref{fig:1profiles} are consistent with an emerging JET-ILW pedestal paradigm \cite{Leyland2015, Hatch2017, Kotschenreuther2017, Hatch2019}, whereby enhanced gas puffing reduces the edge density gradient \cite{Wolfrum2017} and shifts the density pedestal outwards \cite{Hillesheim2016,Dunne2017}, making microinstabilities more virulent \cite{Maggi2017}. Weaker density gradients also reduce the $\mathbf{ E} \times \mathbf{ B}$ shear, which has often been observed to be important for microinstability suppression in the pedestal \cite{Hatch2017, Kotschenreuther2017, Hatch2019}. It is hypothesized that heat transport from more strongly-driven microinstabilities with less shear suppression is responsible for a reduced temperature at the pedestal top \cite{Hatch2019}.

In this work, the electron temperature and density are determined from the High Resolution Thomson Scattering profiles \cite{Pasqualotto2004,Frassinetti2012}. To improve the data statistics, a composite profile is constructed from profiles collected in a time window of 80-99\% in the ELM interval period. The profiles of the ion temperature and rotation are measured with the edge Charge Exchange Recombination Spectroscopy diagnostic \cite{Delabie2016} for fully stripped carbon-12 (${}_6^{12}C^+$), with a time integration of 7.2ms. These ion profiles are collected on a longer 60-99 \% ELM interval period time window. The ${}_6^{12}C^+$ and ion temperature and rotation profiles in the pedestal can differ substantially, as found in some recent DIII-D experiments \cite{Haskey2018, Haskey2018b, Camenen2010}. Since the ITG instability is sensitive to $T_{0i}$ and $R_0/L_{Ti}$, the ITG instability results in \Cref{sec:ITG} should be viewed in the context of potentially large uncertainties in ion temperature measurements, which might significantly underestimate the ion temperature gradient. For this reason, while we have mainly used $T_{0i} > T_{0e}$ and $R_0/L_{Te} > R_0/L_{Ti}$ in our simulations and theory, we have also explored the impact on gyrokinetic microinstabilities of assuming $T_{0e} = T_{0i}$ and $R_0/L_{Ti} = R_0/L_{Te}$, which can be found in \Cref{sec:ITG}. However, unless explicitly mentioned otherwise, we use the measured ion temperature profiles.

To obtain an estimate for the radial electric field, we use the most general ion flow \cite{Hinton1976,Helander2002},
\begin{equation}
\mathbf{ u}_i = - c \frac{\partial \phi_0 }{\partial \psi} R^2 \nabla \zeta - \frac{c}{Z_i e n_{0 i}} \frac{\partial p_{0i}}{\partial \psi} R^2 \nabla \zeta + \frac{ \mathbf{B}}{n_{0i} } K_i (\psi) \frac{\partial T_{0i}}{\partial \psi}.
\label{eq:lowflowequation}
\end{equation}
Here, $R$ is the major radius, and the flux function, $K_i(\psi)$, is determined by neoclassical theory \cite{Hinton1976,Helander2002}. Based on the experimental data in \Cref{fig:1profiles}, we find that $u_{\zeta C} \lesssim (\rho_{P i}/L_{Ti}) v_{ti}$. The quantity $\rho_{P s} = (B/B_P) \rho_s$ is the poloidal gyroradius for a species $s$, where $B_P$ is the poloidal magnetic field strength. Thus, the flow velocity of the ${}^{12}_6C^+$ impurity species is comparable to the size of the ion diamagnetic flow, $u_{\zeta i p}$,
\begin{equation}
\frac{u_{\zeta i p}}{v_{ti}} = - \frac{R c}{Z_i e n_{0i} v_{ti}} \frac{\partial p_{0i}}{\partial \psi} \sim \frac{\rho_{Pi}}{L_{pi}}  \sim \frac{1}{3}.
\label{eq:uzeta}
\end{equation}
Note that this implies that there are only several poloidal gyroradii in a pressure length scale, $L_{pi}$. To obtain a rough estimate of the radial electric field, we use the fact that the measurement of $u_{\zeta_i}$ suggests that the overall flow, the $\mathbf{ E} \times \mathbf{ B} $ flow, the diamagnetic flow in \Cref{eq:uzeta}, and the term proportional to $K_i (\psi)$ are all of the same order. Thus,
\begin{equation}
- \frac{\partial \phi_0}{\partial \psi} \approx \frac{1}{Z_i e n_{0i}} \frac{\partial p_{0i}}{\partial \psi}.
\label{eq:pressgradientradEfield}
\end{equation}
Then, the radial shear in the $\mathbf{ E} \times \mathbf{ B} $ rotation, $\gamma_E (\psi)$, is approximately
\begin{equation}
\gamma_E \equiv - \frac{c r}{q} \frac{\partial}{\partial r} \bigg{(} \frac{\partial \phi_0}{\partial \psi} \bigg{)} \approx  \frac{r}{q} \frac{\partial }{\partial r} \bigg{(} \frac{c}{Z_i e n_{0i}} \frac{\partial p_{0i}}{\partial \psi} \bigg{)}.
\label{eq:gammaEapproximation}
\end{equation}

The location of the simulations was chosen to have equilibrium length scales characteristic of the steep gradient region in the pedestal, and an $\mathbf{ E} \times \mathbf{ B}$ shear value close to the maximum possible for a given equilibrium, using the estimate in \Cref{eq:gammaEapproximation}. The radial location for JET shot 92174, shown in \Cref{fig:1profiles}, is $r/a = 0.9743$. To simulate this discharge, we use the following simulation parameters: $\rho_i = 0.27 \; \mathrm{cm}, \; \nu_{ee} a/ v_{ti} = \textcolor{black}{0.83}, \; a / L_{Te} = 42, \; a/ L_{Ti} = 11, \; a / L_{n} = 10, \; \rho_i / L_{Te} = 0.12, \; T_{0e}/T_{0i} = 0.56, \; \hat{s} = 3.36, \; q = 5.1, \; R_0 = 2.86$ m, $a = 0.91$ m, $R_c = 2.91$ m, and $r_c = 0.89$ m, where $\nu_{ss'} = \sqrt{2} \pi n_{0s'} Z_s^2 Z_{s'}^2 e^4 \ln (\Lambda_{ss'}) / \sqrt{m_s} T_{0s}^{3/2} $, $\ln (\Lambda_{ss'})$ is the Coulomb logarithm, and $\hat{s} = (r / q) \partial  q / \partial  r$ is the magnetic shear. In the instances where we included $\mathbf{ E} \times \mathbf{ B}$ shear and electromagnetic effects, we used $\gamma_E a / v_{ti} = 0.56$ and $\beta = 0.0031$. Here, the quantity $\beta = 8 \pi (p_{0i} + p_{0e})/B_a^2$, where $B_a = 1.99$ T for this equilibrium.

\subsection{Gyrokinetic Simulation Results} \label{subsec:gyroresults}

In this section, we present results obtained from linear gyrokinetic simulations (both electromagnetic and electrostatic) for this radial location and pedestal. For the chosen pedestal and radial location, we will establish that linear \textit{electrostatic} simulations without $\mathbf{ E} \times \mathbf{ B}$ shear give similar growth rate spectra to linear \textit{electromagnetic} simulations with $\mathbf{ E} \times \mathbf{ B}$ shear. The electrostatic limit of \Cref{eq:lowfloweq} is taken by requiring that the turbulent electric field is primarily electrostatic, $|\nabla \overline{\phi}_1^{tb}| \gg (1/c) | \partial \overline{A}_{\parallel 1} ^{tb}/ \partial t | $, and that the turbulent magnetic field is small, $| \mu \overline{B} ^{tb}_1 | \ll | Z_s \overline{\phi} ^{tb}_1| e / m_s$ \footnote{Even though the last term in \Cref{eq:magneticdrift} is formally small in $\beta$ in the electrostatic limit, we keep it in all our electrostatic simulations because the large pressure gradients in the pedestal can make it important.}. It is no coincidence that the electrostatic regime without $\mathbf{ E} \times \mathbf{ B}$ shear and the electromagnetic case with $\mathbf{ E} \times \mathbf{ B}$ shear give similar results; electromagnetic modes are suppressed by $\mathbf{ E} \times \mathbf{ B}$ shear, leaving electrostatic modes that are unaffected by $\mathbf{ E} \times \mathbf{ B}$ shear as the dominant instabilities. Therefore, it is reasonable to study this pedestal with linear electrostatic simulations without $\mathbf{ E} \times \mathbf{ B}$ shear. We will choose to study the electrostatic limit without $\mathbf{ E} \times \mathbf{ B}$ shear rather than an electromagnetic case with $\mathbf{ E} \times \mathbf{ B}$ shear because the former is analytically and numerically simpler. We now proceed to give an overview of gyrokinetic results for the electrostatic pedestal.

We performed these local simulations in ballooning space, which can be represented in a flux-tube \cite{Beer1995}. Because the novel toroidal ETG instability we have found is often driven at large distances along the field line from $\theta = 0$, we require a large range of $\theta$ values, and hence we typically choose a flux-tube with 64 gridpoints in each $2\pi$ period in $\theta$, with nine periods. This is equivalent to a ballooning space calculation extending to nine poloidal turns in the extended ballooning angle. The standard velocity space grid has 20 passing pitch angles, \textcolor{black}{33} trapped pitch angles, and 12 energy gridpoints \cite{Barnes2010}. Resolution scans were performed in all of these parameters \textcolor{black}{by doubling each of them independently; there was no significant difference in the frequencies or the character of these modes.}

\begin{figure}[t]
\centering
    \includegraphics[width=0.9\textwidth]{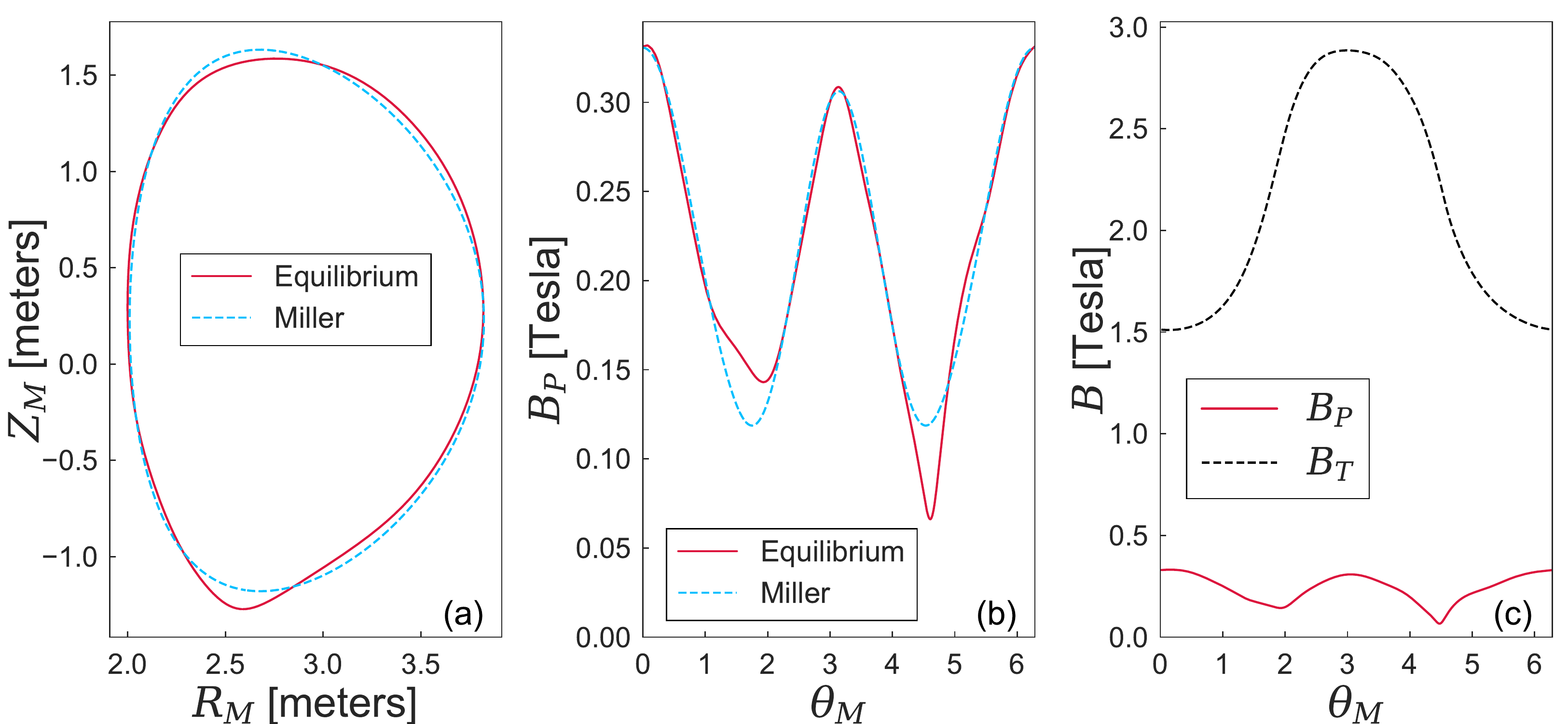}
 \caption{The Miller equilibrium and numerical equilibrium for JET shot 92174 used for gyrokinetic simulations. (a): Equilibrium and Miller flux surfaces in $R_M, Z_M$ space, (b): Equilibrium and Miller poloidal magnetic field versus $\theta_M$, (c): Equilibrium toroidal and \textcolor{black}{poloidal} magnetic fields.}
  \label{fig:Millerfits}
\end{figure}

While GS2 is capable of reading in numerical equilibria, we fit the magnetic equilibrium with Miller geometry. A Miller equilibrium is a prescription to generate flux surfaces that satisfy the Grad Shafranov equation locally by fitting to nine parameters \cite{Miller1998}. \textcolor{black}{The shape of the flux surface $r_c$ and its neighbors is determined by $R = R_M(r, \theta_M)$ and $Z = Z_M(r, \theta_M)$, where $\theta_M$ is the Miller poloidal angle, which is in general not equal to the poloidal angle $\theta$ defined in \Cref{eq:thetaequation}}. In \Cref{fig:Millerfits} we show the difference between the exact flux surface at $r/a = 0.9743$ and the Miller fits that we use. The Miller parameters for this radial location are \textcolor{black}{$dR_c /d r = -0.345$}, $\kappa = 1.55$,
$a (d\kappa/dr) = 0.949$, $ \delta = 0.263$, $a (d \delta/ dr) = 0.737$, $\beta' = \beta a (d \ln p_0 / d r) = -0.161$, where $\kappa$ is the flux surface elongation and $\delta$ is the triangularity.

\begin{figure}
\centering
    \includegraphics[width=1.0\textwidth]{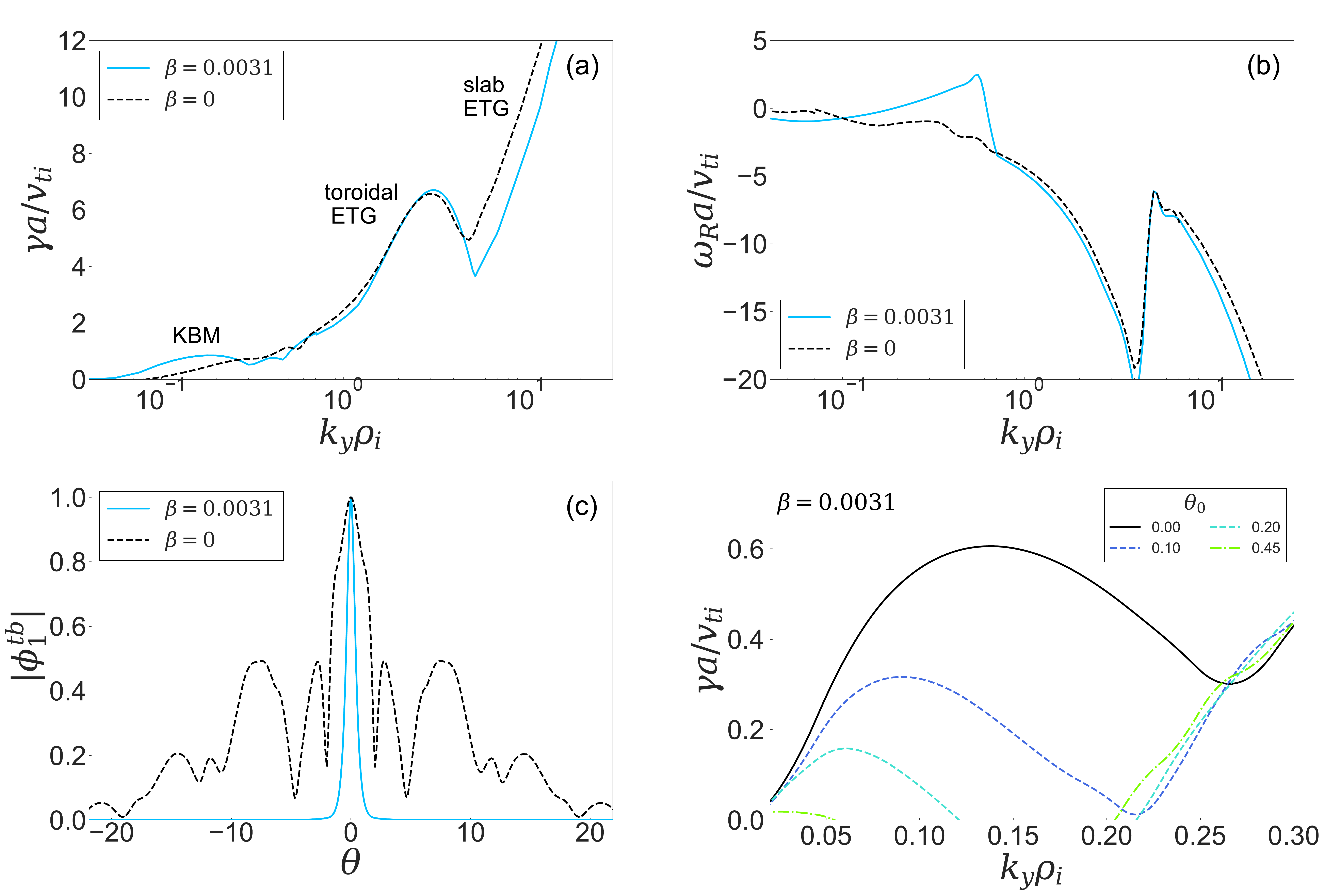}
 \caption{(a): GS2 growth rate ($\gamma$) and (b): \textcolor{black}{R}eal frequency ($\omega_R$) for JET shot 92174 with $\theta_0 = 0$ with and without finite $\beta$. (c): \textcolor{black}{E}igenmodes for $k_y \rho_i  = 0.2$. (d): \textcolor{black}{G}rowth rates for an electromagnetic simulation with different $\theta_0$ values at $k_y \rho_i \sim 0.1$. \textcolor{black}{All of these simulations are performed without $\mathbf{ E} \times \mathbf{ B}$ shear.}}
\label{fig:electromagneticspectraearly}
\end{figure}

Electromagnetic effects have been shown to be important for microinstability in the pedestal \cite{Dickinson2012,Hatch2017,Hatch2016,Hatch2019,Kotschenreuther2019}. While we have neglected electromagnetic effects in most of this study, we have scoped out the potential effects of nonzero $\beta$. As an initial study, this is well-justified since we will show that a linear electromagnetic gyrokinetic simulation with $\mathbf{ E} \times \mathbf{ B}$ shear gives similar results to a linear \textit{electrostatic} gyrokinetic simulation without $\mathbf{ E} \times \mathbf{ B}$ shear. To demonstrate this equivalence, we first show the results of gyrokinetic simulations with and without finite $\beta$ effects in \Cref{fig:electromagneticspectraearly}. To include finite $\beta$ effects, we included values of $\beta$ and $\beta'$ consistent with the Miller equilibrium.

In \Cref{fig:electromagneticspectraearly}, we show the effect of finite $\beta$ on the growth rates (a), real frequencies (b), and eigenmodes (c) for $\theta_0 = 0$, where $\theta_0$ is the ballooning angle, defined as $\theta_0 = k_x / \hat{s} k_y$. Throughout this paper, the eigenmodes are separately normalized such that $|\phi ^{tb} _1|$ has a maximum of 1. When finite $\beta$ effects are included, a KBM appears, as shown by the small bump at $k_y \rho_i \sim 0.1$ in \Cref{fig:electromagneticspectraearly}(a) of the growth rates. This KBM has a standard ballooning eigenmode structure, centered at $\theta = \theta_0 = 0$. However, when $\beta = 0$, there is no KBM, and instead at $k_y \rho_i \sim 0.1$ there are modes with a much lower growth rate and a complicated mode structure in $\theta$ (see \Cref{fig:electromagneticspectraearly}(c)). These eigenmodes \textcolor{black}{ tend to} have maxima in bad curvature regions and \textcolor{black}{can} have \textcolor{black}{either ballooning or tearing parity} in both $\mathrm{Re}(\phi ^{tb} _1)$ and $\mathrm{Im}(\phi ^{tb} _1)$. More details regarding these long wavelength electron modes can be found in \cref{sec:smallkrimodes}.

Much of the rest of the growth rate spectrum is quite unaffected by finite $\beta$ effects. At $k_y \rho_i \approx 1-5$ for $\theta_0 = 0$, there is a peculiar bump in \Cref{fig:electromagneticspectraearly}(a), whose corresponding instability will be the focus of much of this paper. We identify this mode as toroidal ETG. We have undertaken extensive tests described later in \Cref{sec:GKsims} to confirm that it is a novel type of toroidal ETG; for now, we will refer to it as a toroidal ETG mode without justification. Finally, for $k_y \rho_i \gtrsim 5$ and $\theta_0 = 0$, the fastest growing mode becomes a slab ETG mode, which again, we will justify later in \Cref{sec:GKsims}. Clearly the toroidal ETG mode is \textcolor{black}{almost entirely} unaffected by finite $\beta$, and the slab ETG growth rates decrease by roughly 20\%, but the mode structure is qualitatively the same. Thus, apart from the KBM, the electromagnetic and electrostatic growth rates and modes are very similar.

Once $\mathbf{ E} \times \mathbf{ B}$ shear is included in the simulations, the electromagnetic and electrostatic growth rate spectra become qualitatively the same. This is because $\mathbf{ E} \times \mathbf{ B}$ shear is found to easily suppress the KBM. Recall that the KBM is the main difference between the electromagnetic and electrostatic simulations without $\mathbf{ E} \times \mathbf{ B}$ shear. Further evidence for the effectiveness of the $\mathbf{ E} \times \mathbf{ B}$ shear for suppressing the KBM is that the KBM is stable for all $|\theta_0| > \theta_{0c} \approx 0.5$, as shown in \Cref{fig:electromagneticspectraearly}(d), where we show the growth rates for a range of $\theta_0$ values at scales $ 0.01 <k_y \rho_i < 0.3$ in a simulation with finite $\beta$. The dependence on $\theta_0$ is important because $\mathbf{ E} \times \mathbf{ B}$ shear causes a mode's radial wavenumber to vary with time as $\Delta k_x = k_y \gamma_E t$, giving a change of $\theta_0$ of $\Delta \theta_0 = \gamma_E t / \hat{s}$. If a mode is shown to be unstable only for a very narrow range of $\theta_0$ values, $|\theta_0| < |\theta_{0c}|$, it is highly susceptible to $\mathbf{ E} \times \mathbf{ B} $ shear because in a time of order $1/\gamma_E$ its $\theta_0$ changes significantly. After a time $t_{C} \sim \hat{s} \theta_{0c}/ \gamma_E$, $\mathbf{ E} \times \mathbf{ B}$ shear will have suppressed the KBM; in our simulations, $t_C \approx 3$. Thus, to suppress instability we require $\gamma t_C \lesssim 1$, leading to $\gamma_E / \hat{s} \gamma \gtrsim \theta_{0c}\approx 0.5$. We will discuss the $\mathbf{ E} \times \mathbf{ B}$ shear and its effects on all the other instabilities we find in more detail in \Cref{sec:ExBeffects}. \textcolor{black}{Until then, all simulations are performed without $\mathbf{ E} \times \mathbf{ B}$ shear.}

Finally, the perpendicular wavenumber of the KBM is close to the limit where local simulations are valid, which is when $k_{\perp} \rho_i \gg 0.12$. Hence, results from our KBM simulations should be viewed in the context of uncertainties that are present due to the \textcolor{black}{value of $k_{\perp} \rho_i$ for the KBM being close to this limit.}

\subsection{Linear Features of the Electrostatic Pedestal} \label{subsec:linearfeatures}

In this section, we describe the most prominent features of the electrostatic growth rate spectrum.

\begin{figure}
\centering
    \includegraphics[width=1.0\textwidth]{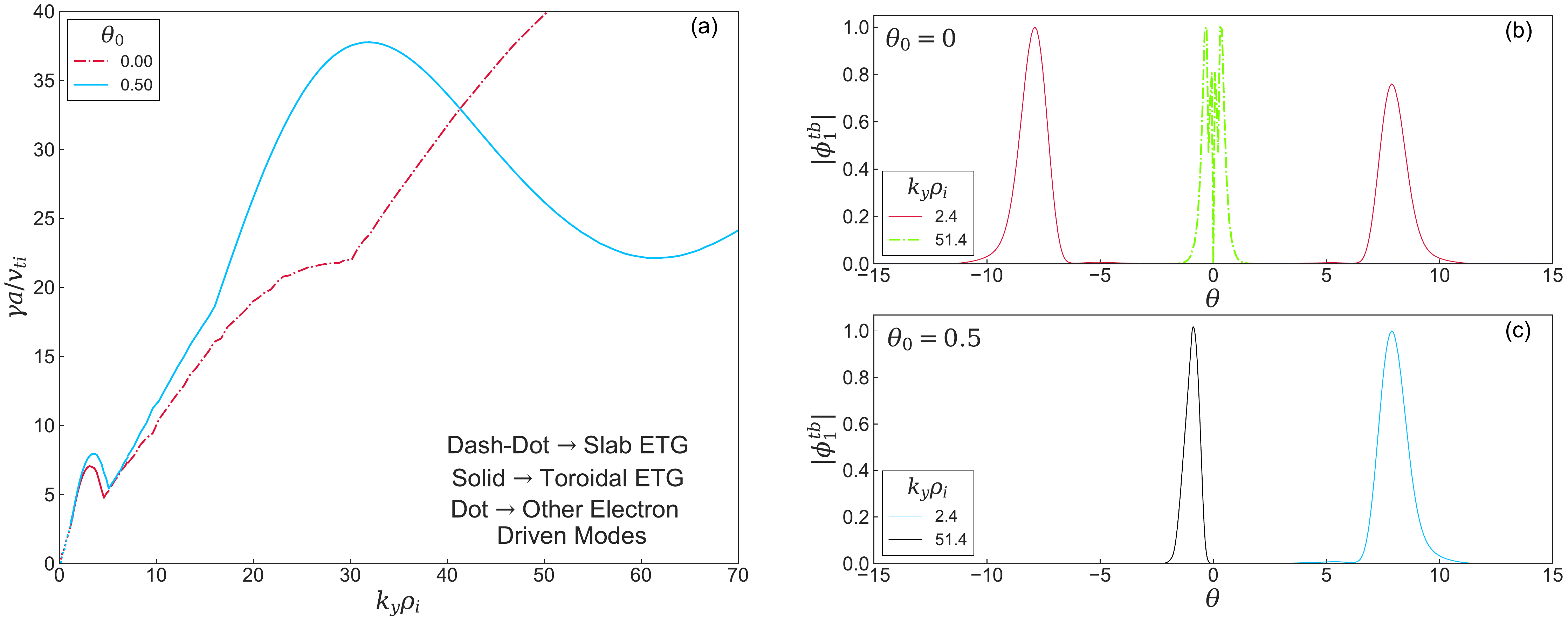}
 \caption{(a): \textcolor{black}{E}lectrostatic growth rates for 2 values of $\theta_0$. (b): \textcolor{black}{E}igenmodes for 2 values of $k_y \rho_i$ at $\theta_0 = 0$. (c): \textcolor{black}{E}igenmodes for 2 values of $k_y \rho_i$ at $\theta_0 = 0.5$.}
\label{fig:electrostaticspectra}
\end{figure}

A notable feature of the growth rate spectrum shown earlier in \Cref{fig:electromagneticspectraearly} is the bump at $k_y \rho_i \approx 1 - 5$ in \Cref{fig:electromagneticspectraearly}(a), which we claimed was a novel toroidal ETG instability. In \Cref{fig:electrostaticspectra}(a), we show the growth rates for two values of $\theta_0$. Focusing first on $\theta_0 = 0$, we again identify the bump at $k_y \rho_i \approx 1 - 5$, which has a peak growth rate at $k_y \rho_i \simeq 3$. Once $k_y \rho_i \gtrsim 5$, the mode switches to a slab ETG instability. In \Cref{fig:electrostaticspectra}(b), we show the eigenmodes for two $k_y \rho_i$ values in the $\theta_0 = 0$ growth rate spectrum, one at $k_y \rho_i = 2.4$ (near the top of the toroidal ETG bump) and one at $k_y \rho_i = 51.4$. The eigenmode associated with $k_y \rho_i = 2.4$ is fairly localized at large $\theta$, whereas the eigenmode associated with $k_y \rho_i = 51.4$ is centered at $\theta = 0$ and has a large parallel wavenumber. The $k_y \rho_i = 2.4$ mode is the novel toroidal ETG mode, and the $k_y \rho_i = 51.4$ mode is a slab ETG mode. In our up-down symmetric equilibrium fit, there is a subtlety for the novel toroidal ETG eigenmodes when $\theta_0 = 0$: there are two independent modes that grow at the same rate, and are localized at opposite signs of $\theta$. Indeed, for toroidal ETG, there must be two independent modes with $\theta_0 = 0$, since the \textcolor{black}{linear} gyrokinetic equation is invariant under the transformation $\theta \to - \theta$, $\theta_0 \to - \theta_0$ \cite{Peeters2005}. Thus, henceforth, when plotting the eigenmodes for $\theta_0 \simeq 0$, we choose a small value of $\theta_0$, $\theta_0 = 0.05$, which causes the mode at one location to grow slightly faster than the mode at the other, but barely changes the growth rate compared with $\theta_0 = 0$. This results in a well-defined single eigenmode, like the one in \Cref{fig:krionetofiveplot}(a), rather than two separate modes, like the ones shown in \textcolor{black}{ \Cref{fig:electrostaticspectra}}(b). The relative size and phase of the modes at opposite values of $\theta$ depend on the initial condition.

To distinguish between the toroidal and slab ETG modes in \Cref{fig:electromagneticspectraearly}(a) and \Cref{fig:electrostaticspectra}(a), we used a set of criteria discussed extensively in \Cref{subsec:toroidalETG}. Briefly, the toroidal ETG eigenmodes are localized far along the field line for smaller $k_y \rho_i$ values, and are at a $\theta$ location with the opposite sign of $\theta_0$ for larger $k_y \rho_i $ values. Sensitivity scans to equilibrium parameters, shown in \Cref{fig:krionetofivesensitivityscan}, reveal that the slab and toroidal ETG branches have different dependences on parameters such as $R_0/L_{Ti}$ and $R_0/L_{n}$. For a given $k_y \rho_i$, slab ETG modes also tend to have a much larger $k_{\parallel}$ than toroidal ETG modes.

While the novel toroidal ETG mode is the fastest growing instability for $1 \lesssim k_y \rho_i \lesssim 5$ when $\theta_0 = 0$, we find that when $\theta_0$ differs slightly from 0, the toroidal ETG mode is the fastest growing for $1 \lesssim k_y \rho_i \lesssim 400$. We show a simple example of the growth rate spectrum for $\theta_0 =  0.5$ in \Cref{fig:electrostaticspectra}(a), where the toroidal ETG mode is the fastest growing mode for that particular value of $\theta_0$ for all $k_y \rho_i \gtrsim 1$. In \Cref{fig:electrostaticspectra}(c), we show the eigenmodes for $\theta_0 = 0.5$ for $k_y \rho_i = 2.4$ and $k_y \rho_i = 51.4$. For $k_y \rho_i = 2.4$, the eigenmodes for $\theta_0 = 0$ and $\theta_0 = 0.5$ have a similar structure, both being localized at $|\theta| \simeq 8$. However, the eigenmode at $k_y \rho_i = 51.4$ is dramatically different to the $\theta_0 = 0$ mode at $k_y \rho_i = 51.4$; the eigenmode for $\theta_0 = 0.5$ is localized at $\theta \simeq -1$, and has, in fact, the same novel toroidal ETG character we identified earlier. In \Cref{sec:GKsims} we will explain these toroidal ETG modes in much more detail, including the reasons why they move in $\theta$ for different values of $k_y \rho_i$, as evidenced by the eigenmodes for $\theta_0 = 0.5$ at $k_y \rho_i = 2.4$ and $k_y \rho_i = 51.4$.
 
For completeness, we briefly describe the modes we find at larger scales. For this JET discharge and the surface $r/a = 0.9743$, we find that the instabilities are electron-driven between $0.005 \lesssim k_y \rho_i \lesssim 400$. For $0.005 \lesssim k_y \rho_i \lesssim \textcolor{black}{0.07}$ the modes have electron tails similar to those described in \cite{Hallatschek2005}, and for $0.1 \lesssim k_y \rho_i \lesssim 1.0$, there are complicated modes that appear to be a form of ETG we do not yet fully understand. Both the electron tails and complicated ETG modes will be excluded from in-depth analysis in the main text, but are described in \cref{sec:smallkrimodes}.

In the next section, we introduce the theory needed to understand these novel toroidal ETG modes as well as the slab ETG modes at $k_y \rho_i \gtrsim 1$. We will see that the existence of these modes follows naturally from the steep temperature gradients in pedestals.

\section{Linear Gyrokinetics With Large Gradients} \label{sec:linGKwithlargegrads}

In this section, we analyze the consequences of large equilibrium gradients for linear collisionless electrostatic gyrokinetic stability, which will considerably change the character of the toroidal ETG instability. We have already motivated the local and linear limits in \Cref{sec:gyrokineticreviews}, and the electrostatic limit in \Cref{subsec:gyroresults}. We now motivate the \textit{collisionless} limit of the electron gyrokinetic equation, which will be used for the theoretical analysis.

The collisionless limit for electrons is justified by the small electron collision frequency, \textcolor{black}{ $\nu_{ee} \ll \gamma$. For JET shot 92174 at $r/a = 0.9743$, $\nu_{ee} \simeq 2.4 \times 10^{5}$ Hz, and $\gamma \simeq 1.6 \times 10^{6}$ Hz for $k_y \rho_i = 2$.}  In gyrokinetic simulations, we found ETG instabilities to be \textcolor{black}{relatively} insensitive to whether collisions are kept. However, for ITG scale instabilities at lower frequencies, electron collisions can decrease the ITG growth rates and cause electrons to be non-adiabatic, as we will see in \Cref{sec:ITG}.

Using the equations laid out in \Cref{sec:gyrokineticreviews}, we take the linear electrostatic collisionless local limit of the gyrokinetic equation in \Cref{sec:electrostatcollisionlesslocal}. Analytically and computationally, this limit is more straightforward, and includes key elements of the pedestal microinstability linear physics that we wish to explain. Motivated by the steep pedestal gradients, we explore the implications of steep equilibrium temperature gradients on ETG instability in \Cref{sec:slabversustoroidalETG}. Simple arguments based on balancing terms with the same order of magnitude reveal how these steep gradients affect the perpendicular scales of the instability and how magnetic shear determines the parallel toroidal ETG mode structure, allowing the toroidal ETG mode to compete with the slab ETG mode. In \Cref{subsec:dispersionrelation}, we convert the gyrokinetic equation derived in \Cref{sec:electrostatcollisionlesslocal} to an algebraic equation in order to analyze slab and toroidal ETG instabilities in the presence of large equilibrium gradients. This is then used to derive an analytical ETG dispersion relation that supports our simplified arguments.

\subsection{Electrostatic Collisionless Local Limit} \label{sec:electrostatcollisionlesslocal}

In this section, we take the electrostatic, linear, collisionless form of the gyrokinetic equation. In this limit, \Cref{eq:lowfloweq} is
\begin{equation}
\fl
\eqalign{
 \frac{\partial \overline{ h}_s}{\partial t}  & +  v_{\parallel} \hat{\mathbf{b} } \cdot  \nabla_{\mathbf{R}_s} \overline{ h}_s + \mathbf{v}_{Ms} \cdot \nabla_{\mathbf{R}_s} \overline{ h}_s =  \frac{Z_s e F_{Ms}}{T_{0s}} \frac{\partial \langle \overline{\phi} ^{tb}_1 \rangle}{\partial t} \\ & + \frac{c}{B} (\nabla_{\mathbf{R}_s } \langle \overline{  \phi}^{tb}_1 \rangle \times \hat{\mathbf{b}} ) \cdot \nabla r \bigg{[} \frac{\partial  \ln n_s}{\partial r} + \frac{\partial \ln T_s}{\partial r} \bigg{(} \frac{m_s  \mathcal{E}}{T_{0s}} - \frac{3}{2} \bigg{)} \bigg{]} F_{Ms}.
}
\label{eq:GKgeneral}
\end{equation}
We have absorbed the toroidal mean flow in the convective derivative as a constant Doppler shift, and neglected the equilibrium $\mathbf{ E} \times \mathbf{ B}$ shear, which is consistent with the low flow ordering in \Cref{eq:pressgradientradEfield}, and is justified in \Cref{sec:ExBeffects} with simulation results.

Substituting the expressions for $\phi^{tb}_1$ and $h_s$ in \Cref{eq:phifourier,eq:hfourier} into \Cref{eq:GKgeneral} gives a Fourier-analyzed gyrokinetic equation,
\begin{equation}
\fl
\eqalign{
 - i \omega h_s & + \frac{2\pi v_{\parallel }}{L_{\theta}} \frac{\partial h_s}{\partial \theta} +  i \mathbf{v}_{Ms} \cdot \mathbf{k}_{\perp} h_s =  - i \omega \frac{Z_s e F_{Ms}}{T_{0s}} \phi ^{tb}_1 J_0 \bigg{(} \frac{k_{\perp} v_{\perp}}{\Omega_s} \bigg{)} \\ & + i \omega_{*s} \bigg{[} 1 + \eta_s \bigg{(} \frac{m_s \mathcal{E}}{T_{0s}}  - \frac{3}{2} \bigg{)} \bigg{]} \frac{Z_s e F_{Ms}}{T_{0s}} \phi ^{tb}_1 J_0 \bigg{(} \frac{k_{\perp} v_{\perp}}{\Omega_s} \bigg{)},
}
\label{eq:GKgeneralFourierAnalyzednotkpll}
\end{equation}
where $J_0$ is a Bessel function of the first kind that comes from gyroaveraging $\overline{ \phi }^{tb} _1$. The perpendicular wavenumber $\mathbf{k}_{\perp}$ is
\begin{equation}
\eqalign{
\mathbf{k}_{\perp} = k_x \nabla x + k_y \nabla y = & \bigg{[}k_x - k_y\bigg{(}\hat{s} \theta - \frac{r}{q} \frac{\partial \nu}{\partial r} \bigg{)} \bigg{]} \nabla x \\ & + \frac{\partial \psi}{\partial r} \frac{1}{B_a} k_y \bigg{[} \nabla \zeta + \bigg{(}\frac{\partial \nu }{\partial \theta} - q\bigg{)} \nabla \theta  \bigg{]},
}
\label{eq:kperpxywithshear}
\end{equation}
where every function is evaluated at $r_c$. We have also introduced the drift frequency, $\omega_{*s}$, and the stability parameter, $\eta_s$,
\begin{equation}
\omega_{*s} \equiv - \frac{c}{B} \frac{T_{0s}}{Z_s e L_{ns}} (\mathbf{ k}_{\perp} \times \hat{\mathbf{b} } )\cdot \nabla r = \frac{c}{B_a} \frac{T_{0s}}{Z_s e L_{ns}} k_y, \;\;\;\; \eta_s \equiv \frac{ L_{ns}}{ L_{Ts}}.
\label{eq:omstedef}
\end{equation}
Note that the factor $ (\mathbf{ k}_{\perp} \times \hat{\mathbf{b} } )\cdot \nabla r$ in $\omega_{*s} $ is only proportional to $k_y$. The system of equations is closed by the first order turbulent quasineutrality condition in \Cref{eq:turbQN},
\begin{equation}
\eqalign{
\frac{e \phi^{tb}_1 n_{0e}}{T_{0e}} \bigg{(} \frac{Z_i T_{0e}}{T_{0i}} +1 \bigg{)} & + 2\pi  \int \frac{B}{|v_{\parallel }|} h_e J_0 \bigg{(} \frac{k_{\perp} v_{\perp}}{\Omega_e} \bigg{)}  d \mathcal{E} d \mu \\
& - 2\pi \int \frac{B}{|v_{\parallel }|} h_i J_0 \bigg{(} \frac{k_{\perp} v_{\perp}}{\Omega_i} \bigg{)} d \mathcal{E} d \mu = 0,
}
\label{eq:turnQNexplicit}
\end{equation}
where we used that the Jacobian of the gyrokinetic transformation is $\mathcal{J} = \partial (\mathbf{r}, \mathbf{v}  )/ \partial(\mathbf{R}, \mathcal{E}, \mu, \varphi ) \simeq B/ |v_{\parallel }|$ \cite{Parra2008}. 

We proceed to demonstrate how the presence of large equilibrium gradients changes the perpendicular scales at which ETG can be strongly driven, and how in the presence of these steep gradients, magnetic shear can act to determine the poloidal location where the ETG mode has its maximum amplitude.

\subsection{Slab Versus Toroidal ETG In Large Gradient Regions} \label{sec:slabversustoroidalETG}

In this section, we describe a novel type of toroidal ETG with anisotropic perpendicular wavenumbers. \Cref{eq:turnQNexplicit} contains two branches of electron temperature gradient driven instability, slab \cite{Rudakov1961,Coppi1967} and toroidal \cite{Coppi1977,Horton1981}. These modes have been covered extensively \cite{Jenko2000,Rudakov1961,Coppi1967,Coppi1977,Horton1981,Cowley1991}. Here, we give a very brief overview. In the slab branch, the density perturbation is caused by a competition between the parallel streaming and the radial $\mathbf{ E} \times \mathbf{ B} $ drift. For sufficiently large $\eta_s$, a large parallel compression causes $\phi ^{tb}_1$ to grow in time. For smaller values of $\eta_s$, the radial $\mathbf{ E} \times \mathbf{ B} $ drift term dominates and we obtain stable electron drift waves. The toroidal instability is caused by magnetic drifts, rather than parallel streaming, creating a compression that again, gives rise to a destabilizing electric field for sufficiently large $\eta_s$. In both cases, at the onset of instability, increasing the temperature gradients causes the linear instability to be more virulent. 

Motivated by the large temperature gradients in \Cref{fig:1profiles}(d), we proceed to demonstrate that
\begin{equation}
\frac{R_0}{L_{Te}}, \frac{R_0}{L_{Ti}} \gg 1,
\end{equation}
has major implications for ETG stability. First, we present an intuitive, albeit non-rigorous argument that will turn out to be incorrect. We then develop a more careful argument, which reveals the distinctive new character of ETG modes in steep gradients, which is very different to the more familiar lower gradient regime typical of the core. Throughout this section, we shall assume that $\theta_0 = 0$. We will investigate the physics of $\theta_0 \neq 0$ in \Cref{sec:effectsofkx}.

First, we present the intuitive, albeit incorrect argument. For the electrons, since $R_0 / L_{Te} \gg 1$, we naively expect that the ratio determining the relative strength of the drive frequency to the magnetic drift frequency to be large. Therefore, in the pedestal, one might naively think that the drive for toroidal ETG is weak and independent of $k_{\perp}$,
\begin{equation}
\frac{\omega_{*e} \eta_e }{\mathbf{ v}_{Me} \cdot \mathbf{  k}_{\perp}} \sim \frac{R_0}{L_{Te}} \gg 1.
\label{eq:drivetodriftstrength}
\end{equation}
Here, we use $\mathbf{ v}_{Me} \cdot \mathbf{  k}_{\perp} \sim k_{\perp} v_{te}^2/ \Omega_e R_0$ and $k_y \sim k_{\perp}$. Comparing the size of the drive frequency to the parallel streaming frequency, we obtain
\begin{equation}
\frac{\omega_{*e} \eta_e}{k_{\parallel} v_{te}} \sim \frac{k_y}{k_{\parallel} } \frac{\rho_e}{L_{Te}}.
\label{eq:drifttoparallelstreamelectron}
\end{equation}
As we will show in \Cref{subsec:dispersionrelation}, the ratios in \Cref{eq:drivetodriftstrength,eq:drifttoparallelstreamelectron} must be of order unity for a large toroidal and slab ETG growth rate, respectively (see \Cref{fig:toroidaltoslab}). Thus, \Cref{eq:drivetodriftstrength} suggests that the magnetic drifts are small for every $k_{\perp}$, whereas in \Cref{eq:drifttoparallelstreamelectron}, $k_{\parallel} $ can become large to drive slab instability. One would therefore expect slab ETG to be the dominant electron microinstability at all scales.

The above argument, however, suffers from a deficiency. It is naive to make the assumption $\omega_{*e} \eta_e / \mathbf{ v}_{Me} \cdot \mathbf{  k}_{\perp} \sim R_0/L_{Te}$ (see \Cref{eq:drivetodriftstrength}) in the presence of magnetic shear, because $\mathbf{k}_{\perp}$ varies along a field line (see \Cref{eq:kperpxywithshear}). At large values of $|\theta|$, the \textit{radial} component of the magnetic drift frequency becomes increasingly large and can compete with the linear drive $\omega_{*e} \eta_e$, to allow the toroidal branch to become unstable. Toroidal modes, with $\mathbf{ v}_{Me} \cdot \mathbf{  k}_{\perp} \sim \omega_{*e} \eta_e$, are therefore possible because the competition between the slab and toroidal modes has a $\mathbf{k}_{\perp}$ dependence, which arises from the fact that $\mathbf{ v}_{Me} \cdot \mathbf{  k}_{\perp}$ depends on both $k_x$ and $k_y$, whereas $\omega_{*e} $ only depends on $k_y$. For convenience, we define the radial component of $\mathbf{k}_{\perp} $ in \Cref{eq:kperpxywithshear} as
\begin{equation}
K_{x} = k_x - k_y \bigg{(} \hat{s} \theta - \frac{r}{q} \frac{\partial \nu}{\partial r}
\label{eq:effectivewavenumber}
\bigg{)}.
\end{equation}
We now show that toroidal ETG modes with $k_{\perp} \sim K_x \gg k_y$ can indeed compete with the slab ETG at sufficiently small $k_y \rho_i $. Motivated by the eigenmodes in \Cref{fig:electrostaticspectra} that are localized far along a field line, we will make $K_x$ large by taking $\hat{s} \theta \gg k_x/k_y = \hat{s} \theta_0$ and $\hat{s} \theta \gg (r/q) \partial \nu / \partial r$. Thus, for $\hat{s} \theta$ large, we find 
\begin{equation}
k_{\perp} \sim K_{x} \sim k_y \hat{s} \theta.
\label{eq:kperpscalingbasic}
\end{equation}
\textcolor{black}{When we compare the size of $\hat{s} \theta$ to other terms, we are actually comparing $|\hat{s} \theta|$; for ease of notation, we will drop the absolute value brackets, but will continue to compare the absolute value. According to \Cref{eq:kperpscalingbasic},} for $\hat{s} \theta \gg 1$, the magnetic drift term that drives toroidal ETG can become comparable to the drive term,
\begin{equation}
\frac{\omega_{*e} \eta_e }{\mathbf{ v}_{Me} \cdot \mathbf{  k}_{\perp}} \sim \frac{k_y}{k_{\perp}} \frac{R_0}{L_{Te}} \sim \frac{1}{\hat{s} \theta} \frac{R_0}{L_{Te}} \sim 1.
\label{eq:sizeofomsteetaetoomegakappae}
\label{eq:shatrlte}
\end{equation}
Thus, for sufficiently small $k_x$, the toroidal mode must be driven far along the field line, 
\begin{equation}
\hat{s} \theta \sim \frac{R_0}{L_{Te}} \gg 1.
\label{eq:shatthetaRLTerln}
\end{equation}
Through detailed analysis in later sections, we will indeed see that this explains the toroidal ETG modes, which are often unstable at large distances along the field line (see \Cref{fig:electrostaticspectra}). Recall that here $\theta$ is the ballooning angle, which has a range $-\infty<\theta<\infty$.

When \Cref{eq:sizeofomsteetaetoomegakappae} is satisfied, we will demonstrate with a local gyrokinetic dispersion relation in \Cref{subsec:dispersionrelation} that the toroidal ETG growth rate becomes comparable to the slab ETG growth rate. This would seem to suggest that toroidal ETG exists for all $k_y$. However, for large $k_y$ and small $k_x$, $k_{\perp} \rho_e \sim \hat{s} \theta k_y \rho_e $ becomes so large that finite Larmor radius (FLR) effects from the electron gyromotion become important. Thus, if $R_0/L_{Te} \gg 1$ and $\hat{s} \theta \gg 1$, for strongly driven toroidal ETG, $K_x$ has a maximum of the order of
\begin{equation}
K_x \rho_e \sim \hat{s} \theta k_y \rho_e \sim 1.
\label{eq:effectiveKXsize}
\end{equation}
If $K_x \rho_e $ is much larger than in \Cref{eq:effectiveKXsize}, the growth rate will be strongly electron FLR damped. Motivated by \Cref{eq:effectiveKXsize}, for a toroidal mode we expect ion FLR damping to be very strong at $k_y \rho_e \ll 1$ with $k_{\perp} \rho_e \sim 1$. Thus, our analytic treatment of toroidal ETG will assume $h_i =  0$ because $|J_0 (k_{\perp} \rho_i)| \ll 1$.
Using \Cref{eq:shatrlte,eq:effectiveKXsize}, we obtain a scale for $k_y$,
\begin{equation}
k_y \rho_e \sim \frac{L_{Te}}{R_0}.
\label{eq:kyrhoeLTeRscale}
\end{equation}
Given that the pedestal profiles have $R_0/L_{Te} \gtrsim \rho_i / \rho_e$ in the steep pedestal regions, toroidal ETG can be unstable even at scales as large as $k_y \rho_i \lesssim 1$. Therefore, $R_0/L_{Te} \gg 1$ extends the minimum $k_y$ scale at which toroidal ETG modes can be strongly driven to ion gyroradius scales or larger.

To obtain the parallel width of a toroidal ETG mode $\Delta \theta$, we balance the parallel streaming term with the change in the magnetic drift over the mode width,
\begin{equation}
\frac{v_{te}}{qR_0} \frac{\partial h_e}{\partial \theta} \sim \Delta \theta \frac{\partial}{\partial \theta} (\mathbf{k}_{\perp} \cdot \mathbf{v}_{Me}) h_e.
\label{eq:parallellengthbalance} 
\end{equation}
This is based on the conjecture that the magnetic drift profiles limit the parallel width of the mode. The quantity $\Delta \theta$ captures the width of the mode envelope, rather than the oscillations within it, which would be captured by $k_{\parallel}$. For the Taylor expansion of the magnetic drift frequency in \Cref{eq:parallellengthbalance} to be valid, $\Delta \theta$ must be small, and as a result, any scalings that we obtain from \Cref{eq:parallellengthbalance} will only be valid as long as $\Delta \theta \ll 1$. 
Assuming that
\begin{equation}
\frac{\partial h_e}{\partial \theta} \sim \frac{h_e}{\Delta \theta}, \;\;\;\; \frac{\partial}{\partial \theta} (\mathbf{k}_{\perp} \cdot \mathbf{v}_{Me}) \sim \mathbf{k}_{\perp} \cdot \mathbf{v}_{Me},
\end{equation}
and that magnetic drifts balance the drive frequency, as in \Cref{eq:sizeofomsteetaetoomegakappae},
\begin{equation}
\mathbf{v}_{Me} \cdot \mathbf{k}_{\perp} \sim \omega_{*e} \eta_e,
\end{equation}
we obtain a scaling for the mode width,
\begin{equation}
\Delta \theta \sim \sqrt{\frac{v_{te}}{q R_0 \omega_{*e} \eta_e} } \sim \sqrt{\frac{1}{q k_y \rho_e } \frac{L_{Te}}{R_0} },
\label{eq:deltathetascaling} 
\end{equation}
where we use $\omega_{*e} \eta_e \sim k_y \rho_e v_{te} / L_{Te}$. Hence, higher values of $R_0/L_{Te}$, $k_y \rho_e $, and $q$ make the mode narrower. Using $\hat{s} \theta \sim R_0 / L_{Te}$, we obtain
\begin{equation}
\frac{\Delta \theta}{\theta}  \sim \hat{s} \sqrt{\frac{1}{q k_y \rho_e } } \left( \frac{L_{Te}}{R_0} \right)^{3/2}.
\end{equation}
In the pedestal, the quantity $\Delta \theta / \theta $ is small, whereas in the core, $\Delta \theta / \theta$ is of order unity. Results from gyrokinetic scans in $q$, $R_0 / L_{Te}$ and $k_y \rho_e$ are in fair agreement with the scalings in \Cref{eq:deltathetascaling}. We report these scans in \Cref{sec:GKsims}.

To summarize thus far, pedestal toroidal ETG --- where $R_0/L_{Te} \gg 1$ --- has a very different character to core toroidal ETG --- where $R_0/L_{Te} \sim 1$. In the pedestal, toroidal ETG can be driven strongly at wavenumbers as small as $k_y \rho_e \sim L_{Te} / R_0 \ll 1$, but with a large effective radial wavenumber $K_x \rho_e \sim 1$, due to the mode being driven far along the field line, $\hat{s} \theta \sim R_0/L_{Te} \gg 1$. For pedestal toroidal ETG, the radial component of the magnetic drift is essential for instability. In contrast, core toroidal ETG only becomes unstable at much larger poloidal wavenumbers $k_y \rho_e \sim 1$, and has a much smaller radial wavenumber $K_x \rho_e \ll 1$ due to $\theta \approx 0$. For core toroidal ETG, the in-surface poloidal magnetic drift is essential to the instability drive.

Slab ETG is also shifted to larger perpendicular scales by $R_0/L_{Te} \gg 1$. Re-examining \Cref{eq:drifttoparallelstreamelectron}, and requiring a strong slab drive,
\begin{equation}
\frac{\omega_{*e} \eta_e}{k_{\parallel} v_{te}} \sim \frac{k_y \rho_e }{k_{\parallel} R_0} \frac{R_0}{L_{Te}} \sim 1.
\end{equation}
Thus, the scale for which slab ETG can be strongly driven is
\begin{equation}
k_y \rho_e \sim k_{\parallel} R_0 \frac{L_{Te}}{R_0}. 
\label{eq:krescaleslab}
\end{equation}
We place bounds on $k_y \rho_e $ for the `pure' slab ETG branch by considering two linear effects that can constrain the parallel mode extent.  The first constraint on the slab ETG mode is that the mode is not too strongly FLR damped, which according to \Cref{eq:effectiveKXsize}, requires
\begin{equation}
\theta \lesssim \frac{1}{\hat{s}} \frac{1}{k_y \rho_e}. 
\label{eq:FLRconstrainttheta}
\end{equation}
A mode that oscillates only a few times before reaching the maximum value of $\theta$ in \Cref{eq:FLRconstrainttheta} has a parallel wavenumber $k_{\parallel} \sim k_y \rho_e \hat{s} / q R_0$. Using \Cref{eq:krescaleslab}, we find that such a mode would have $R_0/L_{Te} \sim \hat{s}/q$. Electron temperature gradients smaller than this value would be FLR damped. Since the gradients in the pedestal satisfy $R_0/L_{Te} \gg \hat{s}/q$, we conclude that the FLR damping constraint on the electron temperature gradient for the slab ETG mode is irrelevant in pedestals.

The second constraint on the slab ETG mode determines how far the mode can extend in the parallel direction while still retaining a parallel streaming frequency that is faster than the magnetic drift frequency. From \Cref{eq:shatrlte}, the largest $\theta$ value a mode can have before $\mathbf{v}_{Me} \cdot \mathbf{k}_{\perp}$ and $\omega_{*e} \eta_e$ become comparable is
\begin{equation}
\theta \lesssim \frac{1}{\hat{s}} \frac{R_0}{L_{Te}}.
\end{equation}
A mode that oscillates only a few times before reaching this value of $\theta$ has a parallel wavenumber of order
\begin{equation}
k_{\parallel} \sim \frac{\hat{s}}{qR_0} \frac{L_{Te}}{R_0}.
\label{eq:kplldeltatheta}
\end{equation}
A slab ETG mode with such a $k_{\parallel}$ is the mode with the smallest $k_y \rho_e $ value because, for smaller values of $k_y \rho_e$, the mode would have to extend into the region of $\theta$ where the magnetic drift is large. Thus, due to the magnetic drift condition, slab ETG modes must satisfy
\begin{equation}
k_y \rho_e \gtrsim \frac{\hat{s}}{q} \left( \frac{L_{Te}}{R_0} \right)^2.
\label{eq:magneticdriftconstraintkre}
\end{equation}
\textcolor{black}{Then}, for a fast-growing `pure' slab ETG mode, we require
\begin{equation}
\frac{\hat{s}}{q} \left( \frac{L_{Te}}{R_0} \right)^2 \lesssim  k_y \rho_e \lesssim 1.  
\end{equation}
Even though our simple estimates suggest that slab ETG modes can grow for wavenumbers as small as $k_y \rho_e \sim (\hat{s}/q) (L_{Te}/R_0)^2 \sim 1/30000$, we should point out that kinetic ion physics is important at such large scales, and hence the slab ETG will be modified at these very long wavelengths.

In principle, the above arguments are also valid for toroidal and slab ITG in the collisionless limit with identical gradients. However, in the JET pedestal equilibrium we have studied, $R_0/L_{Te} > R_0/L_{Ti}$, which causes the ITG growth rates to decrease substantially. Furthermore, in the pedestal the electrons are sufficiently collisional to be non-adiabatic on ITG timescales; as we will show in \Cref{sec:ITG}, these electron collisions also decrease the ITG growth rate. Indeed, we will see that the less steep measured ion temperature gradients and collisions result in ITG being the subdominant mode at all scales. For $k_y \rho_i \lesssim 1$, ITG is likely stable, and hence we do not expect ITG to cause significant transport in the equilibrium and radial location studied in this paper. For other JET pedestal equilibria that we studied in less detail, it was also true that $R_0/L_{Te} > R_0 / L_{Ti}$ in the steep gradient region; these equilibria had qualitatively similar growth rate spectra to the equilibrium studied in this paper (see \textcolor{black}{ \cref{app:dischargeparams}}).

We now proceed to obtain an ETG dispersion relation using the approximations in the previous sections. Its solutions will provide useful insights on toroidal ETG stability, which will be used heavily in subsequent sections. 

\subsection{ETG Dispersion Relation} \label{subsec:dispersionrelation}

\begin{figure}[t]
\centering
    \includegraphics[width=0.55\textwidth]{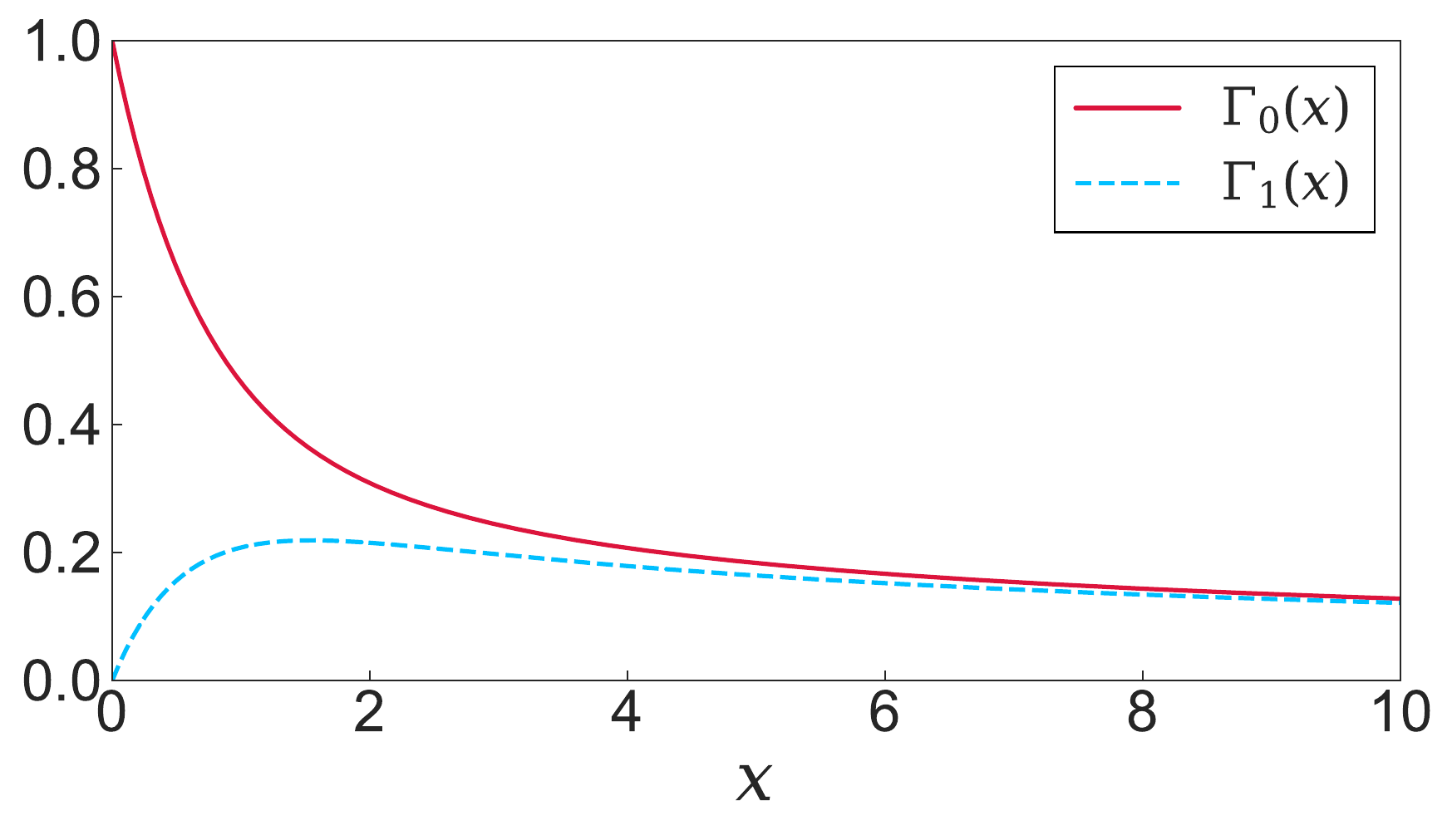}
 \caption{The functions $\Gamma_0$ and $\Gamma_1$ that appear in \Cref{eq:newDR}.}
 \label{fig:gammafunsreminder}
\end{figure}

Formally solving \Cref{eq:GKgeneralFourierAnalyzednotkpll} for $h_s$ gives
\begin{equation}
h_s = \frac{- \arc{ \omega}_s + \arc{ \omega}_{*s} \bigg{[} 1 + \eta_s \left( \hat{v}_{\parallel }^2 + \hat{v}_{\perp}^2 - 3/2 \right) \bigg{]} }{- \arc{ \omega}_s + \arc{ k}_{\parallel s} \hat{v}_{\parallel } + \sigma \hat{v}_{\parallel}^2 + \arc{\omega}_{\nabla B s} \hat{v}_{\perp}^2/2} \frac{Z_s e}{T_{0s}} \phi ^{tb} _1  F_{Ms} J_0 \left( \sqrt{2 b_s} \hat{v}_{\perp} \right),
\label{eq:hsomegakappanormalized}
\end{equation}
where the parallel wavenumber is the operator
\begin{equation}
i k_{\parallel} h_s \equiv \hat{\mathbf{ b} } \cdot \nabla h_s,
\end{equation}
and we define $b_s$ and $\hat{v}$ as
\begin{equation}
b_s = \frac{k_{\perp}^2 T_{0s}}{m_s \Omega_s^2}, \;\;\;\;\; \hat{v} = \frac{v}{v_{ts}}.  
\end{equation}
We have non-dimensionalized quantities using the modulus of the curvature magnetic drift frequency $\omega_{\kappa s}$,
\begin{equation}
\eqalign{
\sigma \equiv \frac{\omega_{\kappa s}}{|\omega_{\kappa s} |}, \;\; \arc{\omega} \equiv \frac{\omega }{|\omega_{\kappa s} |}, \;\; \arc{\omega}_{\nabla B s} \equiv \frac{\omega_{\nabla B s} }{|\omega_{\kappa s} |}, \;\; \arc{\omega}_{*s} \equiv \frac{\omega_{*s}}{|\omega_{\kappa s} |}, \;\; \arc{k }_{\parallel} \equiv \frac{k_{\parallel} v_{ts}}{|\omega_{\kappa s} |},
}
\end{equation}
where \begin{equation}
\fl \omega_{\kappa s} \equiv \frac{v_{ts}^2 \mathbf{ k}_{\perp}}{\Omega_s} \cdot \left(\hat{\mathbf{b} } \times \left( \nabla \ln B + \frac{4\pi}{B^2} \frac{\partial p_0}{\partial r} \nabla r \right) \right), \;\;\; \omega_{\nabla B s} \equiv \frac{v_{ts}^2 \mathbf{ k}_{\perp}}{\Omega_s} \cdot (\hat{\mathbf{b} } \times  \nabla \ln B ).
\end{equation}
We write the total magnetic drift frequency as
\begin{equation}
\mathbf{v}_{Ms} \cdot \mathbf{k}_{\perp} = \omega_{\kappa s} \hat{v}_{\parallel}^2 + \omega_{\nabla B s} \frac{\hat{v}_{\perp}^2}{2}. 
\end{equation}
It is important to note that \Cref{eq:hsomegakappanormalized} is valid for any value of $\theta_0$, since in this work we are paying particular attention to the radial component of \textcolor{black}{$\mathbf{k}_{\perp}$} (see \Cref{eq:kperpxywithshear}) due to its importance for the toroidal ETG instability in steep temperature gradient regions. Thus, $b_s$, $\omega_{\kappa s}$, and $\omega_{\nabla B s}$ depend on $\theta_0$; this differs from many previous works where only the $\nabla y$ component of the magnetic drift frequency was retained.

\begin{figure}[t]
\centering
    \includegraphics[width=\textwidth]{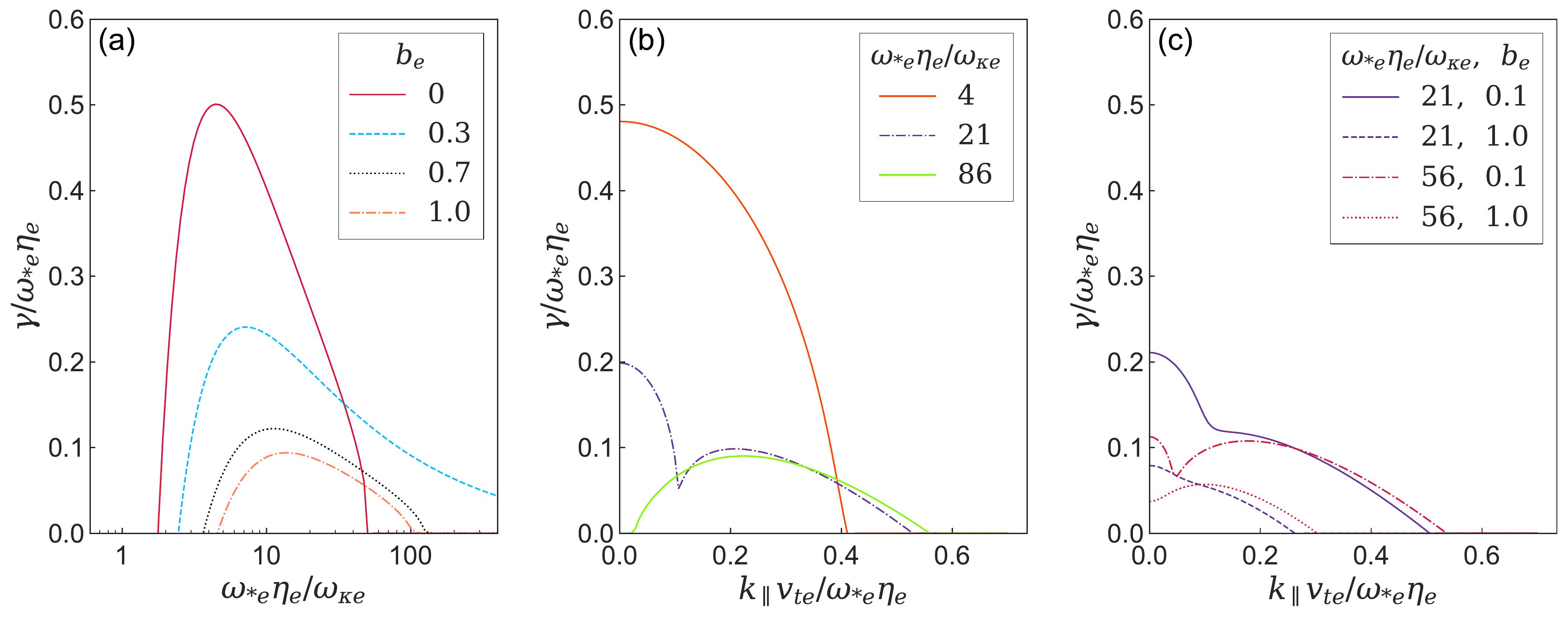}
 \caption{Solutions to \Cref{eq:toroidalETGsolved} with $\eta_e = 4.28$. (a): \textcolor{black}{G}rowth rates for different $\omega_{*e} \eta_e$ and $b_e$ with $k_{\parallel } =0$. (b): \textcolor{black}{G}rowth rates versus $k_{\parallel }$ for different values of $\omega_{*e} / \omega_{\kappa e}$ with $b_e = 0$ and $\omega_{\kappa e} > 0$. (c): \textcolor{black}{G}rowth rates \textcolor{black}{versus} $k_{\parallel }$ for different values of $\omega_{*e} / \omega_{\kappa e}$ and $b_e$. Here, we set $\omega_{\kappa e} = \omega_{\nabla B e}$. \textcolor{black}{In (a), we only plot the growth rate for $\omega_{*e} \eta_e / \omega_{\kappa e} > 0$ because we find that all solutions are damped for $\omega_{*e} \eta_e / \omega_{\kappa e} < 0$.}}
\label{fig:toroidaltoslab}
\end{figure}

As a simplified model, we will take $k_{\parallel}$ to be a number. We obtain the ETG dispersion relation by substituting \Cref{eq:hsomegakappanormalized} into quasineutrality, as demonstrated in \cref{app:disprlnfull}. For a single ion species, this gives
\begin{equation}
\frac{T_{0e}}{T_{0i}} Z_i + 1 - \sum_s D_s = 0,
\label{eq:localdispreln}
\end{equation}
where $D_s$ is given by
\begin{equation}
\fl
\eqalign{
D_s = & i Z_s^2 \frac{T_{0e} n_{0s}}{T_{0s} n_{0e}} \int_0^{\infty} d \lambda \frac{ \Gamma_0(\hat{b}_s^{\sigma}) }{(1 + i \sigma \lambda)^{1/2}} \frac{ 1}{(1 + i \arc{\omega}_{\nabla B s} \lambda / 2)} \exp \bigg{(} i \lambda \arc{\omega} - \frac{(\lambda \arc{k }_{\parallel})^2}{4(1 + i \sigma \lambda)} \bigg{)}  
\\ & \times \Bigg{[} -\arc{\omega}
 + \arc {\omega}_{*s}  \Bigg{(} 1 + \eta_s \bigg{\{} \frac{1}{1 + i \arc{\omega}_{\nabla B s} \lambda / 2} - \frac{3}{2}
\\ & \;\;\;\;\;\; + \frac{2(1 + i \sigma \lambda) - (\arc{k}_{\parallel}\lambda)^2}{4 (1+i \sigma \lambda)^2} - \hat{b_s^{\sigma}} \frac{1-\Gamma_1(\hat{b}_s^{\sigma})/\Gamma_0(\hat{b}_s^{\sigma})}{1 + i \arc{\omega}_{\nabla B s} \lambda / 2}  \bigg{\}} \Bigg{)} \Bigg{]}.
}
\label{eq:newDR}
\end{equation}
The quantities $\Gamma_{\nu}$ and $\hat{b}_s^{\sigma}$ are defined as
\begin{equation}
\Gamma_{\nu} (x) = I_{\nu} (x) \exp(-x), \;\;\;\; \hat{b}^{\sigma}_s \equiv \frac{b_s}{1 + i \arc{\omega}_{\nabla B s} \lambda/2},
\end{equation}
where $I_{\nu}$ is a modified Bessel function of the first kind. We plot $\Gamma_0$ and $\Gamma_1$ in \Cref{fig:gammafunsreminder}; the function $\Gamma_0$ will be used extensively in this work.

We have numerically solved \Cref{eq:localdispreln} in the adiabatic ion limit, $h_i = 0$,
\begin{equation}
\frac{T_{0e}}{T_{0i}} Z_i + 1 - D_e = 0,
\label{eq:toroidalETGsolved}
\end{equation}
which is justifed by $k_{\perp} \rho_i \gg 1$. For information on the numerical techniques used to solve \Cref{eq:toroidalETGsolved}, refer to \cref{app:disprlnfull}. In \Cref{fig:toroidaltoslab}, we solve \Cref{eq:toroidalETGsolved}, performing a scan in $\omega_{*e} \eta_e / \omega_{\kappa e}$ and $k_{\parallel} v_{te} / \omega_{*e} \eta_e$. Note that while for \Cref{fig:toroidaltoslab} we have set $\omega_{\kappa e} = \omega_{\nabla B e}$, when we solve \Cref{eq:toroidalETGsolved} with the geometry for the discharge 92174 in forthcoming sections, we use the correct values of $\omega_{\kappa e}$ and $\omega_{\nabla B e}$ (for example, see \Cref{fig:kymaxmovement,fig:omegaMeflip,fig:growthversussimulation}). For the toroidal ETG mode, we observe two stability limits in $\omega_{*e} \eta_e / \omega_{\kappa e}$. \Cref{fig:toroidaltoslab}(a) shows that for $b_e = 0$, toroidal ETG instability only occurs when $1.4 \lesssim \omega_{*e} \eta_e / \omega_{\kappa e} \lesssim 42$, and we found no instability when $\omega_{*e} \eta_e / \omega_{\kappa e} < 0$.

We observe in \Cref{fig:toroidaltoslab}(b) and (c) that increasing $k_{\parallel} $ causes the ETG instability to transition from the toroidal ETG branch to the slab ETG branch for the values of $\omega_{*e} \eta_e / \omega_{\kappa e}$ where the toroidal mode is unstable. Generally, increasing $b_e$ strongly decreases the growth rate for both the toroidal and slab branches, although small increasing values of $b_e$ can increase the growth rate, shown by comparing the $\omega_{*e} \eta_e / \omega_{\kappa e} = 21$ values in \Cref{fig:toroidaltoslab}(b) and (c).

The $h_i = 0$ limit is generally an accurate description of toroidal and slab ETG instability in the JET pedestal discharges we analyzed, as will be described in \Cref{sec:GKsims}. This is not surprising given that for the toroidal ETG instability we require $K_x \rho_e \sim 1$, which means that $h_i \approx 0$ because of the large argument of $J_0$ (see \Cref{eq:hsomegakappanormalized}). For the fastest growing slab ETG instability we usually find that $k_y \rho_i \gg 1$, again resulting in $h_i \approx 0$. However, the $h_i = 0$ approximation might not always be justified for $k_y \rho_i \sim 1$ slab ETG instability, where FLR damping has not substantially decreased the size of the ion kinetic response.

In the next section, we proceed to use gyrokinetic simulations to study ETG stability in the pedestal. Of particular interest, consistent with the predictions of this section, we will find both toroidal and slab ETG modes at scales $k_y \rho_i \sim (\rho_i/\rho_e) L_{Te}/R_0 \lesssim 1$, and long poloidal wavelength toroidal ETG being unstable at $\hat{s} \theta \sim R_0 / L_{Te}$ (for $\theta_0 = 0$).

\begin{figure}[t]
\centering
    \includegraphics[width=\textwidth]{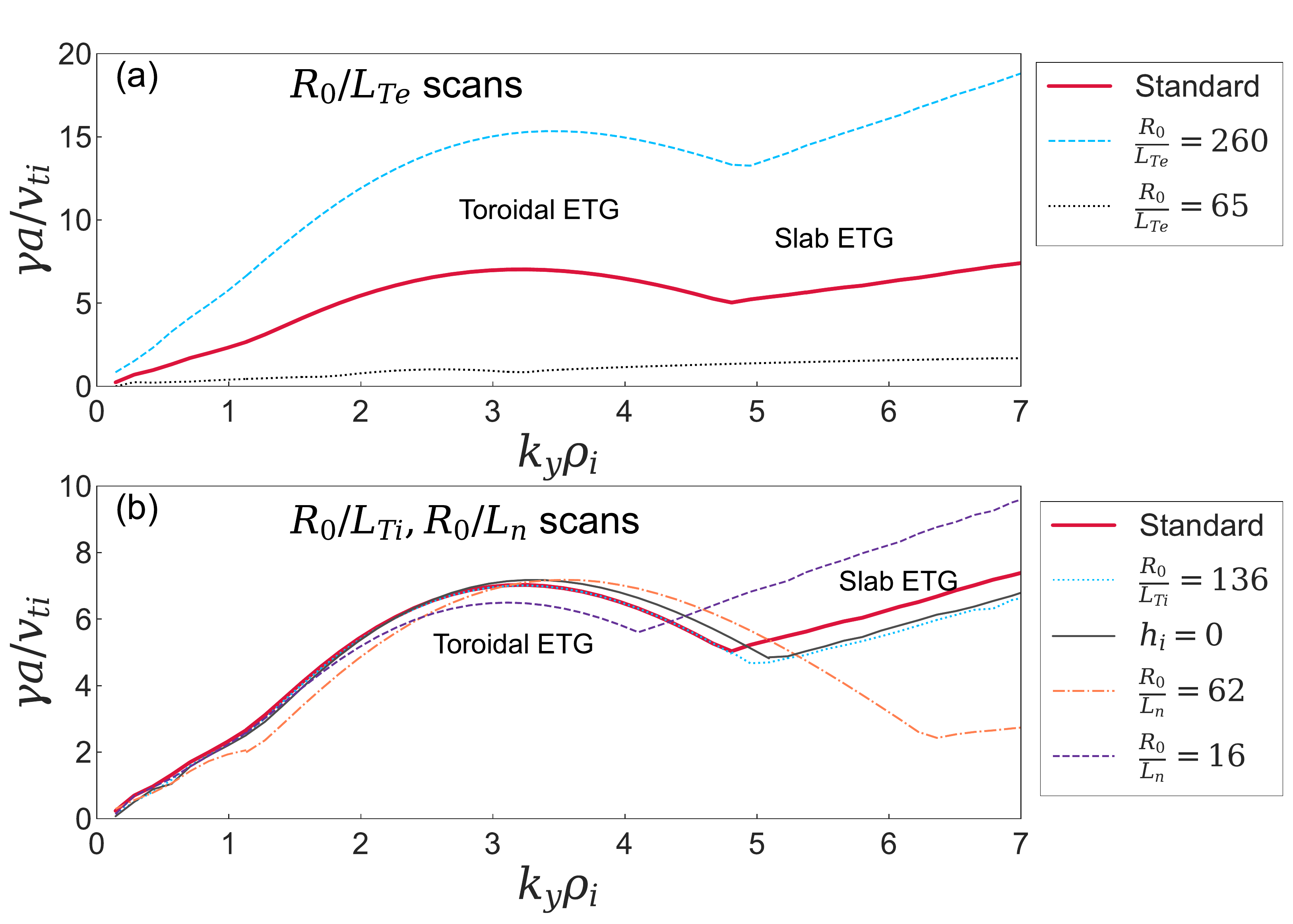}
 \caption{Electrostatic GS2 growth rates for JET shot 92174 for $0.15 \leq k_y \rho_i \leq 7.0$ and sensitivity scans, all with $\theta_0 = 0$. (a): $R_0/L_{Te}$ scans. (b): $R_0/L_{Ti}$ and $R_0/L_n$ scans. `Standard' denotes simulations performed with the following parameters: $R_0/L_{Te} = 130,\; R_0/L_{Ti} = 34,\; R_0/L_n = 31$. All of the fastest growing `Standard' modes at scales $k_y \gtrsim 0.1$ are ETG-like instabilities.}
 \label{fig:krionetofivesensitivityscan}
\end{figure}

\section{ETG Stability in JET Shot 92174} \label{sec:GKsims}

In this section, we describe ETG instability in electrostatic gyrokinetic simulations of JET shot 92174 at $r/a = 0.9743$.

The layout of this section is as follows. We first discuss the character of the toroidal and slab ETG instability in the pedestal in \Cref{subsec:toroidalETG}. In \Cref{sec:locwidth}, we describe the parallel dynamics of the toroidal ETG mode, detailing how its parallel location and mode width are determined. In \Cref{sec:effectsofkx}, the effects of a nonzero $\theta_0$ for the toroidal ETG mode are analyzed, including an estimate for the quasilinear diffusion coefficient. Then in \Cref{sec:critR0LTe}, we study the critical temperature gradient for the toroidal ETG mode described in \Cref{sec:linGKwithlargegrads}.

\subsection{Toroidal ETG Versus Slab ETG Instability} \label{subsec:toroidalETG}

\begin{figure}[t]
\centering
    \includegraphics[width=\textwidth]{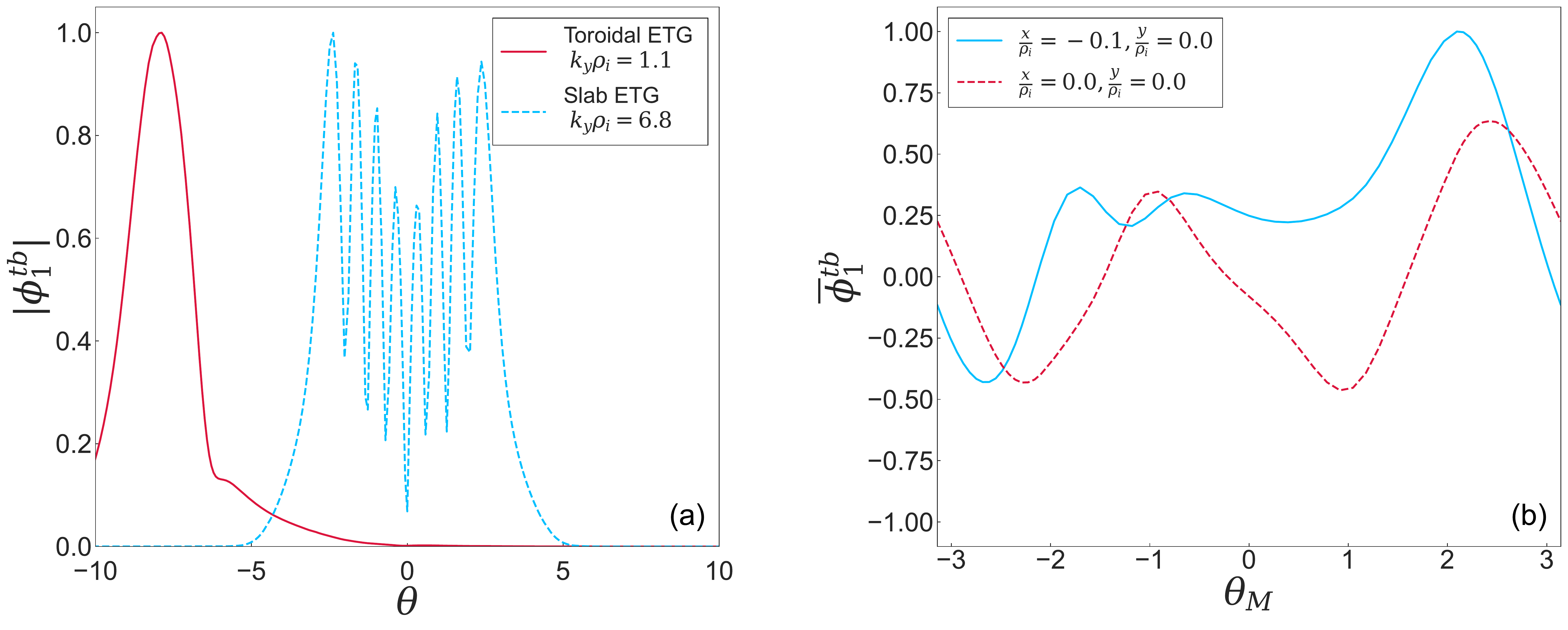}
 \caption{(a): Ballooning eigenmodes for toroidal and slab ETG in GS2 simulations. (b): Toroidal ETG eigenmodes in $\theta_M$ space with $k_y \rho_i = 1.1$, using the transformation in \Cref{eq:undoballooningtransform} at two locations: (1) $x/\rho_i = -0.1, y/\rho_i = 0.0$, and (2) $x/\rho_i = 0 , y/\rho_i = 0$. Location (1) is where the mode amplitude is maximum.}
 \label{fig:krionetofiveplot}
\end{figure}

Gyrokinetic simulations show toroidal and slab ETG instability as the fastest growing modes for $k_y \rho_i \gtrsim 0.1$ for JET shot 92174. Unlike ETG instability in the core, where the linear growth rate typically peaks at $k_y \rho_e \sim 1$, we find instances of maximum toroidal ETG growth rates at spatial scales as large as $k_y \rho_i \sim (\rho_i/\rho_e) L_{Te}/R_0 \lesssim 1$, strongly supporting the arguments in \Cref{sec:linGKwithlargegrads}. We emphasize that very similar modes have been seen in previous works \cite{Told2008,Jenko2009,Told2012,Fulton2014,Baumgaertel2011, Kotschenreuther2019}, but have not been explained until now. For $\theta_0 \neq 0$, we find toroidal ETG as the fastest growing mode at all spatial scales between $k_y \rho_i \sim 1$ and $k_y \rho_e > 1$, which we will discuss in \Cref{sec:effectsofkx}. In \Cref{fig:krionetofivesensitivityscan}, we show the growth rates of modes with $\theta_0 = 0$, where we find two dominant ETG modes: for this specific pedestal location, the toroidal ETG branch is the fastest growing mode for $1 \lesssim k_y \rho_i \lesssim 5$. Once $k_y \rho_i$ is sufficiently large ($k_y \rho_i \approx 5$)\textcolor{black}{,} the toroidal ETG is FLR damped, and the slab ETG branch grows faster. The slab ETG branch is not FLR damped as quickly as the toroidal branch because the slab branch generally satisfies $K_x \sim k_y$.

\begin{figure}[t]
\centering
    \includegraphics[width=\textwidth]{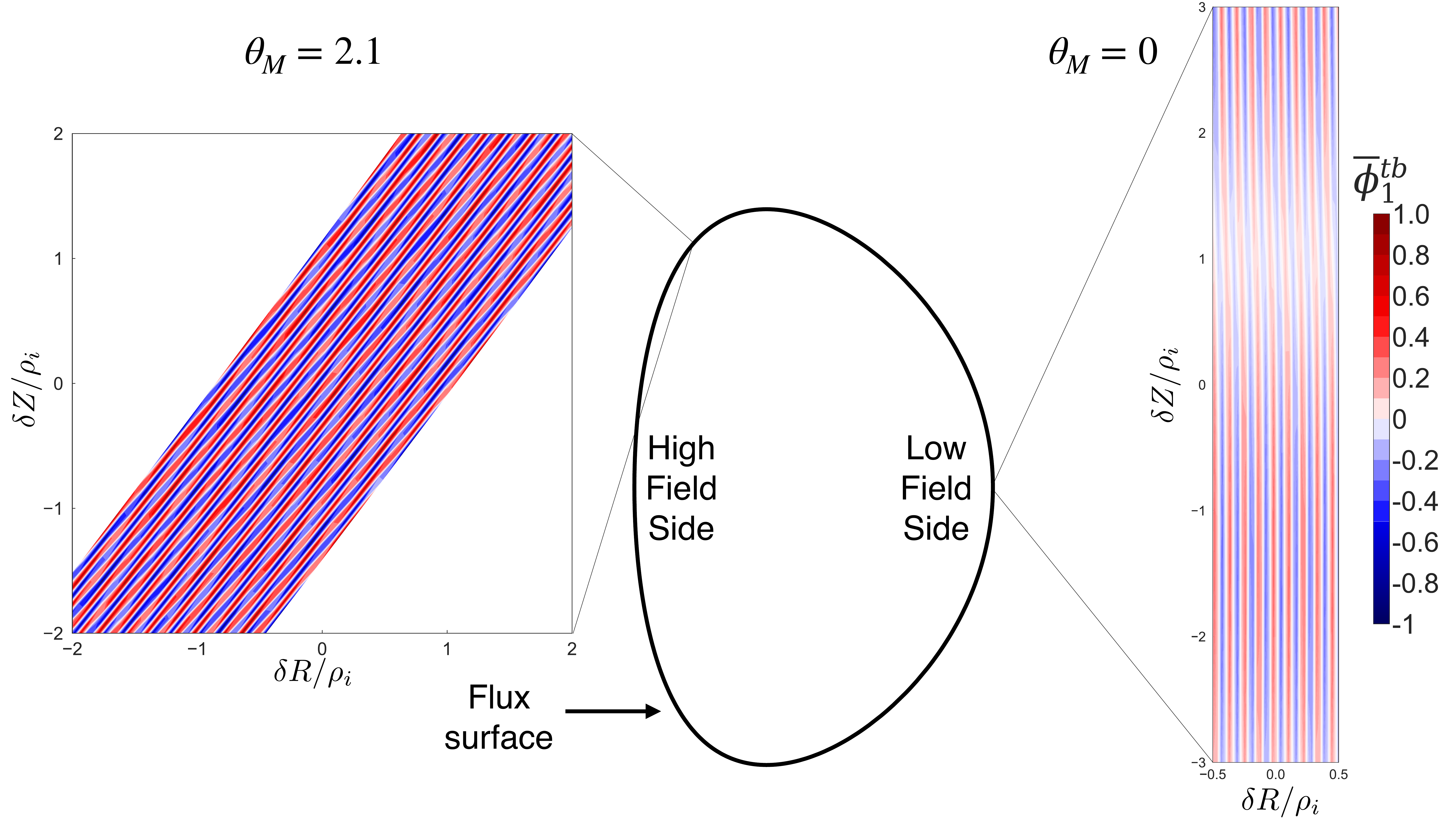}
 \caption{Real space images at the outboard midplane ($\theta_M = \vartheta = 0$), and at $\vartheta = 1.6$, $\theta_M = 2.1$, of a single toroidal ETG ballooning mode with $k_y \rho_i = 1.1$ and $\theta_0 = 0.0$ from GS2 simulations, demonstrating a relatively large radial wavenumber at both $\theta_M$ locations, and that the mode has a larger amplitude at $\theta_M = 2.1$ than at the outboard midplane. These were obtained using the transformation in \Cref{eq:undoballooningtransform}. We define the coordinates $\delta R = R - R_M(r_c, \theta_r)$ and $\delta Z = Z - Z_M(r_c, \theta_r)$, where $\theta_r = -0, 2.1$ is the Miller poloidal angle of the image. The gyroradius $\rho_i$ is evaluated on the usual $r/a = 0.9743$ flux surface \textcolor{black}{at the outboard midplane}. Both plots are normalized to the same colorbar. The maximum absolute mode amplitude at $\theta_M = 0$ is about 25\% of the mode amplitude at $\theta_M = 2.1$. The specific $\theta_M = 2.1$ location was chosen as this was the location of the maximum value of $\overline{ \phi } ^{tb} _1$, which can be seen in \Cref{fig:krionetofiveplot}(b). \textcolor{black}{The diameter of the flux surface normalized to the ion gyroradius is large: $2 r_c / \rho_i \simeq 660$. Hence, the box at $\theta_M = 2.1$ has a radial width roughly equal to 1/165 of the flux surface diameter, and the box at $\theta_M = 0.0$ has a radial width roughly equal to 1/660 of the flux surface diameter.}}
 \label{fig:krionetofiveplotrealspace}
\end{figure}

We use several criteria to distinguish between the toroidal and slab ETG modes in the pedestal. First, as predicted in \Cref{sec:linGKwithlargegrads}, toroidal ETG modes have $\Delta \theta / \theta \ll 1$, and have a $\theta$ location that satisfies $\hat{s} \theta \sim L_{Te} / R_0$ for $|\theta_0|$ sufficiently small. Parameter scans can also be used to determine whether the location along a field line of a suspected toroidal ETG mode changes as predicted by \Cref{eq:shatthetaRLTerln}. In contrast, slab ETG modes tend to have a much larger $k_{\parallel}$ (at a fixed $k_y \rho_i$), and to have eigenmodes that are centered around $\theta = 0$. In \Cref{fig:krionetofiveplot}, we show both toroidal and slab ETG eigenmodes in (a). To go from ballooning angle $\theta$ to the physical poloidal angle $\vartheta$, where $-\pi \leq \vartheta \leq \pi$, we use the ballooning transform,
\begin{equation}
\fl
\eqalign{
\overline{\phi} ^{tb}_1 (\vartheta, x&, y) = \sum_{p = -\infty}^{\infty} \phi ^{tb}_1  (\vartheta - 2\pi p) \exp \left( i k_y x \hat{s} \left(\vartheta - 2\pi p - \frac{r}{\hat{s} q} \frac{\partial \nu}{\partial r} \right) - i k_y y \right) \\ 
& + \sum_{p = -\infty}^{\infty} {\phi^{tb}_1}^{*} (\vartheta - 2\pi p) \exp \left(- i k_y x \hat{s} \left(\vartheta - 2\pi p - \frac{r}{\hat{s} q} \frac{\partial \nu}{\partial r}\right) + i k_y y \right),
}
\label{eq:undoballooningtransform}
\end{equation}
where $*$ denotes a complex conjugate. In \Cref{fig:krionetofiveplot}(b), the toroidal ETG eigenmode is plotted against the Miller angle $\theta_M$ for $x/\rho_i = 0, \; y/\rho_i = 0$ and for $x/\rho_i = -0.1, \; y/\rho_i = 0.0$. We have normalized the mode such that the maximum of $\overline{\phi } ^{tb} _1 $ is 1, and we have chosen the mode's phase such that the maximum is located at $y = 0$. The maximum value of $\overline{ \phi} ^{tb} _1$ occurs at $x/\rho_i = -0.1$. In \Cref{fig:krionetofiveplotrealspace}, we show the real space picture of the mode at the outboard midplane ($\theta_M = 0$) and where the amplitude is maximum, at $\theta_M = 2.1$. As expected, the toroidal ETG modes have $K_x \gg k_y$ at both the outboard midplane and at $\theta_M = 2.1$, and the maximum amplitude is far away from the outboard midplane. To make the plots in \Cref{fig:krionetofiveplotrealspace}, we first evaluated \Cref{eq:undoballooningtransform} for $k_y \rho_i = 1.1$ on a uniform $x,y$ grid. We then performed a change of variables from $x,y$ to $R,Z$ using the Miller formulas for $R_M$ and $Z_M$. Finally, we changed from $\vartheta$ to $\theta_M$ variables. \textcolor{black}{\Cref{fig:krionetofiveplotrealspace},  where we have plotted a toroidal ETG mode with $\theta_0 = 0$, demonstrates how the wavenumbers $K_x$ and $k_x$ can differ dramatically due to the presence of magnetic shear. At both the outboard midplane and at $\theta_M = 2.1$, this mode has $\lambda_x \simeq 0.1 \rho_i$, and so $K_x \rho_i \simeq 65$, which is consistent with the requirement that $K_x \rho_e \sim 1$ for the toroidal ETG mode. Here, $\lambda_x$ is the radial mode wavenumber. Since $k_{\perp}$, which is non-trivial (see \Cref{eq:kperpxywithshear}), enters the Bessel function arguments and not simply $k_x$ and $k_y$, the distinction between $k_x$ and $K_x$ is crucial for the character of the mode.}

\begin{figure}[t]
\centering
    \includegraphics[width=\textwidth]{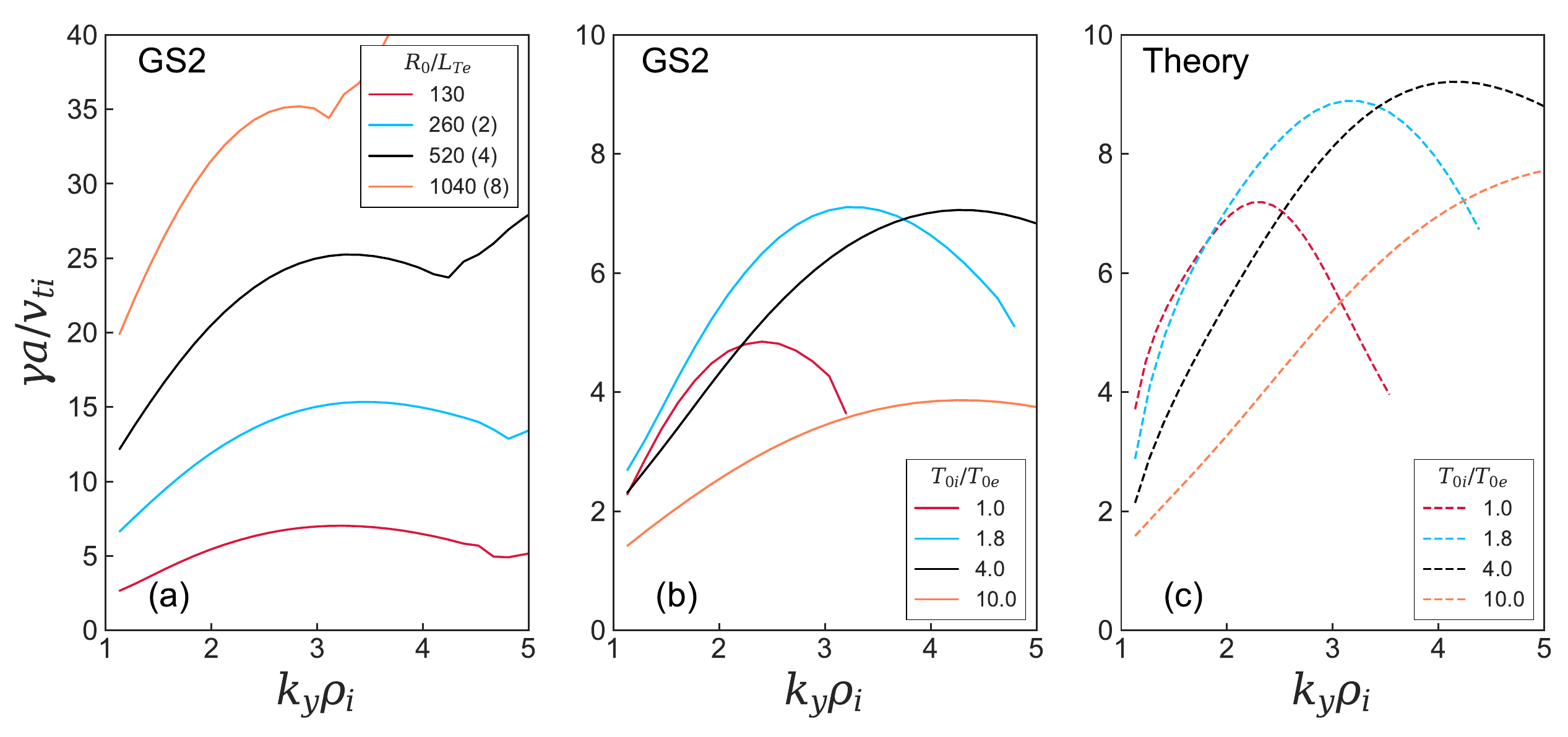}
 \caption{Growth rates. (a): GS2 scan in $R_0/L_{Te}$, (b): GS2 scan in $T_{0i}/T_{0e}$, and (c): \textcolor{black}{T}heory scan in $T_{0i}/T_{0e}$. These scans show the value of $k_y \rho_i$ for the peak growth rate of the toroidal ETG mode shifting. For $T_{0i}/T_{0e}$ scans, $T_{0i}$ was fixed and $T_{0e}$ was allowed to vary. (b): \textcolor{black}{G}rowth rates from GS2 simulations with consistent collisionality. (c): \textcolor{black}{T}he collisionless dispersion relation in \Cref{eq:toroidalETGsolved} was solved, along with a Fourier-transformed value of $k_{\parallel}$ for each $k_y \rho_i$ mode, described in \Cref{sec:locwidth}. The numbers in parentheses in the legend for (a) are the multiples of the correct $R_0/L_{Te}$ value.}
 \label{fig:kymaxmovement}
\end{figure}

To investigate the character of the toroidal and slab ETG modes, we have performed a scan in equilibrium gradients, as shown in the linear gyrokinetic spectrum in \Cref{fig:krionetofivesensitivityscan}. Our simulations indicate that the fastest growing toroidal ETG modes are driven strongly by $R_0/L_{Te}$ because they depend strongly on this parameter, as shown in \Cref{fig:krionetofivesensitivityscan}(a). Conversely, these modes are relatively insensitive to $R_0/L_n$, and do not depend on $R_0/L_{Ti}$. Modifying $R_0/L_n$ mainly affects the slab ETG growth rate, determining at which $k_y \rho_i $ it will exceed the toroidal ETG growth rate. Kinetic ion physics is usually unimportant for toroidal ETG instability because $k_{\perp} \rho_i \gg 1$. This is demonstrated by the linear spectrum for the toroidal ETG being unchanged when the non-adiabatic part of the ion distribution function is artificially set to zero, $h_i = 0$, shown \textcolor{black}{in} \Cref{fig:krionetofivesensitivityscan}(b). The simulation results in \Cref{fig:kymaxmovement} also show higher $R_0/L_{Te}$ and smaller $T_{0i}/T_{0e}$ shifting the maximum growth rate of the toroidal ETG instability to a smaller $k_y \rho_i$, as predicted by \Cref{eq:kyrhoeLTeRscale}. Unlike the wavenumber of the fastest growing modes, the size of the maximum growth rate in the range of wavelengths shown depends \textcolor{black}{on} $T_{0e}/T_{0i}$ in a non-trivial way. We show in \Cref{fig:kymaxmovement} that this dependence is consistent with a theory that we describe in \Cref{sec:locwidth}. 

\textcolor{black}{For $k_y \rho_i \gtrsim 1$, the modes are unlikely to be a trapped electron mode (TEM) since $\omega_{\mathrm{b} e} \ll \gamma_{\mathrm{ETG} }$ for $k_y \rho_i \gtrsim 1$, where $\omega_{\mathrm{b} e} = v_{te} \sqrt{r_c / q_c^2 R_c^3}$ is the electron bounce frequency and $\gamma_{\mathrm{ETG} }$ is the ETG growth rate. In this equilibrium, we find that $\omega_{\mathrm{b} e} a / v_{ti} \simeq 1.5$, which is comparable to $\gamma_{\mathrm{ETG} } a / v_{ti}$ only when $k_y \rho_i \simeq 0.5$. Furthermore, $\omega_{be}/ \nu_{ee} \simeq 1.9$, and so we expect the passing and trapped electron particle distributions to be fairly well equilibrated.}

\begin{figure}[t]
\centering
    \includegraphics[width=\textwidth]{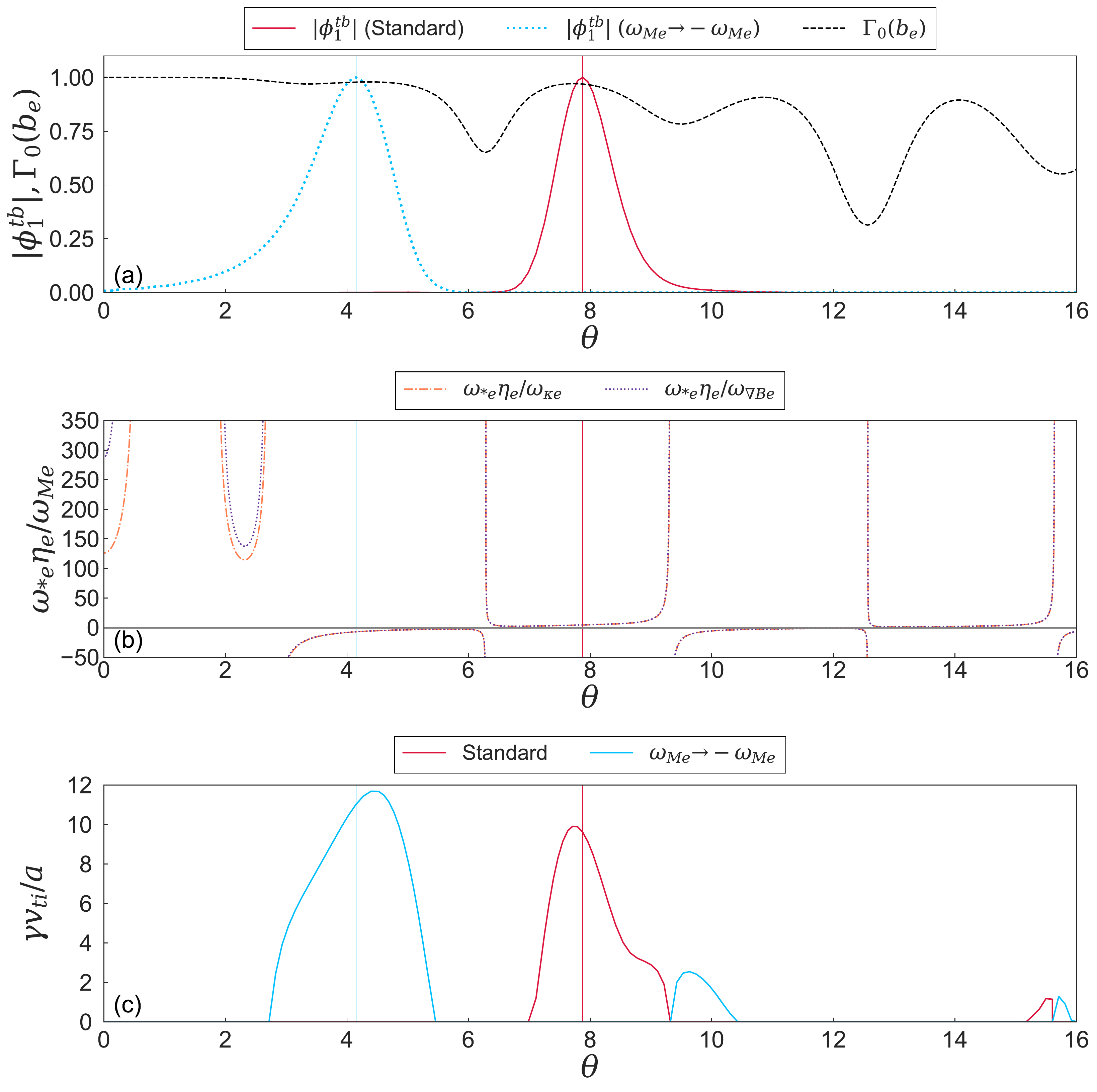}
 \caption{(a): \textcolor{black}{T}wo eigenmodes obtained from two separate GS2 simulations, and the function $\Gamma_0(b_e)$ for $k_y \rho_i = 3.4$. When $\omega_{M e} \to - \omega_{M e}$, the mode moves to a location where the sign of $\omega_{M e} $ allows instability, where $\omega_{M e}$ refers to \textcolor{black}{both} $\omega_{\kappa e}$ \textcolor{black}{and} $\omega_{\nabla B e}$. (b)\textcolor{black}{:} The quantities $\omega_{*e} \eta_e / \omega_{\kappa e}$ and $\omega_{*e} \eta_e / \omega_{\nabla B e}$. The eigenmodes in (a) have their maxima in bad curvature regions, corresponding to $\omega_{*e} \eta_e / \omega_{M e} > 0$. (c): \textcolor{black}{F}inding the growth rates for the ETG dispersion relation in \Cref{eq:toroidalETGsolved} for two signs of $\omega_{*e} \eta_e / \omega_{M e} $ in JET shot 92174. Note how the maximum growth rates in (c) roughly align with the eigenmode maxima in (a). Vertical red and blue lines denote the eigenmode location for the two signs of $\omega_{Me}$ in (a). Here, $\omega_{*e} < 0$, $\eta_e = 4.28$, $k_y \rho_i = 3.4$, $k_{\parallel} = 0$, $\theta_0 = 0$.}
 \label{fig:omegaMeflip}
\end{figure}

To understand the $\theta$ location of the toroidal ETG eigenmodes, we solve the dispersion relation in \Cref{eq:toroidalETGsolved} locally for JET shot 92174 at each value of $\theta$ by choosing $k_y$ and \textcolor{black}{ setting $k_{\parallel} = 0$}, and by using $\omega_{\kappa e}, \omega_{\nabla B e}$ and $b_e$ from the Miller equilibrium. This is an approximation that assumes the mode's growth rate is local in $\theta$. Note that $\mathbf{k}_{\perp}$ in \Cref{eq:kperpxywithshear} is a function of $\theta$. By solving the dispersion relation, we obtain a set of frequencies as a function of $\theta$. \Cref{fig:omegaMeflip}(c) shows the growth rates along $\theta$ with $k_{\parallel} = 0$ (for the present discussion, consider only the curve labeled 'Standard'; the curve labeled `$\omega_{Me} \rightarrow - \omega_{Me}$' will be discussed in \Cref{sec:locwidth}). For $\theta_0 = 0$, we find that the maximum growth rates are at $|\theta| \simeq 7.7$ with the standard sign of $\omega_{\kappa e}$ and $\omega_{\nabla B e}$. This $\theta$ location is very close to the $\theta$ where GS2 toroidal ETG eigenmodes have their maximum amplitude, as shown by comparison of \Cref{fig:omegaMeflip}(a) and (c). Therefore, the parallel location of the toroidal ETG is fairly well described by our model.

\begin{figure}[t]
\centering
    \includegraphics[width=0.9\textwidth]{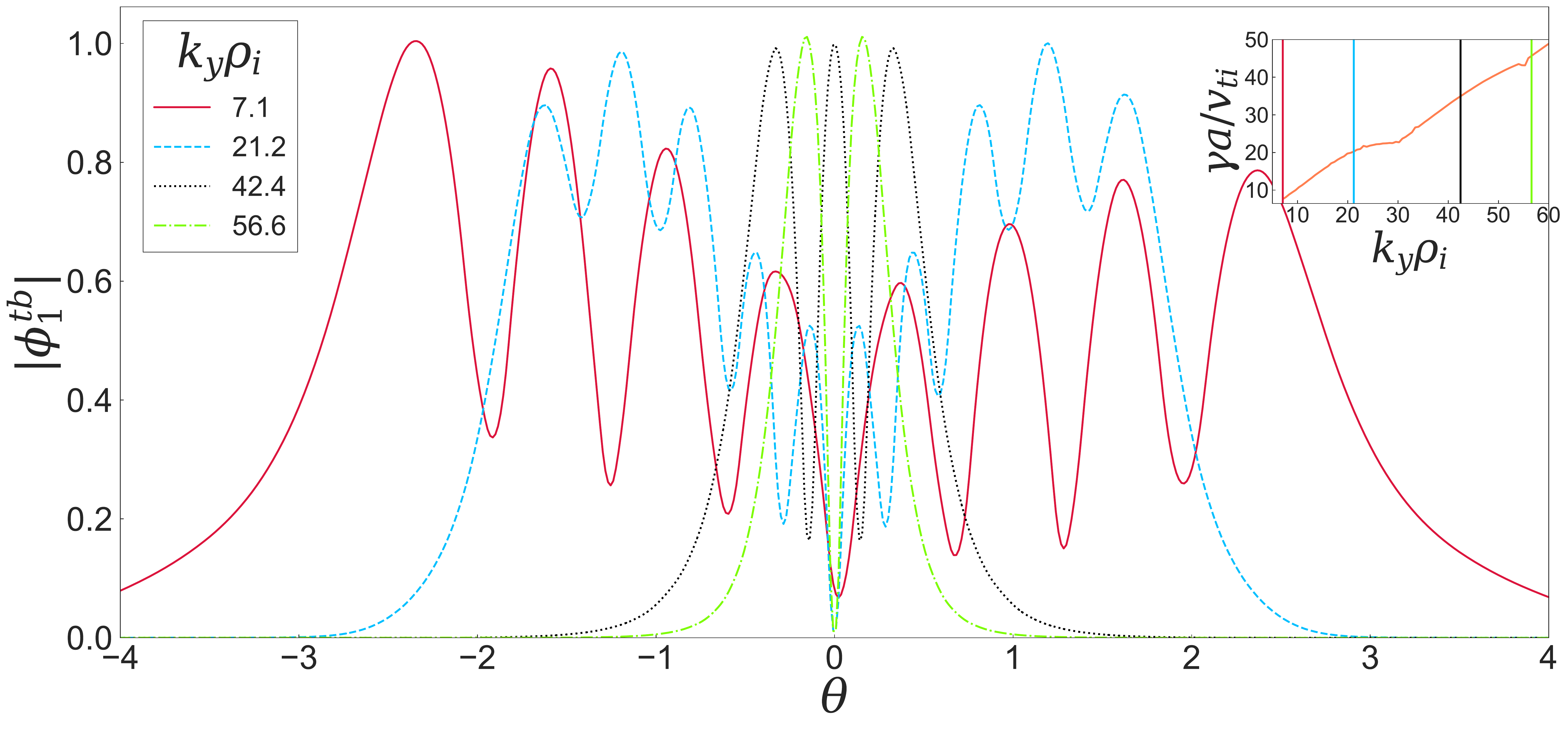}
 \label{fig:torEmodes}
 \caption{Electrostatic slab eigenmodes from GS2 for $ k_y \rho_i > 7.0$ instabilities at $\theta_0 = 0.05$. The corresponding linear growth rates are shown in the inset.}
 \label{fig:kribiggerthanfive}
\end{figure}

One prediction of \Cref{sec:linGKwithlargegrads} was that the toroidal ETG mode is driven most strongly at $\hat{s} \theta \gg 1$ when $R_0/L_{Te} \gg 1$. This causes the $k_y \hat{s} \theta \nabla x$ term in $\mathbf{k}_{\perp} $ in \Cref{eq:kperpxywithshear} to become particularly large. In \Cref{fig:krionetofiveplot}, we show that the toroidal ETG eigenmodes are indeed driven at $\hat{s} \theta \gg 1$. As an experiment, we set the $k_y \hat{s} \theta \nabla x$ component of $\mathbf{v}_{Me}$ to zero. As expected, the toroidal ETG mode was not driven, and slab ETG was the fastest growing mode.

In JET shot 92174,  slab ETG instability is the fastest growing mode for $k_y \rho_i \gtrsim 5$ when $\theta_0 = 0$ --- however, the `slab' ETG we observe is not always the conventional slab ETG with $\omega_{\kappa e} = \omega_{\nabla B e} = 0$. By artificially turning the magnetic drift off in gyrokinetic simulations, we observed that the slab ETG growth rate was reduced by factors of order unity. As shown in \Cref{fig:kribiggerthanfive}, the slab ETG eigenmodes have quite a wide $\theta$ extent, especially for smaller $k_y \rho_i$ where FLR effects are less strong, and hence the magnetic drift, which increases for increasing $\theta$, can have a strong impact on the character of the slab ETG in the pedestal. As $k_y \rho_i $ increases, FLR effects become stronger and the slab ETG eigenmode becomes more localized near $\theta = 0$. Hence, when we refer to the `slab' ETG in the pedestal simulations described in this paper, we refer to the modes with a $k_{\parallel}$ much larger than the toroidal ETG, but also sometimes with a significant magnetic drift contribution.

\begin{figure}[t]
\centering
    \includegraphics[width=0.9\textwidth]{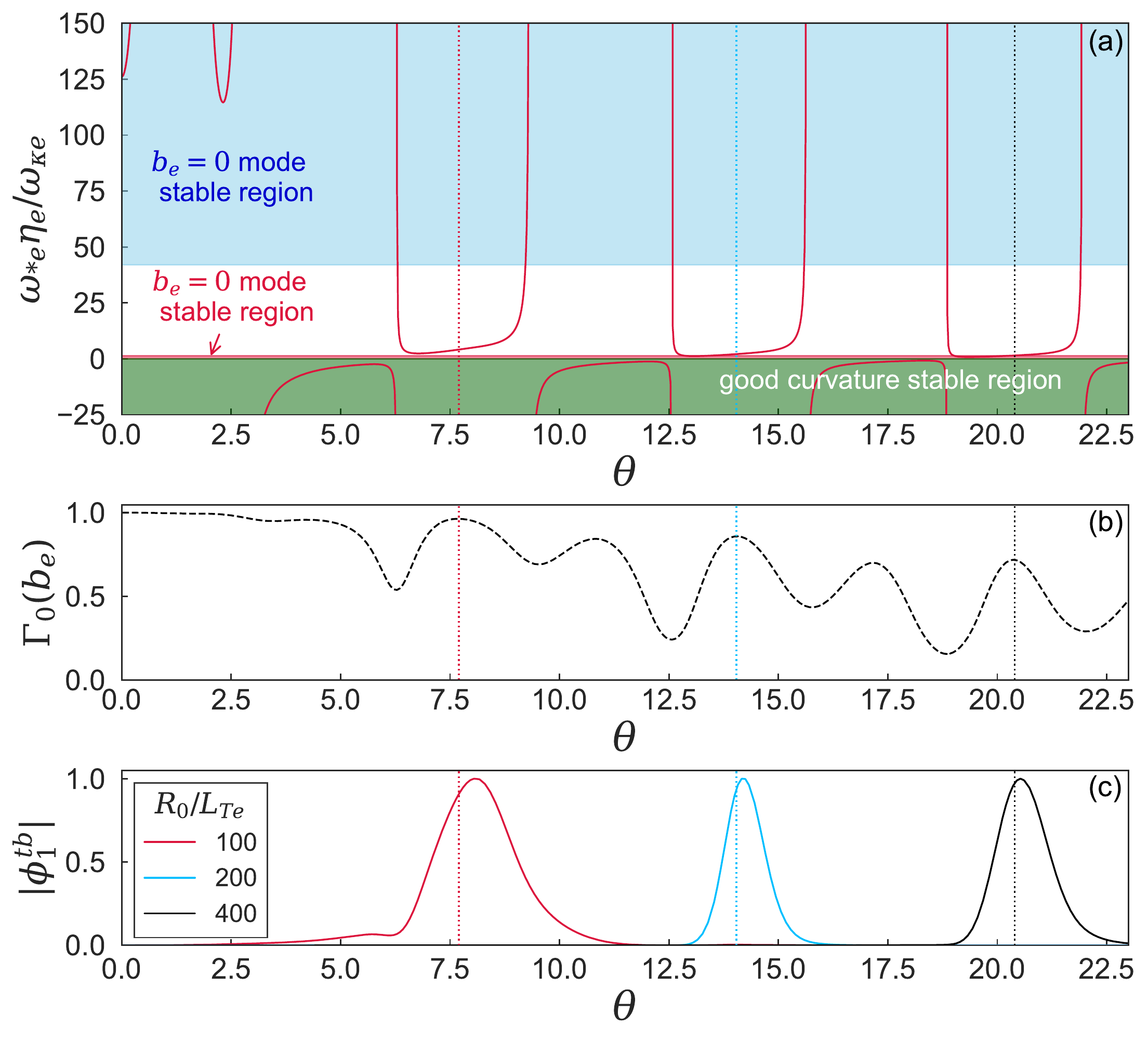}
 \caption{A stability plot for the toroidal ETG mode, combining theory and GS2 simulations. For (a), the small red $b_e = 0$ stable region corresponding to $0 < \omega_{*e} \eta_e / \omega_{\kappa e}  \lesssim 1.8$, is obtained from \Cref{fig:toroidaltoslab}. The blue $b_e = 0$ stable region is also obtained from \Cref{fig:toroidaltoslab}, and corresponds to $\omega_{*e} \eta_e / \omega_{\kappa e} \gtrsim 42$. This is valid for $\theta_0 = 0$ and $k_y \rho_i = 1.1$. (b): \textcolor{black}{Q}uantity $\Gamma_0(b_e)$ versus $\theta$ for $k_y \rho_i = 1.1$. (c): \textcolor{black}{T}he associated eigenmodes from GS2 with different temperature gradients, demonstrating that these modes are centered close to local maxima in $\Gamma_0(b_e)$, and that increasing $R_0/L_{Te}$ moves the mode to larger $\hat{s} \theta$, predicted in \Cref{eq:kyrhoeLTeRscale}. Only for (c), we artificially lowered $\hat{s} \to 1.68$ to make the mode more mobile in $\theta$. Dashed vertical lines show the local maxima of $\Gamma_0(b_e)$ in bad curvature regions.}
 \label{fig:generalstabilityplot}
\end{figure}

The toroidal ETG modes are not affected by kinetic ion physics due to their large radial wavenumber $K_x \rho_i \gg 1$, but the ions can modify the slab ETG modes slightly when $k_y \rho_i \sim 1$, as we demonstrate in \Cref{fig:krionetofivesensitivityscan}, where we show results with the full ion kinetic response and with $h_i = 0$. This is consistent with the fact that slab modes with $k_y \rho_i \sim 1$ have $K_x \rho_i \sim 1$. We have checked that $h_i$ becomes unimportant at larger values of $k_y \rho_i$.

Note that the slab ETG modes in \Cref{fig:kribiggerthanfive} are asymmetric. This asymmetry is not a result of our choice of $\theta_0$ because we observe it in modes with $\theta_0 = 0$. Due to the symmetry of the \textcolor{black}{linear} gyrokinetic equation described in \cite{Peeters2005}, for $\theta_0 = 0$, if one obtains an asymmetric mode, there must be two modes with opposing asymmetry that grow at the same rate. We have run our simulations with a small value of $\theta_0$ to avoid getting a linear combination of these two modes --- the final result would depend on the initial conditions in this case.

Thus far, using the method described above to solve the dispersion relation in \Cref{eq:toroidalETGsolved}, we found we could predict the parallel location of the toroidal ETG modes. We next describe the physics that determines the parallel location and width of the toroidal ETG mode in more detail.

\subsection{Location And Width Of The Toroidal ETG Mode} \label{sec:locwidth}

We now discuss the parallel location and width of the toroidal ETG mode. The parallel location of the toroidal ETG mode is subject to four main constraints:
\begin{enumerate}
\item \textbf{The mode can only be driven in bad curvature regions,} $ \omega_{*e} \eta_e / \omega_{\kappa e} > 0$, which eliminates roughly half of the parallel domain.
\item \textbf{The mode is only unstable when} $\mathbf{  A > \omega_{*e} \eta_e /\omega_{\kappa e} > C}$. According to the results in \Cref{fig:toroidaltoslab}(a), for toroidal ETG instability the value of $\omega_{*e} \eta_e /\omega_{\kappa e}$ must be above some critical value $C$ for instability, but not larger than another critical value $A$. Consistent with \Cref{fig:toroidaltoslab}(a), we observe that no toroidal ETG modes with $\theta_0 = 0$ can exist at $|\theta| \lesssim 6$; this is because $\omega_{*e} \eta_e / \omega_{\kappa e} $ is too large and the bad curvature region is too narrow, as shown in \Cref{fig:generalstabilityplot}(a) (note that for smaller values of $R_0/L_{Te}$, the $\theta_0 = 0$ toroidal ETG mode \textit{can} have its maximum amplitude at $|\theta| \lesssim 6$ because $\omega_{*e} \eta_e/ \omega_{\kappa e} $ is smaller --- see \Cref{sec:critR0LTe}). Note that we discuss `good' and `bad' curvature using the quantity $\omega_{*e} \eta_e / \omega_{\kappa e} $ rather than $\omega_{*e} \eta_e / \omega_{\nabla B e} $ because in the regions where the toroidal ETG mode is typically most unstable (at large $|\theta|$), $\omega_{\kappa e}/ \omega_{\nabla B e} \simeq 1$ (see \Cref{fig:omegaMeflip}(b), for example). There are important exceptions, which occur for $\theta_0 \neq 0$ with larger values of $k_y \rho_i$, which we discuss briefly in \Cref{sec:critR0LTe}.
\item \textbf{The parallel extent of bad curvature regions must be sufficiently wide.} We require that the `bad curvature' regions not be too narrow in the parallel direction; if this is the case, the mode acquires a large value of $k_{\parallel}$ and becomes damped.
\item \textbf{The mode maximum is close to a local maximum in $\Gamma_0 (b_e)$.} The maximum amplitude for the fastest growing toroidal ETG mode (at a given $k_y \rho_i$) is usually centered close to a local maximum in $\Gamma_0 (b_e)$ (or equivalently a local minimum in $b_e$) to limit FLR damping. We choose to plot the quantity $\Gamma_0 (b_e)$ rather than $b_e$ to demonstrate the importance of FLR damping at different $\theta$ locations. This is because $\Gamma_0 (b_e) \in [0,1]$, and therefore it is easier to convey the size of FLR damping, whereas $b_e$ is unbounded and can become extremely large. Furthermore, the term $\Gamma_0 (b_e)$ appears directly in the dispersion relation in \Cref{eq:newDR}, and thus is a good measure of the size of FLR effects.
\end{enumerate}

\begin{figure}[t]
\centering
    \includegraphics[width=\textwidth]{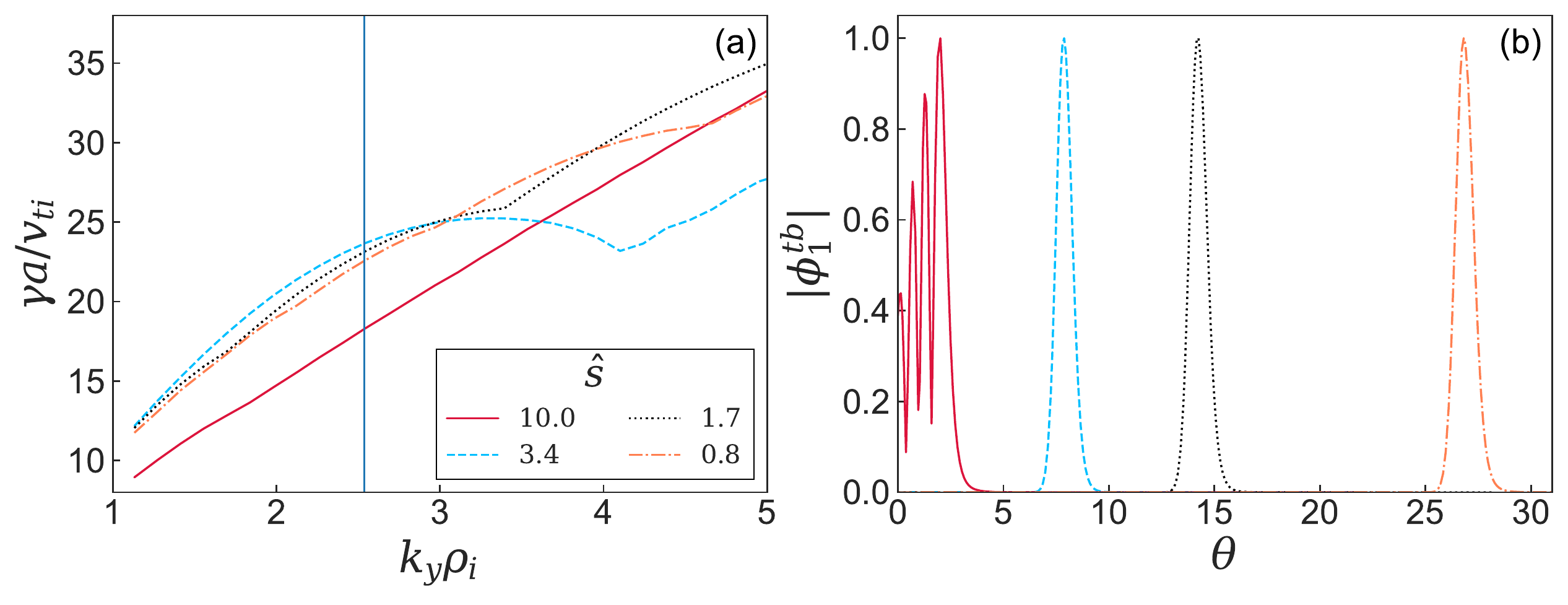}
 \caption{(a): \textcolor{black}{L}inear growth rates from GS2 for different $\hat{s}$ values with $R_0/L_{Te} = 520$. \textcolor{black}{This is likely not an experimentally relevant temperature gradient; it was used to test the scaling of $\theta$ with $L_{Te}$.} (b): \textcolor{black}{C}orresponding eigenmodes for $k_y \rho_i = 2.6$.}
 \label{fig:shiftingthetawithshat}
\end{figure}

As an experiment, we artificially reversed the signs of the magnetic drifts in GS2. As expected, the toroidal ETG modes only grew in regions that were previously `good curvature' regions, which due to the sign reversal of $\omega_{\kappa e}$, are turned into `bad curvature' regions. This is shown in \Cref{fig:omegaMeflip}, being substantiated both by GS2 simulations (\Cref{fig:omegaMeflip}(a)) and the results of our model ETG dispersion relation (\Cref{fig:omegaMeflip}(c)).

Since $\omega_{*e} \eta_e$ is fixed for a given $k_y \rho_i $, the $\theta$ location will be such that $\omega_{\kappa e}$ and $b_e$ have the right value for maximum growth subject to FLR and curvature constraints. These constraints are shown in \Cref{fig:generalstabilityplot}(a) and (b). According to \Cref{fig:generalstabilityplot}(a) and the above arguments, the smallest $|\theta|$ that a mode with $\theta_0 = 0$ can occupy is $|\theta| \simeq 6.5$. We denote this minimum $\theta$ location as $\theta_{\mathrm{min} }$. The toroidal ETG mode cannot occupy a smaller $|\theta|$ value because either $\omega_{*e} \eta_e / \omega_{\kappa e} < 0$, $\omega_{*e} \eta_e /\omega_{\kappa e}$ is too large, or the bad curvature region is too narrow.

\begin{figure}[t]
\centering
    \includegraphics[width=\textwidth]{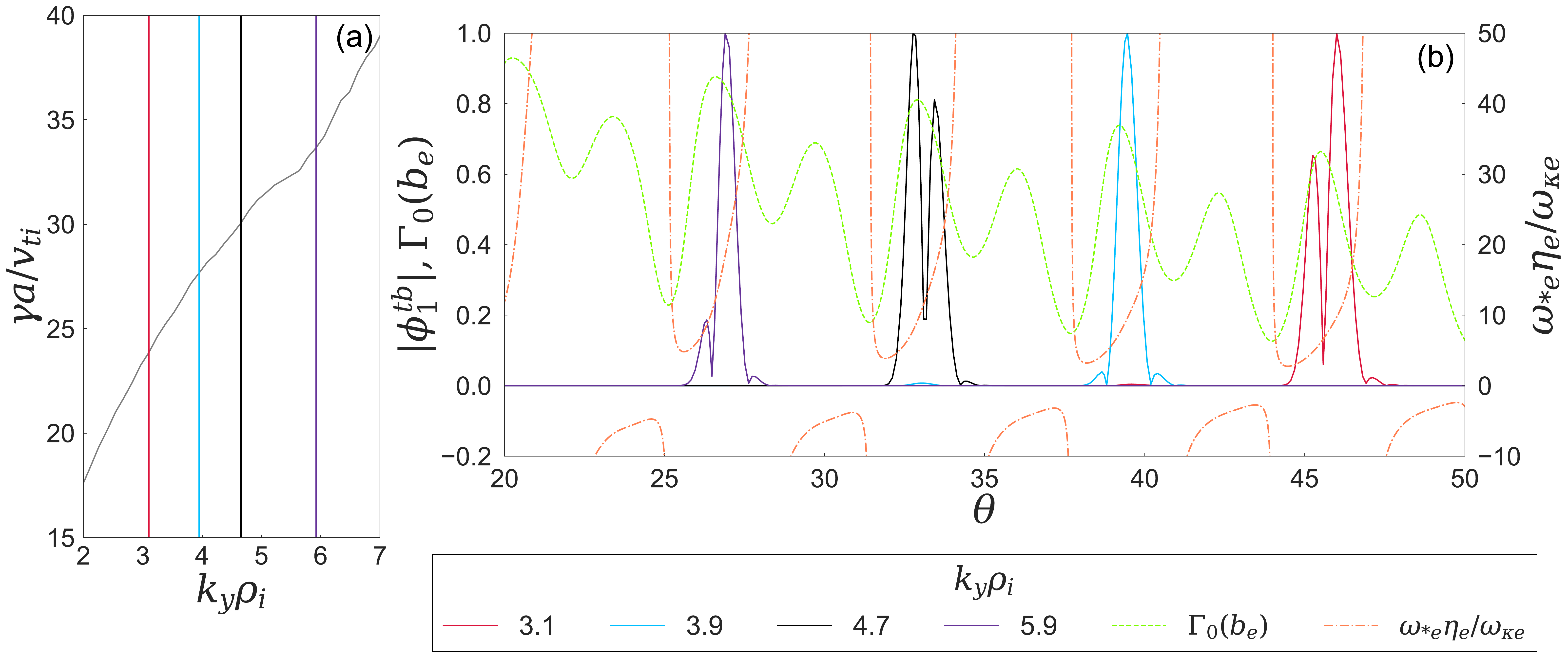}
 \caption{(a): \textcolor{black}{G}rowth rates versus $k_y \rho_i$. (b): \textcolor{black}{C}orresponding eigenmodes and the functions $\Gamma_0(b_e)$ and $\omega_{*e} \eta_e / \omega_{\kappa e} $. The toroidal ETG mode shifts due to changing $k_y \rho_i $, predicted by \Cref{eq:kyrhoeLTeRscale}. Here we have set $\hat{s} = 0.45$ and $R_0/L_{Te} = 520$, allowing the mode to be very mobile in $\theta$. \textcolor{black}{This is likely not an experimentally relevant temperature gradient; it was used to test the scaling of $\theta$ with $L_{Te}$.} The values of $\Gamma_0(b_e)$ are evaluated for $k_y \rho_i = 5.9$.}
 \label{fig:krishiftingthemode}
\end{figure}

From these considerations, there are several obvious parameters that can change where the mode is located. As already predicted in \Cref{eq:kyrhoeLTeRscale}, a larger $R_0/L_{Te}$ causes a mode to be unstable at larger $\theta$ values; in \Cref{fig:generalstabilityplot}(c) we show that increasing $R_0/L_{Te}$ increases the $\theta$ location of the mode. In \Cref{fig:generalstabilityplot}(c), we use a smaller value of $\hat{s}$ (1.68 instead of 3.36), since we found that, for larger values of $\hat{s}$, increasing $R_0/L_{Te}$ was not particularly effective at shifting the mode to larger values of $|\theta|$ --- this is because $b_e$ increases nonlinearly with $\hat{s}$, and once $\hat{s}$ is sufficiently large, a toroidal ETG mode becomes significantly more FLR damped as it moves along $\theta$. The parallel location of the modes with different values of $R_0/L_{Te}$ agrees well with the curvature and FLR constraints discussed above. Smaller $\hat{s}$ and $k_y \rho_i$ also force the mode to larger $\theta$ --- as predicted in \Cref{eq:kyrhoeLTeRscale}, the shifting of modes due to $\hat{s}$ and $k_y \rho_i$ is shown in \Cref{fig:shiftingthetawithshat,fig:krishiftingthemode}, respectively. 

\Cref{fig:shiftingthetawithshat}(a) illustrates that the toroidal ETG growth rate is relatively insensitive to $\hat{s}$, until $\hat{s}$ exceeds a threshold value. Recall that $\omega_{*e} \eta_e / \omega_{\kappa e} \sim R_0 / L_{Te} \hat{s} \theta$. This implies that if $\hat{s}$ changes, a toroidal mode would move in $\theta$ to have a $R_0 / L_{Te} \hat{s} \theta$ that maximizes its growth rate. As $\hat{s}$ increases, the $|\theta|$ location will decrease. However, the mode cannot be driven linearly unstable below $\theta_{\mathrm{min}}$, so at a critical value of $\hat{s}$ the mode will become increasingly stabilized by FLR effects while the mode maximum remains at fixed $\theta = \theta_{\mathrm{min} }$. In \Cref{fig:shiftingthetawithshat}(a), we show that increasing $\hat{s}$ beyond some critical $\hat{s}$ indeed decreases the growth rate of the toroidal ETG mode. This increase in $\hat{s}$ once the mode was at $\theta_{\mathrm{min} }$ increased $k_{\perp}$, and hence caused its growth rate to be lower than the slab ETG mode --- this occurred for a value of $\hat{s}$ somewhere between $\hat{s} = 3.4$ and $\hat{s} = 10$ in \Cref{fig:shiftingthetawithshat}(b).

The $\theta$ location of the mode also depends strongly on $k_y \textcolor{black}{ \rho_i}$, as shown in \Cref{fig:krishiftingthemode}(b) where we ran GS2 simulations with a smaller value of $\hat{s} = 0.45$ and an increased value of $R_0/L_{Te}$, which makes the location of the mode more sensitive to changes in $k_y$. Clearly, the eigenmodes are centered very close to a local minimum in $b_e$. The toroidal ETG modes are close to this minimum because of a competition between the size of the magnetic drift and FLR effects; as shown in \Cref{fig:toroidaltoslab}, the growth rates are very sensitive to $b_e$. Careful inspection of the growth rates in \Cref{fig:krishiftingthemode}(a) reveals that there is a change in mode type as the mode jumps to a new $\theta$ location --- this can be seen by discontinuities in $\partial \gamma / \partial k_y$.

\begin{figure}[t]
\centering
    \includegraphics[width=\textwidth]{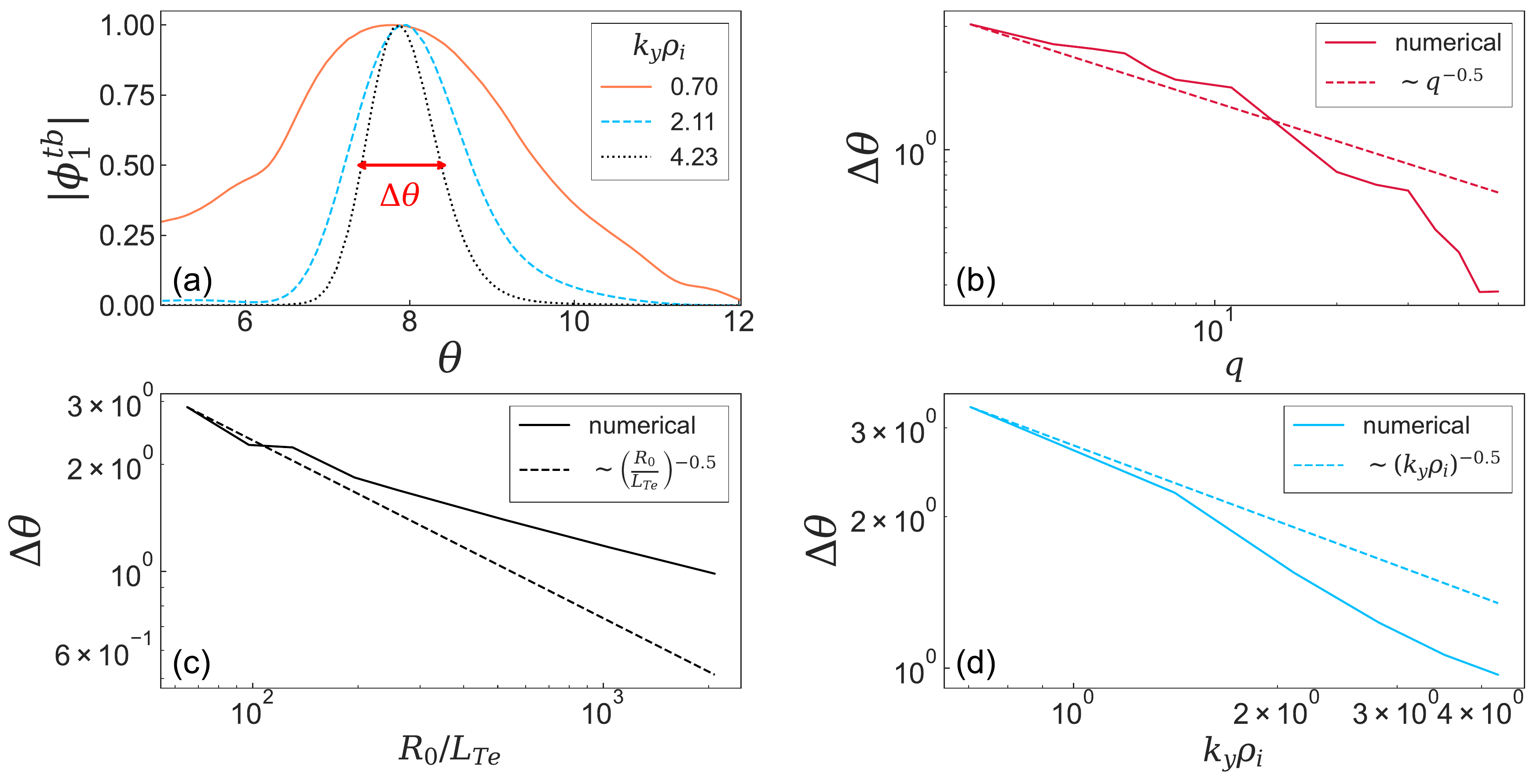}
 \caption{(a): \textcolor{black}{T}oroidal ETG eigenmodes for different values of $k_y \rho_i $, and numerical definition of $\Delta \theta$ used in subsequent subplots. (b): Numerical (solid) and predicted (dashed) $\Delta \theta$ versus $q$ scaling, (c): $\Delta \theta$ versus $R_0/L_{Te}$ scaling, (d): $\Delta \theta$ versus $k_y \rho_i $ scaling.}
 \label{fig:narrowingthemode}
\end{figure}

We now examine the scalings for the mode width from \Cref{eq:deltathetascaling} by comparing them with toroidal ETG eigenmodes from GS2 simulations. We calculate the width $\Delta \theta$ as the length in $\theta$ for the half height of the mode; this is shown in \Cref{fig:narrowingthemode}(a). \Cref{eq:deltathetascaling} predicts that the mode width $\Delta \theta$ scales with $R_0/L_{Te}$, $k_y \rho_i$, and $q$ as $\Delta \theta \sim \sqrt{L_{Te} / R_0 k_y \rho_e q}$. Scans in these quantities, shown in \Cref{fig:narrowingthemode}, demonstrate increasing $R_0/L_{Te}$, $k_y \rho_i$, and $q$ narrows the toroidal ETG mode structure. However, the scaling exponents do not appear to be quantitatively correct. The theoretical scaling $\Delta \theta \sim \sqrt{L_{Te} / R_0 k_y \rho_e q}$ in \Cref{eq:deltathetascaling} is not perfect because the mode changes location. Indeed, since the parallel location of the mode is sensitive to $q, k_y \rho_i$, and $R_0/L_{Te}$, changing the location of the mode by changing these parameters changes the local derivative of $\mathbf{v}_{Me} \cdot \mathbf{k}_{\perp}$, and hence changes $\Delta \theta$. Additionally, because we have used a Taylor expansion assuming that the variation in $\mathbf{v}_{Me} \cdot \mathbf{k}_{\perp}$ is proportional to $\Delta \theta$, this expansion breaks down when $\Delta \theta$ becomes too large.

As the toroidal ETG instability is FLR damped at increasing $k_y$, the mode switches to the slab branch, with an accompanying increase in $k_{\parallel} $. The switch from toroidal to slab at fixed $k_y$ is shown in the simple dispersion relation used to plot \Cref{fig:toroidaltoslab}(c). At this transition, $k_{\parallel} $ for the slab mode is much larger than for the toroidal mode and the eigenmodes move from being quite localized around a large value of $\theta$, to oscillating rapidly about smaller $\theta$, as shown in \Cref{fig:krionetofiveplot}(a).

To demonstrate this transition, we need to define $k_{\parallel}$. Our choice of $\theta$ in \Cref{eq:thetaequation} is such that $\theta$ is proportional to the length along the magnetic field line. Thus, Fourier analyzing in $\theta$ is equivalent to obtaining the spectrum in $k_{\parallel}$.

\begin{figure}[t]
\centering
    \includegraphics[width=0.9\textwidth]{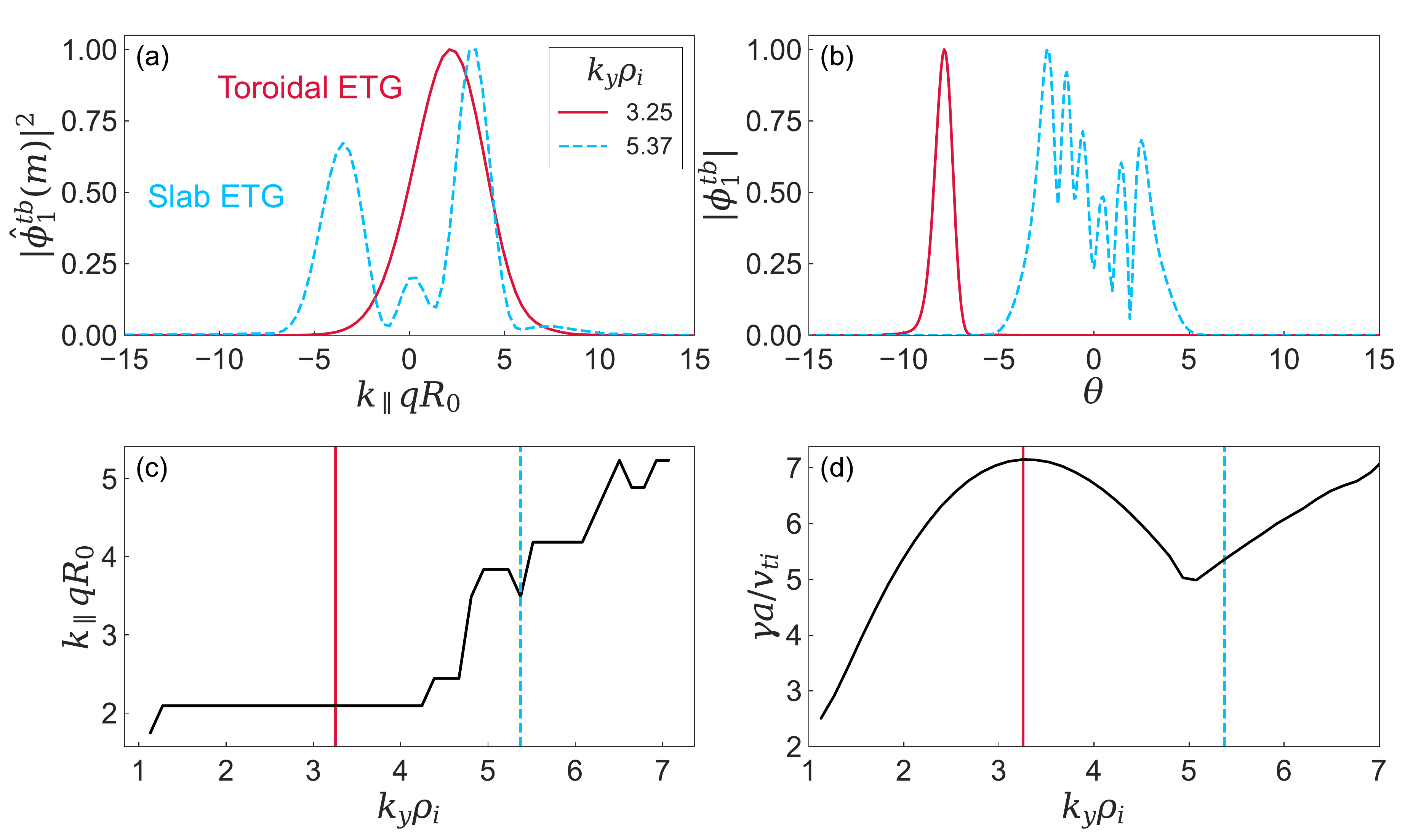}
 \caption{(a): \textcolor{black}{T}he Fourier transformed coefficient $|\hat{\phi} ^{tb} _1 (m)|^2$ spectrum for 2 modes from GS2 with different values of $k_y \rho_i$. \textcolor{black}{The coefficient $|\hat{\phi} ^{tb} _1 (m)|^2$ is normalized so that its maximum value is 1.} (b): \textcolor{black}{E}igenmodes. (c): \textcolor{black}{T}he $k_{\parallel} $ associated with the largest coefficient $|\hat{\phi} ^{tb} _1 (m)|^2$ in (a). (d): \textcolor{black}{G}rowth rates. All of these plots have $\theta_0 = 0.02$.}
 \label{fig:transitionfromtoroidaltoslab}
\end{figure}

To carry out the Fourier transform, we first interpolate $\phi ^{tb}_1 (\theta)$ onto a regular $\theta$ grid, since GS2's $\theta$ grid is not usually regularly spaced. Next, we apply a Fast Fourier Transform \cite{Cooley1965} to obtain the Fourier transform of \textcolor{black}{$\phi ^{tb}_1$},
\begin{equation}
\hat{\phi} ^{tb}_1 (m) = \int_{-\infty}^{\infty} \phi ^{tb}_1 (\theta) \exp(-i m \theta) d \theta. 
\label{eq:phifourierkpll}
\end{equation}
The relation between $m$ and $k_{\parallel} $ is
\begin{equation}
k_{\parallel} = \frac{2\pi}{L_{\theta}}m. 
\end{equation}
\Cref{fig:transitionfromtoroidaltoslab}(a) shows that the power spectrum $|\hat{\phi} ^{tb}_1|^2 $ changes significantly at the transition between toroidal and slab ETG. The toroidal ETG spectrum is Gaussian whereas the slab spectrum is more complicated, with at least two peaks. It is noteworthy that the toroidal ETG has a non-zero $k_{\parallel}$ for its fastest growing mode, since theory predicts toroidal ETG with the highest growth rate at $k_{\parallel} = 0$, shown in \Cref{fig:toroidaltoslab}. Previous studies of toroidal ETG have also found $k_{\parallel} = 0$ as the fastest growing mode \cite{Dorland2000}.

\begin{figure}[t]
\begin{center}
 \includegraphics[width=0.49\textwidth]{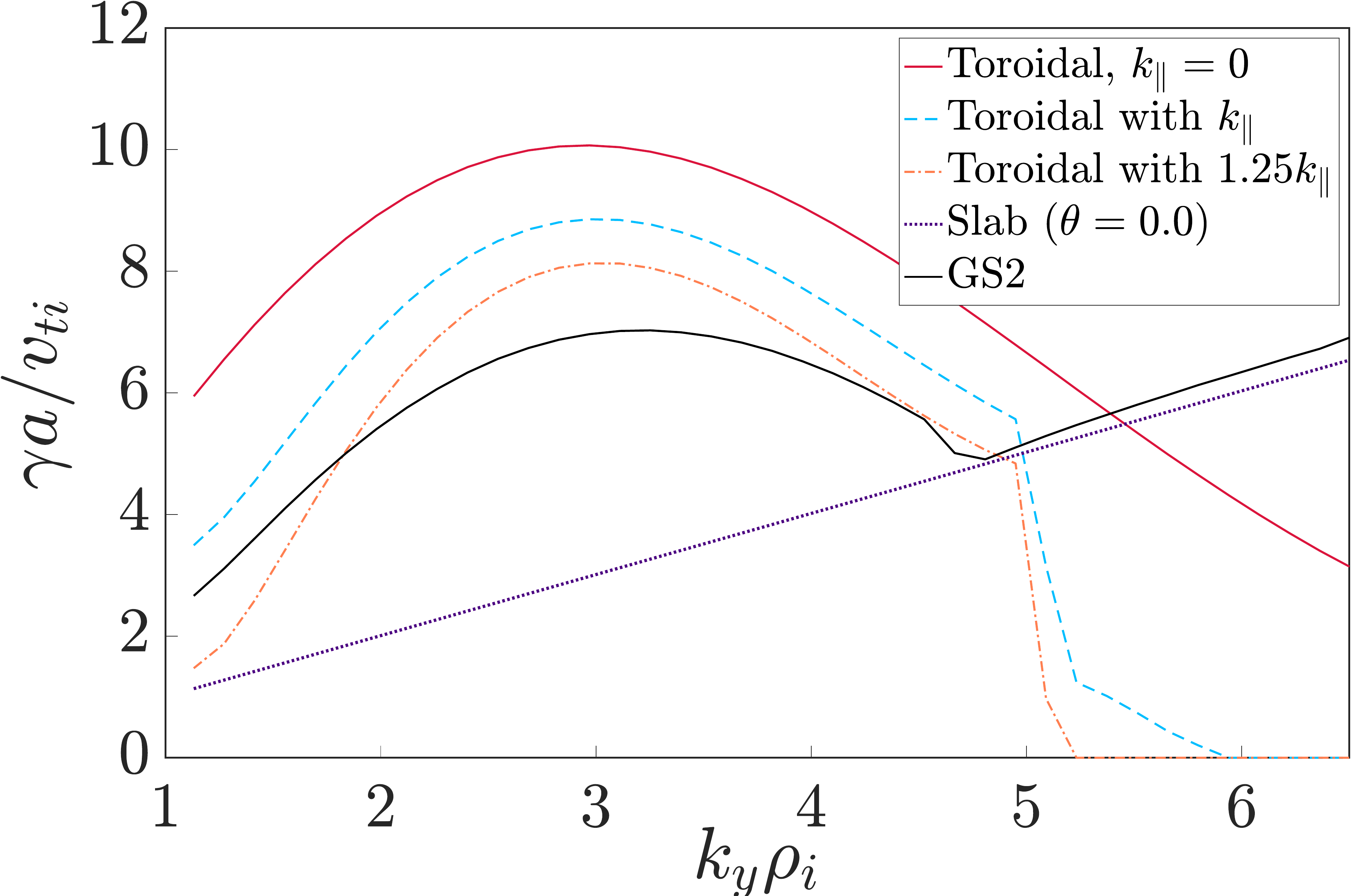}
\end{center}
\caption{The growth rates obtained in theory and in GS2. For the toroidal ETG growth rate, we found the $\theta$ with the highest growth rate for \Cref{eq:toroidalETGsolved}, which occurred at $\theta = 7.7$, and for the slab ETG growth rate, we evaluated the dispersion relation at $\theta = 0.0$ (note that $\omega_{\kappa e}$ is nonzero at $\theta=0$). The $k_{\parallel} $ input for the toroidal ETG was obtained by Fourier transforming the GS2 eigenmodes for each $k_y$, and for the `$1.25 k_{\parallel}$' series, we multiplied all $k_{\parallel} $ values by 1.25.}
\label{fig:growthversussimulation}
\end{figure}

We now use \Cref{eq:phifourierkpll} to calculate the toroidal ETG growth rates for a range of $k_y \rho_i $. Our analytic model requires $k_{\parallel}$ as an input, which we obtain from GS2 by choosing the value of $k_{\parallel}$ that corresponds to the largest amplitude in the poloidal Fourier transform $\hat{\phi} ^{tb} _1$. Once we have obtained $k_{\parallel} $ from the GS2 data for each value of $k_y \rho_i$, we solve the model dispersion relation in \Cref{eq:toroidalETGsolved} for each value of $\theta$, inputting the correct value of $k_{\perp}$, $\omega_{\kappa e}$, and $\omega_{\nabla B e}$ at each $\theta$ location. For each $k_y \rho_i$ value, we take the growth rate from the $\theta$ location with the highest growth rate to be the growth rate of the toroidal ETG mode for that $k_y \rho_i $. There is excellent agreement between the $\theta$ location with the highest growth rate by solving \Cref{eq:toroidalETGsolved} and the eigenmode maximum from GS2. This method for calculating $k_{\parallel}$ gave a toroidal ETG growth rate reasonably close to the values obtained from GS2 shown in \Cref{fig:growthversussimulation}, as well as the $k_y \rho_i$ location of the peak. \textcolor{black}{Since the toroidal ETG mode is no longer the fastest growing instability for $k_y \rho_i \gtrsim 5$, the value of $k_{\parallel}$ that we deduce from GS2 and we use to plot the values of the toroidal ETG growth rate in \Cref{fig:growthversussimulation} is not reliable for $k_y \rho_i \gtrsim 5$. To calculate the toroidal ETG growth rate for $k_y \rho_i > 5.0$, we simply evaluated the growth rate at $\theta = 7.7$ with $k_{\parallel}$ given by the slab ETG mode from GS2. To calculate the slab ETG growth rate, we found the value of $k_{\parallel}$ for which the growth rate at $\theta = 0.0$ was maximized.} Surprisingly, this method also gives a very good approximation to the slab ETG growth rate even though slab ETG modes are very extended (see \Cref{fig:kribiggerthanfive}).

The theory presented in this paper cannot self-consistently calculate $k_{\parallel} $ and thus we have used solutions with a $k_{\parallel} $ associated with the numerical simulations. Until now, our analysis has been performed with $\theta_0 = 0$. In the next section, we extend our analysis to toroidal ETG with a nonzero value of $\theta_0$. 

\subsection{Effects of $\theta_0$} \label{sec:effectsofkx} 

We now consider ETG instability for $\theta_0 \neq 0$. The growth rate of microinstabilities and MHD ballooning instabilities has a complicated dependence on $\theta_0$. Previous works have found that nonzero $\theta_0$ can substantially change the growth rates for toroidal ITG \cite{Migliano2013,Kotschenreuther2017}, ETG \cite{Jenko2009, Told2012}, and MTMs \cite{Hatch2016}. For MHD ballooning modes, it was found that for smaller pressure gradients, increasing $|\theta_0|$ is stabilizing, but once the gradients become sufficiently large, increasing $|\theta_0|$ is destabilizing \cite{Miller1995}. 

\begin{figure}
\centering
    \includegraphics[width=\textwidth]{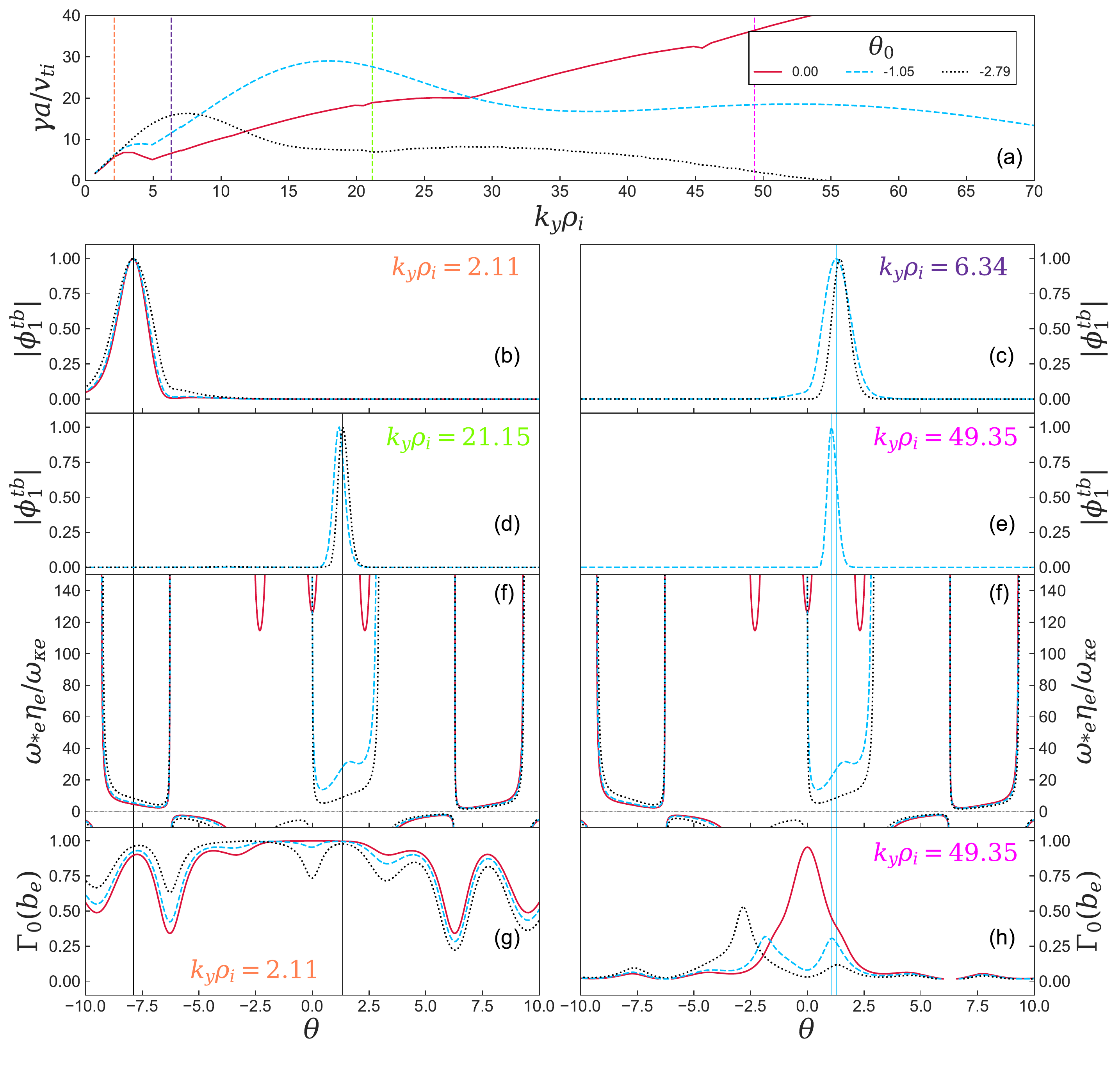}
 \caption{The effect of $\theta_0$ on growth rates and eigenmodes. (a): \textcolor{black}{G}rowth rates with three values of $\theta_0$. Vertical dashed lines indicate the $k_y \rho_i$ values for the eigenmodes that are shown in (b), (c), (d), and (e). (b), (c), (d), and (e): \textcolor{black}{E}igenmodes for $k_y \rho_i = 2.11, 6.34, 21.15, 49.35$ and different $\theta_0$. (f): $\omega_{*e} \eta_e / \omega_{\kappa e}$ for different $\theta_0$; for $|\theta_0|$ sufficiently large, new good curvature regions near $\theta = 0$ appear. (g) and (h): $\Gamma_0(b_e)$ for different $\theta_0$ at two values of $k_y \rho_i$. Vertical solid lines on rows 2 - 5 indicate the maximum amplitude of a selected toroidal ETG eigenmode for a given $\theta_0$; if the eigenmode is not shown for a given $k_y \rho_i$, the fastest growing mode for that $k_y \rho_i$ is not a toroidal ETG mode. Rows 2-5 share the same $\theta$ axis. Consistent coloring and linestyle series is used throughout the plot, determined by the legend in (a).}
 \label{fig:theta0growthrate} 
\end{figure}

As briefly discussed in \Cref{sec:jetprofiles}, we find that increasing $|\theta_0|$ can substantially increase the toroidal ETG growth rate, shown in \Cref{fig:theta0growthrate}(a). For many values of $\theta_0$, the toroidal ETG mode can be the fastest growing mode not only at ion scales, $k_y \rho_i \sim 1$, but at scales smaller than the electron gyroradius: $k_y \rho_e > 1$. To be precise, we find that at low values of $k_y \rho_i $ ($k_y \rho_i \lesssim 2$), the toroidal ETG has a similar growth rate for all values of $\theta_0$, whereas for larger values of $k_y \rho_i $, the toroidal ETG growth rate becomes very strongly dependent on $\theta_0$. We proceed to explain why.

For $k_y \rho_i \lesssim 2$, the location and growth rate of the toroidal ETG mode are fairly independent of $\theta_0$, as shown in \Cref{fig:theta0growthrate}(a) and (b). For such small values of $k_y \rho_i$, FLR damping is weak at many $\theta$ locations, that is, $k_{\perp} \rho_e \ll 1$ (and hence $\Gamma_0 (b_e) \approx 1$) in many distinct bad curvature regions. Since $\Gamma_0 (b_e) \approx 1$ in multiple regions, the fastest growing mode will be located at $\theta$ where $\omega_{*e} \eta_e / \omega_{\kappa e}$ is optimal. The value of $\omega_{*e} \eta_e / \omega_{\kappa e}$ is modified by $\theta_0$, shown in \Cref{fig:theta0growthrate}(f). The modification is particularly noticeable for $|\theta| \lesssim 6$, where there are regions of much smaller values of $\omega_{*e} \eta_e / \omega_{\kappa e}$ when $\theta_0$ is nonzero. For example, for $\theta_0 = -1.05$, \Cref{fig:theta0growthrate}(f) shows that $\omega_{*e} \eta_e / \omega_{\kappa e}$ has values as small as $\omega_{*e} \eta_e / \omega_{\kappa e} \simeq 15-30$ for $1 \lesssim \theta \lesssim 2$. While this value of $\omega_{*e} \eta_e / \omega_{\kappa e}$ is appropriate to have an unstable toroidal ETG mode, at larger values of $|\theta|$ there exists an even smaller value of $\omega_{*e} \eta_e / \omega_{\kappa e}$ (recall that smaller $\omega_{*e} \eta_e / \omega_{\kappa e}$ typically gives higher growth rates as long as $\omega_{*e} \eta_e / \omega_{\kappa e} \gtrsim 2-3$, see \Cref{fig:toroidaltoslab}). Again considering the $\theta_0 = -1.05$ mode, we see that $\omega_{*e} \eta_e / \omega_{\kappa e} \simeq 3-10$ for $-8 \lesssim \theta \lesssim -7$. Because we are currently considering relatively small values of $k_y \rho_i $, the FLR damping at $\theta = -7.7$ is not much stronger than at $\theta = 1.5$ (see \Cref{fig:theta0growthrate}(g)). Therefore, a mode at $\theta \simeq -7.7$ grows faster than a mode at $\theta \simeq 1.5$. The $k_y \rho_i = 2.11$ modes in \Cref{fig:theta0growthrate}(b) (all with $\theta_0 \leq 0$) have their maximum amplitude at $\theta = -7.7$ rather than $\theta = 7.7$ because FLR damping is slightly weaker at $\theta = -7.7$. Because \textit{both} the $\omega_{*e} \eta_e / \omega_{\kappa e}$ profiles and the $\Gamma_0 (b_e)$ profiles are not strongly dependent on $\theta_0$ for $|\theta| \gtrsim 6$ (see \Cref{fig:theta0growthrate}(f) and (g)), the location of the toroidal ETG modes and their associated growth rates are almost independent of $|\theta_0|$ for $k_y \rho_i \lesssim 2$, although the sign of the $\theta$ location does depend on $\mathrm{sign}(\theta_0)$.

\begin{figure}
\centering
    \includegraphics[width=\textwidth]{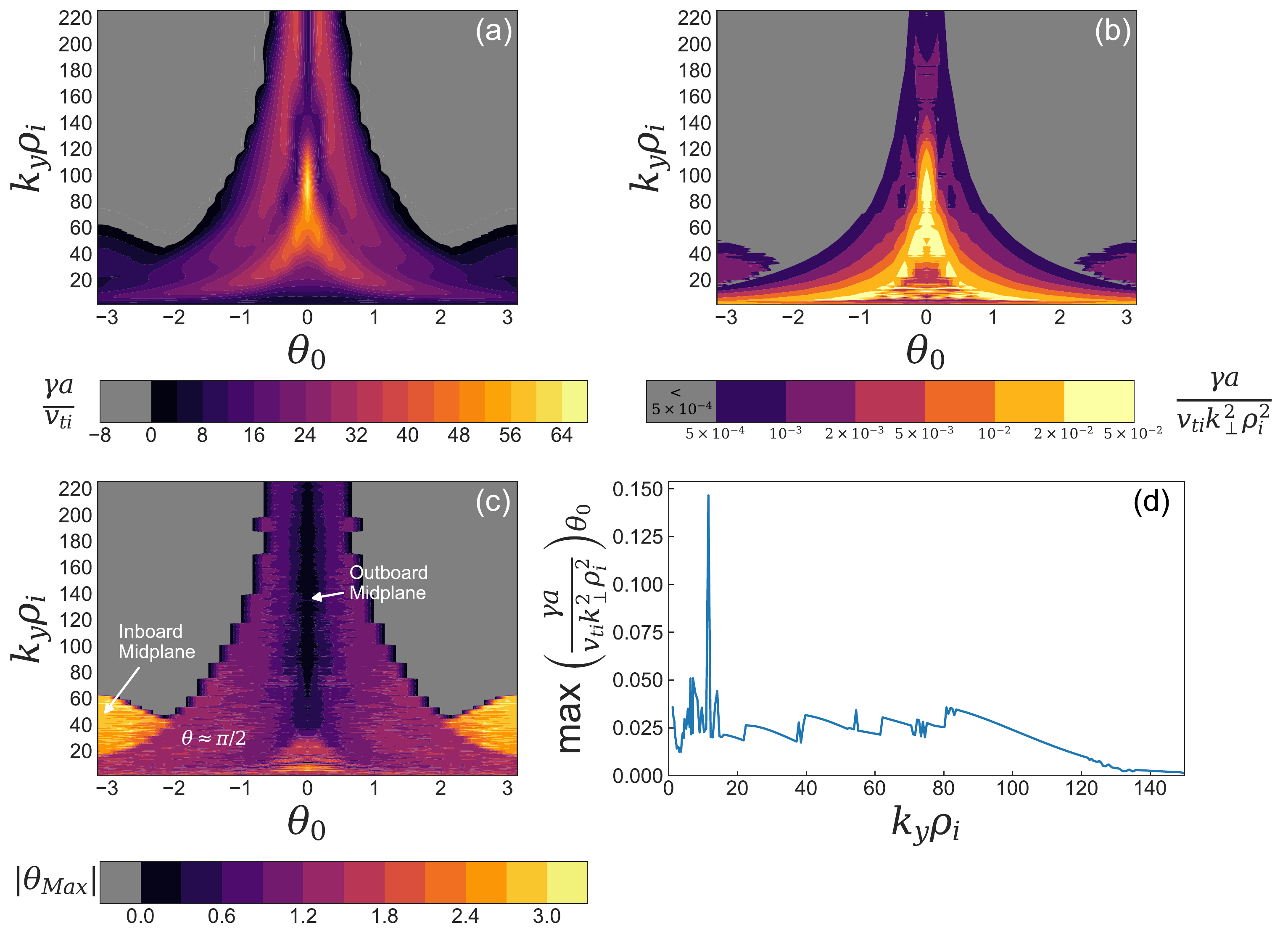}
 \caption{Growth rate-associated quantities from GS2 simulations. (a): \textcolor{black}{C}ontour plot of growth rates versus $\theta_0$ and $k_y \rho_i$. (b): \textcolor{black}{C}ontour plot of $\gamma / k_{\perp}^2$ versus $\theta_0$ and $k_y \rho_i$. (c): \textcolor{black}{L}ocation of the maximum of $|\phi _1 ^{tb} |$, $\theta_{\mathrm{Max} }$. (d): \textcolor{black}{T}he maximum value of $\gamma / k_{\perp}^2$ (over all $\theta_0$ values) for each value of $k_y \rho_i$.}
 \label{fig:theta0growthrateandmodeclassification} 
\end{figure}

We now consider what happens for larger values of $k_y \rho_i$. Here, the $\Gamma_0$ profiles are much more strongly dependent on $\theta_0$, as shown in \Cref{fig:theta0growthrate}(h). For $\theta_0 = 0$, as $k_y \rho_i $ increases the toroidal ETG mode cannot grow at a smaller value of $|\theta|$ because either $\omega_{*e} \eta_e / \omega_{\kappa e}$ is too large, or the bad curvature region is too narrow, causing the mode to have a stabilizing value of $k_{\parallel}$. Hence, the $\theta_0 = 0$ toroidal ETG mode becomes increasingly FLR damped as $k_y \rho_i $ increases and at $k_y \rho_i \simeq 5$, the slab ETG mode overtakes the FLR damped toroidal ETG mode to become the fastest growing mode (see \Cref{fig:theta0growthrate}(a)). However, for nonzero $\theta_0$, the toroidal ETG mode \textit{can} grow at a smaller value of $|\theta|$ where FLR damping is much weaker, and have a high growth rate because $\omega_{*e} \eta_e / \omega_{\kappa e}$ is sufficiently small. A consequence of the toroidal ETG mode moving to a bad curvature region with reduced FLR damping is that modes can be unstable in a wide range of poloidal locations, even close to the inboard midplane of the tokamak, a region that has traditionally been considered to have `good curvature' for all values of $\theta_0$ (see \Cref{fig:omegaMeflip}(b), where even the toroidal ETG mode with $\theta_0 = 0$ is unstable close to the inboard midplane). However, the maximum eigenmode amplitude for the fastest growing mode is typically close to $\theta \; \mathrm{mod} \; 2 \pi \simeq \pm \pi/2$, which is mainly due to local magnetic shear making a local maximum in $\Gamma_0$ at $\theta \; \mathrm{mod} \; 2 \pi \simeq \pm \pi/2$.

As shown in \Cref{fig:theta0growthrate}(c), (d), and (e), for nonzero $\theta_0$ and larger values of $k_y \rho_i$, the mode moves to a $\theta$ location that satisfies $\theta \theta_0 < 0$. This can be explained by including $\theta_0$ in the scaling for $\omega_{*e} \eta_e / \omega_{\kappa e}$,
\begin{equation}
\frac{\omega_{*e} \eta_e}{\omega_{\kappa e}} \sim \frac{k_y}{k_{\perp}} \frac{R_0}{L_{Te}} \sim \frac{1}{\hat{s} (\theta_0 - \theta)} \frac{R_0}{L_{Te}} \sim 1.    
\end{equation}
Hence, at larger values of $k_y \rho_i$ when a mode needs to move to a location with a smaller $|\theta|$ value, it will choose the location where $\theta \theta_0 < 0$ in order to make $\omega_{*e} \eta_e / \omega_{\kappa e}$ small.

To summarize, for smaller values of $k_y \rho_i$ (here $k_y \rho_i \lesssim 2$), FLR effects are relatively weak in multiple bad curvature regions, allowing the toroidal ETG mode to choose between multiple $\theta$ locations in order to find the optimal value of $\omega_{*e} \eta_e / \omega_{\kappa e}$. For the equilibrium considered in this paper, this occurs for $|\theta| \gtrsim 6$. However, when $k_y \rho_i$ is much larger and $\theta_0 = 0$, FLR damping prevents instability at higher values of $|\theta|$, even though bad curvature regions still exist there. For larger $k_y \rho_i $ and $\theta_0 \neq 0$, instability becomes possible at lower $|\theta|$ values due to modest FLR damping in select regions near $\theta = 0$.

To gauge the relative importance of toroidal and slab ETG modes for transport, we calculate the quantity $\gamma / k_{\perp}^2$ for all modes at $1 \lesssim k_y \rho_i \lesssim 230$ and $|\theta_0| < \pi$. The quantity $\gamma / k_{\perp}^2$ is a rough quasilinear estimate for the transport diffusion coefficient of the mode. To estimate $k_{\perp}$ for each mode, we find the $\theta$ location with the largest eigenmode amplitude, and calculate $k_{\perp}$ at that location. In \Cref{fig:theta0growthrateandmodeclassification}(a), we show the growth rates versus $\theta_0$ and $k_y \rho_i$. There is a notable maximum in the growth rate at $k_y \rho_i \approx 80$ and $\theta_0 = 0$ (which corresponds to a slab ETG mode). In \Cref{fig:theta0growthrateandmodeclassification}(b) we show the quantity $\gamma / k_{\perp}^2$ --- normalized and presented as the dimensionless parameter $\gamma a / v_{ti} k_{\perp}^2 \rho_i^2$ --- versus $\theta_0$ and $k_y \rho_i$. We observe that $\gamma / k_{\perp}^2$ has its largest values across a wide range of $k_y \rho_i $ and $\theta_0$ scales, $5 \lesssim k_y \rho_i \lesssim 100$ and $|\theta_0| \lesssim 1.5$. Most of these are toroidal ETG modes, although when $\theta_0 = 0$ and $k_y \rho_i \gtrsim 5$, the fastest growing mode is a slab ETG mode. We stress that the quantity $\gamma / k_{\perp}^2$ is only an approximate measure, and that nonlinear simulations will be needed to ascertain which modes are most important for transport. In \Cref{fig:theta0growthrateandmodeclassification}(c), we plot the $|\theta|$ location of the maximum of $|\phi ^{tb} _1|$, denoted as $|\theta_{\mathrm{Max}}|$; we see that modes with large values of $\gamma / k_{\perp}^2$ tend to have $0 \lesssim |\theta_{\mathrm{Max}}| \lesssim \pi/2$. In \Cref{fig:theta0growthrateandmodeclassification}(d), for each $k_y \rho_i$ we plot the normalized value of $\gamma/ k_{\perp}^2$ that is maximum over $\theta_0$. This plot demonstrates that there is a comparable quasilinear diffusion coefficient estimate for all fastest growing modes between $1 \lesssim k_y \rho_i \lesssim 100$, and hence suggests that a wide range of $k_y \rho_i$ values might be important for transport. \textcolor{black}{In \cref{app:dischargeparams}, we show estimates of $\gamma / k_{\perp}^2$ for the other three JET discharges we have examined, which demonstrate a qualitatively similar dependence of $\gamma / k_{\perp}^2$ on $\theta_0$ and $k_y \rho_i$ as JET shot 92174 in \Cref{fig:theta0growthrateandmodeclassification}(b).}

While significant heat might be transported by toroidal ETG modes, they are unlikely to transport particles because the ions are very close to adiabatic (see \Cref{fig:krionetofivesensitivityscan}(b)). However, since the ions are not fully adiabatic for the slab ETG at lower $k_y \rho_i$ (see \Cref{fig:krionetofivesensitivityscan}(b)), the long wavelength slab ETG instability could cause particle transport. Finally, the `extended ETG' modes, which are the fastest growing modes for $0.1 \lesssim k_y \rho_i \lesssim 1$ (see \cref{sec:smallkrimodes}), can also have a large non-adiabatic ion response, and thus they too, may cause particle transport.

Next, we show how the values of $\theta_0$, $\theta_{\mathrm{min} }$, and $\hat{s}$ determine the critical temperature gradient of the toroidal ETG mode.

\subsection{Critical $R_0 / L_{Te}$} \label{sec:critR0LTe}

\begin{figure}[t]
\centering
    \includegraphics[width=\textwidth]{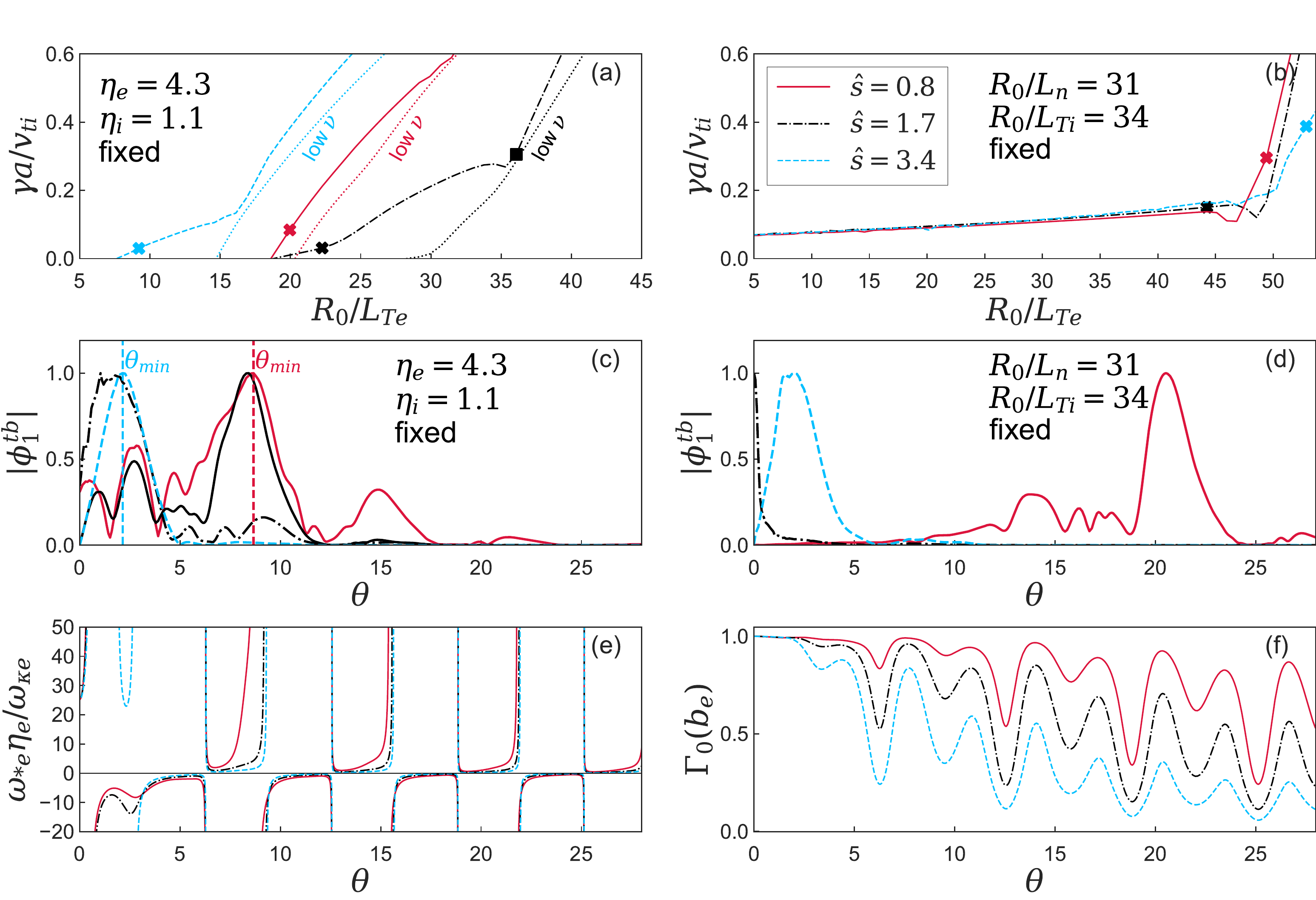}
 \caption{Stability plots of the toroidal ETG mode with $k_y \rho_i = 2.8$. (a): \textcolor{black}{G}rowth rate scan in $R_0/L_{Te}$ with $\eta_e$ and $\eta_i$ fixed for three values of $\hat{s}$ \textcolor{black}{with the standard and lowered collisionality (denoted by dotted lines labelled with `low $\nu$')}. (b): \textcolor{black}{G}rowth rate scan in $R_0/L_{Te}$ with $R_0/L_n$ and $R_0/L_{Ti}$ fixed for three values of $\hat{s}$. (c): \textcolor{black}{E}igenmodes corresponding to values of $R_0/L_{Te}$ denoted by the markers in (a). \textcolor{black}{The black dash dotted eigenmode corresponds to the cross marker for $\hat{s} = 1.7$ in (a), and the black solid eigenmode to the square marker for $\hat{s} = 1.7$ in (a).} (d): \textcolor{black}{E}igenmodes corresponding to values of $R_0/L_{Te}$ denoted by the markers in (b). (e): \textcolor{black}{T}he quantity $\omega_{*e} \eta_e / \omega_{\kappa e}$ for three values of $\hat{s}$, where $R_0/L_{Te} = 26, \; \eta_e = 4.3$. (f): \textcolor{black}{T}he quantity $\Gamma_0(b_e)$ for three values of $\hat{s}$.}
 \label{fig:RLTeandShatThetaminScan}
\end{figure}

\begin{figure}[t]
\centering
    \includegraphics[width=\textwidth]{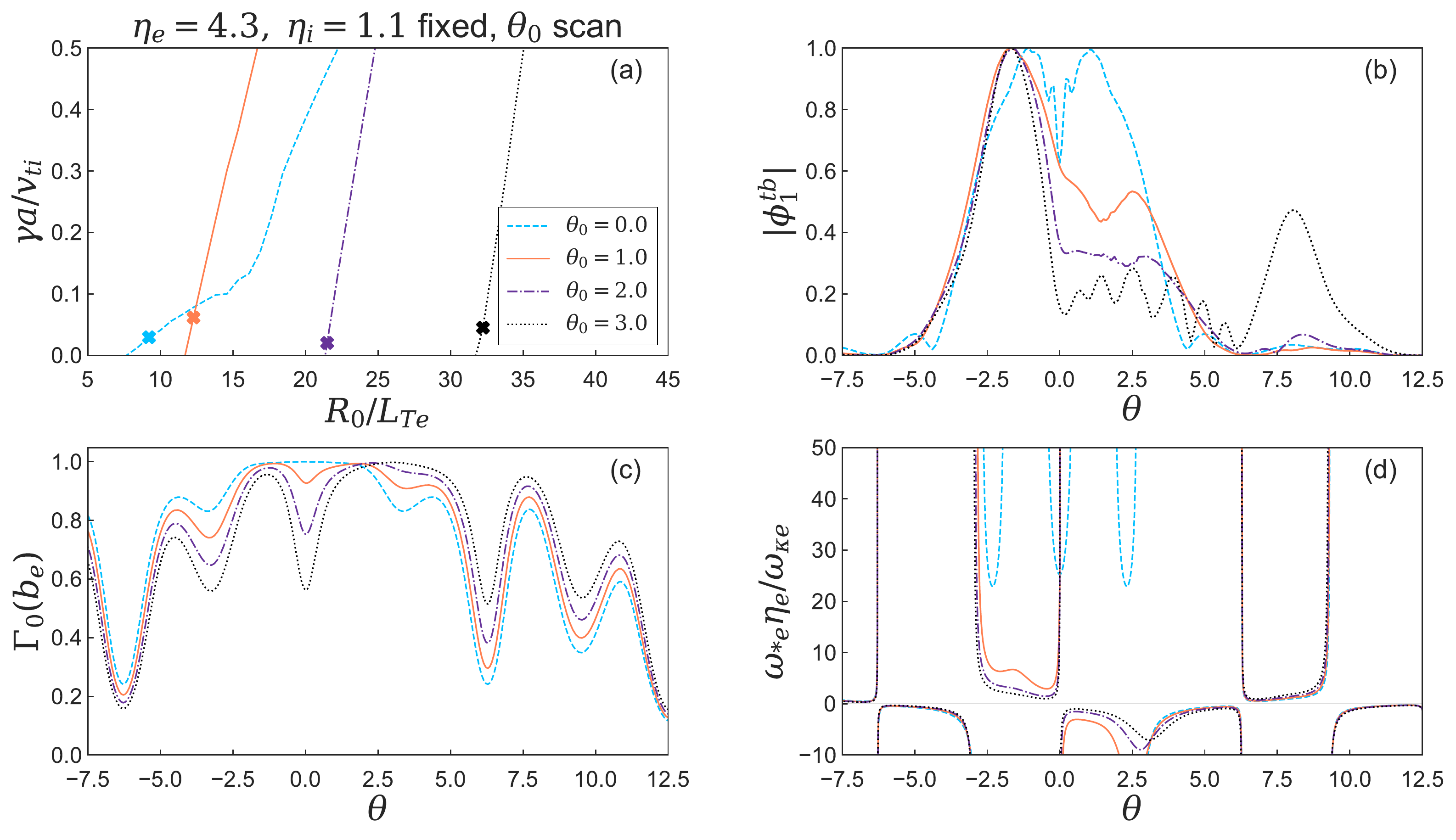}
 \caption{Stability plots of the toroidal ETG mode with $k_y \rho_i = 2.8$. (a): \textcolor{black}{G}rowth rate scan in $R_0/L_{Te}$ with $\eta_e$ and $\eta_i$ fixed for four values of $\theta_0$. (b): \textcolor{black}{C}orresponding eigenmodes at locations indicated by markers in (a). (c): \textcolor{black}{T}he quantity $\Gamma_0(b_e)$ for different values of $\theta_0$. (d): \textcolor{black}{T}he ratio $\omega_{*e} \eta_e / \omega_{\kappa e}$ for different values of $\theta_0$, using $R_0/L_{Te} = 26$.}
 \label{fig:RLTeandShatScanetaifixed}
\end{figure}

We now discuss the critical temperature gradient for the toroidal ETG instability that we are studying. We find critical $R_0/L_{Te}$ values as large as $R_0/L_{Te} \approx \textcolor{black}{32}$ for toroidal ETG modes in the pedestal (see \Cref{fig:RLTeandShatThetaminScan}(a) and \Cref{fig:RLTeandShatScanetaifixed}(a)), significantly larger than in the core. Unless mentioned otherwise, the quantity $\eta_e$ will be kept fixed, to prevent the ETG from becoming stable due to $\eta_e$ being less than its critical value.

We want to understand the dependence of the critical $R_0/L_{Te}$ on different parameters. Recall from \Cref{fig:toroidaltoslab}(a) that there exists a stability boundary $\omega_{*e} \eta_e / \omega_{\kappa e}$ for the toroidal ETG mode; that is, for instability we require
\begin{equation}
\frac{\omega_{*e} \eta_e}{\omega_{\kappa e}} > C.
\label{eq:Cequation}
\end{equation}
For $b_e = 0$, $C \simeq 2$. Given that $\omega_{*e} \eta_e / \omega_{\kappa e} \sim R_0 / \hat{s} \theta L_{Te}$, and that $\hat{s}$ and $R_0/L_{Te}$ are fixed parameters, the only free parameter in our scaling theory for the ratio $\omega_{*e} \eta_e / \omega_{\kappa e}$ for a given equilibrium is $\theta$ (note that $C$ in \Cref{eq:Cequation} is weakly dependent on $\theta$, because $C$ depends on $b_e$, which in turn depends on $\theta$). For the toroidal ETG mode to be unstable we then require 
\begin{equation} 
\frac{R_0}{\hat{s} L_{Te}} \frac{1}{C} \gtrsim \theta \gtrsim \theta_{\mathrm{min} }.
\label{eq:thetaboundsinstability}
\end{equation}
The quantity $\theta_{\mathrm{min}}$ is determined by the profiles of $\omega_{*e} \eta_e/ \omega_{\kappa e}$ and $\Gamma_0$ (see discussion at start of \Cref{sec:locwidth}). If a simulation only resolves up to $\theta < \theta_{\mathrm{min}}$ in ballooning space (or equivalently insufficiently large values of $|K_x|$), a toroidal ETG mode might incorrectly appear to be stable.

\begin{figure}[t]
\centering
    \includegraphics[width=\textwidth]{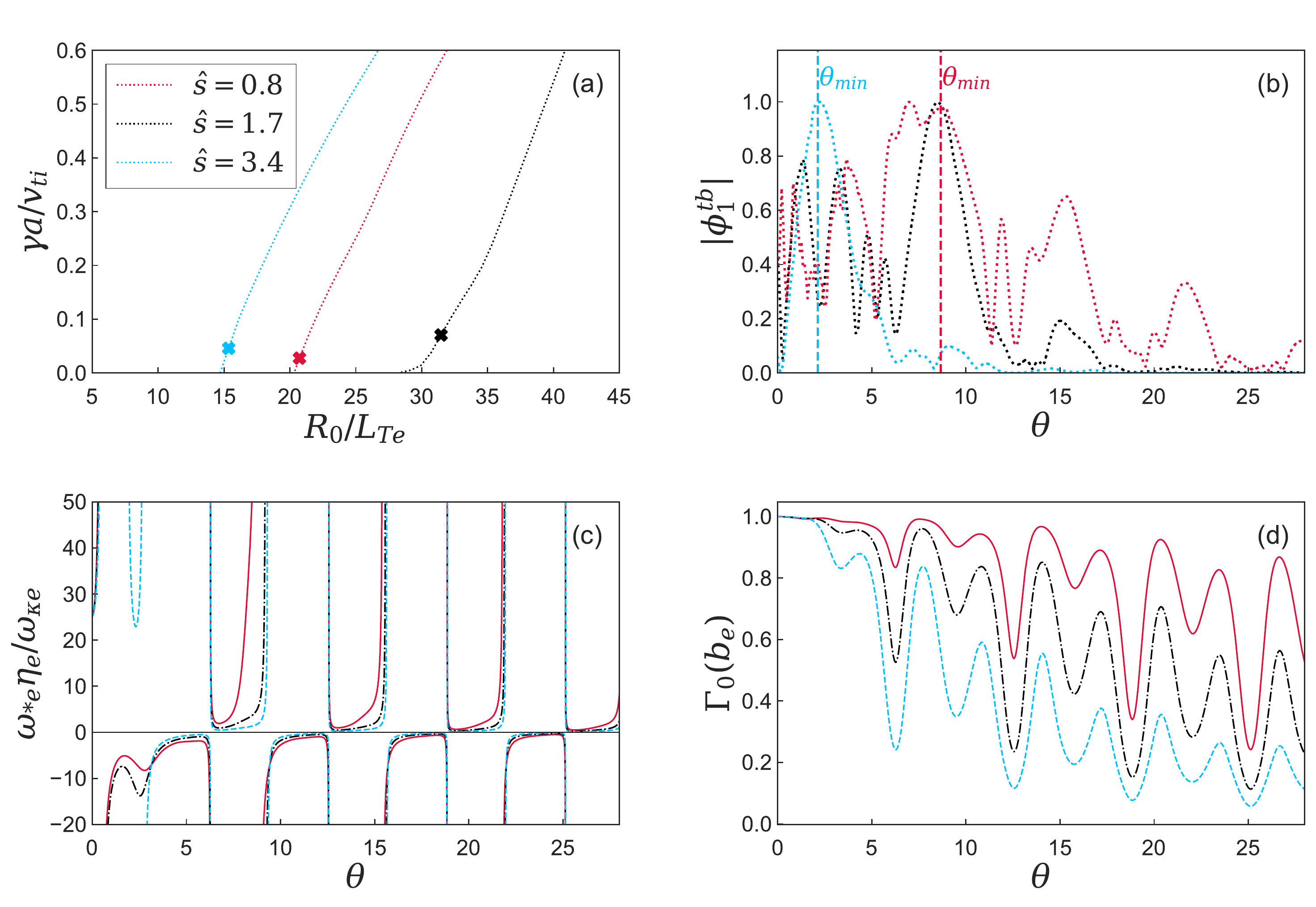}
 \caption{\textcolor{black}{Stability plots of the toroidal ETG mode with $k_y \rho_i = 2.8$ with lower collisionality. (a): Growth rate scan in $R_0/L_{Te}$ with $\eta_e$ and $\eta_i$ fixed for three values of $\hat{s}$. (b): Eigenmodes corresponding to values of $R_0/L_{Te}$ denoted by the colors in (a). (c): The quantity $\omega_{*e} \eta_e / \omega_{\kappa e}$ for three values of $\hat{s}$, where $R_0/L_{Te} = 26, \; \eta_e = 4.3$. (d): The quantity $\Gamma_0(b_e)$ for three values of $\hat{s}$.}}
 \label{fig:RLTeandShatThetaminScanlowcol}
\end{figure}

Numerical results have shown that $ \theta_{\mathrm{min} } $ is only very weakly dependent on $R_0 / L_{Te}$, but can be strongly dependent on $\theta_0$, and on $\hat{s}$ for large values of $\hat{s}$. For now we set $\theta_0 = 0$, but will soon consider the $\theta_0 \neq 0$ case. Thus, from \Cref{eq:thetaboundsinstability} we obtain a critical gradient, $R_0 / L_{Te}^{\mathrm{crit} }$,
\begin{equation}
\frac{R_0}{L_{Te}^{\mathrm{crit}} } \approx \hat{s} \theta_{\mathrm{min} } C.
\label{eq:RLTecrit}
\end{equation}

\textcolor{black}{When the growth rate is relatively small and comparable to $\nu_{ee}$, and $|\theta_0|$ is sufficiently small and $\hat{s}$ is sufficiently large, a mode different from the toroidal ETG modes that we are studying often appears. This means that we are sometimes unable to directly show the toroidal ETG growth rate going to zero. When we artificially decrease the collision frequency (keeping all other parameters fixed) to $\nu_{ee} a / v_{ti} \simeq 0.1$, the toroidal ETG growth rates visibly go to zero. Therefore, we first discuss the low collisionality cases in which we can almost find $R_0/L_{Te}^{\mathrm{crit} }$ for the toroidal ETG mode before another mode (such as the mode due to high collisionality) appears. Following this, we discuss the simulations with the standard collisionality.}

\textcolor{black}{For the low collisionality case, we demonstrate the $\hat{s}$ and $\theta_{\mathrm{min} }$ scaling of the critical temperature gradient by performing a scan in $R_0 / L_{Te}$ for three different values of $\hat{s}$, shown in \Cref{fig:RLTeandShatThetaminScanlowcol}(a). Here, $\eta_e$ and $\eta_i$ are held fixed to avoid the $\eta_s$ stability boundary. This scan is performed in GS2 for $k_y \rho_i = 2.8$ with the standard pedestal equilibrium we have used before, except for changing the value of $\hat{s}$. In \Cref{fig:RLTeandShatThetaminScanlowcol}(a), we see that $\theta_{\mathrm{min} } \simeq 2$ for $\hat{s} = 3.4$, as shown by the eigenmode in \Cref{fig:RLTeandShatThetaminScanlowcol}(b). For this value of $\hat{s}$, the eigenmode can have a relatively small value of $\theta_{\mathrm{min} }$ because of the bad curvature region ($\omega_{*e} \eta_e / \omega_{\kappa e} > 0$) that appears at $\theta \simeq 2$ in \Cref{fig:RLTeandShatThetaminScanlowcol}\textcolor{black}{(c)}. Once $\hat{s}$ is decreased, the smallest possible value for the mode appears to be $\theta_{\mathrm{min} } \simeq 8.5$, as shown in \Cref{fig:RLTeandShatThetaminScanlowcol}\textcolor{black}{(b) and (c)}. Due to the scaling of $R_0/L_{Te}^{\mathrm{crit}}$ in \Cref{eq:RLTecrit}, a much larger value of $\theta_{\mathrm{min} }$ causes $R_0/L_{Te}^{\mathrm{crit}}$ to increase, shown in \Cref{fig:RLTeandShatThetaminScanlowcol}(a). Both the cases $\hat{s} = 0.8$ and $\hat{s} = 1.7$ have the same value of $\theta_{\mathrm{min} } \simeq 8.5$, but the $\hat{s} = 1.7$ case has a much higher $R_0/L_{Te}^{\mathrm{crit}}$ due to its value of $\hat{s}$ being larger. Thus, we have demonstrated that increasing both $\hat{s}$ and $\theta_{\mathrm{min} }$ increases $R_0/L_{Te}^{\mathrm{crit}}$ for the toroidal ETG mode.}

\textcolor{black}{For the standard collisionality case, for $\hat{s} = 1.7, 3.4$ we see that new modes appear at lower values of $R_0/L_{Te}$ due to higher collisionality, shown in \Cref{fig:RLTeandShatThetaminScan}(a). These modes are different from the toroidal ETG instability because these modes can have large amplitudes in good curvature regions (see the eigenmode corresponding to this `collisional' mode for $\hat{s} = 1.7$ in \Cref{fig:RLTeandShatThetaminScan}(c), denoted by the dash dotted black line). These modes merit further investigation, but they are outside the scope of this work. Shown by the eigenmode with the solid black line in \Cref{fig:RLTeandShatThetaminScan}(c) (corresponding to the square marker in \Cref{fig:RLTeandShatThetaminScan}(a)), we see that before the fastest growing mode switches to the collisional mode as $R_0/L_{Te}$ decreases, the toroidal ETG mode indeed has $\theta_{\mathrm{min} } \simeq 8.5$, as one would predict from the profile of $\omega_{*e} \eta_e / \omega_{\kappa e}$ in \Cref{fig:RLTeandShatThetaminScan}(c). In \Cref{fig:RLTeandShatThetaminScan}(a), we see that $\theta_{\mathrm{min} } \simeq 2$ for $\hat{s} = 3.4$, as shown by the corresponding eigenmode in \Cref{fig:RLTeandShatThetaminScan}(c). For this value of $\hat{s}$, the eigenmode can have a relatively small value of $\theta_{\mathrm{min} }$ because of the bad curvature region ($\omega_{*e} \eta_e / \omega_{\kappa e} > 0$) that appears at $\theta \simeq 2$ in \Cref{fig:RLTeandShatThetaminScan}(e).} \textcolor{black}{Once $\hat{s}$ is decreased to a value of $\hat{s} = 1.7$, $\theta_{\mathrm{min} }$ \textit{appears} to also have a value of $\theta_{\mathrm{min} } \simeq 2$ (shown in \Cref{fig:RLTeandShatThetaminScan}(c)), yet $R_0/L_{T_e}^{\mathrm{crit} }$ increases, in apparent contradiction to \Cref{eq:RLTecrit}, which predicts that $R_0/L_{T_e}^{\mathrm{crit} }$ should decrease for smaller values of $\hat{s}$ at fixed $\theta_{\mathrm{min} }$. However, this contradiction is due to the collisional mode that appears for smaller values of $R_0/L_{Te}$, which for $\hat{s} = 1.7$ has a value of $|\theta_{\mathrm{min} }|$ that is much smaller than for the toroidal ETG mode, where $|\theta_{\mathrm{min} }| \simeq 8.5$.}

As mentioned above, there is another critical value of $R_0/L_{Te}$ that occurs due to $\eta_e$ being too small \cite{Jenko2001}. \Cref{fig:RLTeandShatThetaminScan}(b) shows a scan in $R_0/L_{Te}$ and $\hat{s}$ with $R_0/L_{n}$ and $R_0 / L_{Ti}$ fixed, allowing $\eta_e$ to vary; here, we see that the critical value of $\eta_e$ for the toroidal ETG mode is $\eta_e \approx 1.3$ (this cannot be seen directly here, but we have checked in the low collisionality case). Interestingly, for smaller values of $R_0/L_{Te}$ we find a weakly driven slab ITG mode, whose growth rate depends on $R_0/L_{Te}$. The slab ITG mode and the toroidal ETG modes for the scan in $R_0/L_{Te}$ at fixed $\eta_i$ are shown in \Cref{fig:RLTeandShatThetaminScan}(d).

The above arguments assumed that $|\theta_0| \ll |\theta|$. The critical temperature gradient is also modified by $\theta_0$. As discussed previously, larger values of $|\theta_0|$ can allow a new region of bad curvature to appear at small values of $|\theta|$, as shown in \Cref{fig:RLTeandShatScanetaifixed}(d). Allowing $\theta_0 \neq 0$, for instability, we require
\begin{equation}
\frac{R_0}{L_{Te}} \gtrsim \hat{s} |\theta - \theta_0| C.
\label{eq:theta0critscaling}
\end{equation}
We expect that for nonzero $\theta_0$, $\theta$ and $\theta_0$ have opposite signs because the mode will grow faster where $\omega_{*e} \eta_e/ \omega_{\kappa e} \sim R_0/L_{Te} \hat{s} |\theta- \theta_0|$ is smallest, giving the critical temperature gradient
\begin{equation}
\frac{R_0}{L_{Te}^{\mathrm{crit} }} \approx \hat{s}( |\theta_{\mathrm{min} }| + |\theta_0| ) C.
\label{eq:theta0critscalingtheta0large}
\end{equation}
Consistent with this idea, we see that for $|\theta| \lesssim 6$ the only accessible bad curvature regions appear when $\theta \theta_0 < 0$ and when $|\theta_0|$ is sufficiently large. To demonstrate the scaling in \Cref{eq:theta0critscalingtheta0large}, we performed a scan in $\theta_0$ and $R_0/L_{Te}$ at fixed $\hat{s}$, $\eta_e$, and $\eta_i$, shown in \Cref{fig:RLTeandShatScanetaifixed}(a); we observe that $R_0/L_{Te}^{\mathrm{crit} }$ indeed increases with $\theta_0$ as expected. Furthermore, the assumption that $\theta_{\mathrm{min} } \theta_0 < 0$ is also shown to be correct, as seen by the eigenmodes in \textcolor{black}{\Cref{fig:RLTeandShatScanetaifixed}(b)}. \textcolor{black}{Curiously, we note that the collisional mode that we found in \Cref{fig:RLTeandShatScanetaifixed} only appears for $\theta_0 = 0$ at smaller values of $R_0/L_{Te}$. For $\theta_0 = 1.0, 2.0, 3.0$, we cannot find such a mode.}

Finally, we briefly discuss the effect of the difference between $\omega_{\kappa e}$ and $\omega_{\nabla B e}$ on toroidal ETG stability. Throughout this paper, we have exclusively used $\omega_{*e} \eta_e / \omega_{\kappa e}$ for our analysis, which is justifiable if $\omega_{\kappa e} \simeq \omega_{\nabla B e}$ in the parallel vicinity of where the toroidal mode is most unstable. While this is true for $|\theta| \gtrsim \pi$ (see \Cref{fig:omegaMeflip}(b)), for $|\theta| \lesssim \pi$, the value of $\omega_{\kappa e} / \omega_{\nabla B e}$ in bad curvature regions can be as large as 1.5 in a sufficiently-wide parallel region for some values of $\theta_0$. For certain values of $k_y \rho_i$ and $\theta_0$, the stability boundary for the toroidal ETG mode is increased when $\omega_{\kappa e} > \omega_{\nabla B e}$, which is consistent with previous work \cite{Bourdelle2003}.

To summarize, we have demonstrated that the value of $R_0/L_{Te}^{\mathrm{crit}} $ for toroidal ETG depends on $\hat{s}, \theta_{\mathrm{min} }$, and $\theta_0$. Most relevant to the Miller equilibrium of JET discharge 92174, scans in $\theta_0$ at fixed $\hat{s} = 3.4$ showed $R_0/L_{Te}^{\mathrm{crit}} \textcolor{black}{\approx 8 - 32}$, depending on the value of $\theta_0$. This is a much higher value of $R_0/L_{Te}^{\mathrm{crit}} $ than is typically observed in the core (for example, $R_0/\textcolor{black}{ L_{Ti}^{\mathrm{crit}}} \approx 3$ for Cyclone Base Case toroidal ITG). This new type of stability boundary for toroidal ETG directly results from the importance of the radial component of the magnetic drift, in contrast to the core, where the $\nabla y$ component of the drift is usually considered more important.

\section{ITG Instability in JET Shot 92174} \label{sec:ITG}

In this section, we discuss the ITG instability in JET shot 92174. Previous works have emphasized the importance of ITG instability in the pedestal \cite{Hatch2017,Hatch2019,Wang2012,Saarelma2013,Chen2018}. In this work, we find that with the measured $T_{0i}$ profiles, the ITG growth rate is extremely small compared with the ETG instability growth rate. This is due to $R_0 / L_{Ti}$ and $\eta_i$ being relatively small, and electron collisions that decrease the ITG growth rates. If we increase the ion temperature profiles to make them equal to the electron temperature profiles and we ignore the $\mathbf{ E} \times \mathbf{ B}$ shear, the ITG instability is the fastest growing mode at very large scales, $k_y \rho_i \sim L_{Ti}/R_0$. This finding is entirely consistent with \Cref{sec:linGKwithlargegrads}'s results, as the same arguments can equally be applied to ITG (since $R_0/L_{Ti} \gg 1$). While this section will discuss ITG for $\theta_0 = 0$, we also performed a scan in $\theta_0$, to see if any other $\theta_0 \neq 0$ values could be unstable at $k_y \rho_i \lesssim 1$ using the measured ion temperature profile. We found no significant increase in growth rates due to $\theta_0$ with the measured ion profiles. 

\begin{figure}[t]
\centering
    \includegraphics[width=0.9\textwidth]{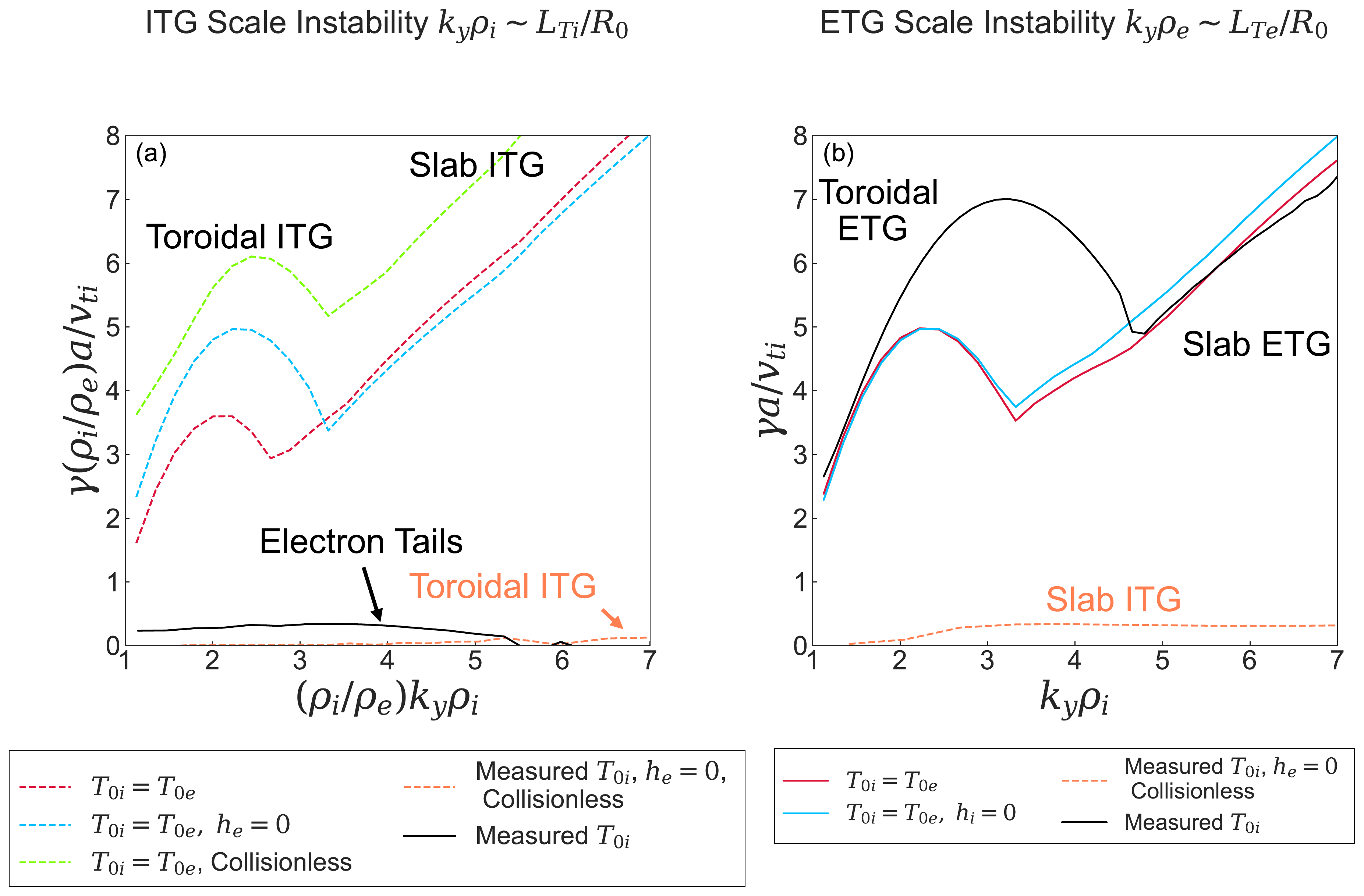}
 \caption{Linear ITG and ETG GS2 growth rates at (a): $k_y \rho_i \sim L_{Ti}/R_0$ (ITG scales) and (b):  $k_y \rho_e \sim L_{Te}/R_0$ (ETG scales). Dashed series indicates an ITG mode, solid is a mode driven by electron temperature gradients. For the ITG scales, the growth rates and $k_y \rho_i $ have been multiplied by $\rho_i / \rho_e$. The series `$T_{0i} = T_{0e}$' indicates that $T_{0e} = T_{0i}$, $L_{Ti} = L_{Te}$; `Measured $T_{0i}$' indicates that values of $T_{0i}$ and $L_{Ti}$ are taken from the measured ion profiles. Here, $\rho_i / \rho_e \approx 82$ for the measured $T_{0i}$ and $T_{0e}$ profiles, \textcolor{black}{and $\rho_i / \rho_e \approx 61$ when $T_{0i} = T_{0e}$.}}
  \label{fig:ITGscan}
\end{figure}

Due to the symmetry of the collisionless ITG and ETG dispersion relations when $h_e = 0$ for ITG and $h_i = 0$ for ETG, the growth rates of ITG and ETG are isomorphic: $\gamma_{\mathrm{ITG}} = \gamma_{\mathrm{ETG} } \rho_e/\rho_i$ at wavenumbers $k_{y \mathrm{ITG} } = k_{y \mathrm{ETG} } \rho_e/\rho_i$. 

Here we investigate how the non-adiabatic electron response and a difference in equilibrium profiles in the pedestal break this isomorphism. According to the isomorphism, ITG instability is driven at $k_y \rho_i \sim L_{Ti}/R_0 \ll 1$, and the ETG instability is driven at $k_y \rho_i \sim (\rho_i/\rho_e) L_{Te}/R_0$, as demonstrated in \Cref{fig:ITGscan}. In \Cref{fig:ITGscan}, we show the growth rates of ITG at `ITG' scales, $k_y \rho_i \sim L_{Ti}/R_0$, and the growth rates of ETG at  `ETG' scales, $k_y \rho_i \sim (\rho_i/\rho_e) L_{Te}/R_0$, for JET shot 92174. The isomorphism between ITG and ETG is confirmed, with the `$T_{0i} = T_{0e}, h_e = 0$' and `$T_{0i} = T_{0e}, h_i = 0$' cases having the same isomorphic growth rates. Here, `$T_{0i} = T_{0e}$' means that both the ion and electron temperatures and their gradients are set equal to each other --- specifically, \textcolor{black}{$T_{0e}$ is increased to match $T_{0i}$, and $R_0/L_{Ti}$ becomes as large as $R_0/L_{Te}$. This affects the electron collision frequencies, which are decreased self-consistently. Note that the difference between the toroidal ETG growth rates in \Cref{fig:ITGscan}(b) is mainly due to a different electron temperature, not a different collisionality.}

Electron collisions have a significant effect on the toroidal and slab ITG growth rates. As shown in \Cref{fig:ITGscan}(a), there is a substantial difference between the collisional and collisionless simulations, indicated by `$T_{0i} = T_{0e}$' and `$T_{0i} = T_{0e}$, Collisionless' cases. In the simulations we have performed, electron collisions reduce the toroidal and slab ITG growth rates. It is not obvious that electron collisions should always decrease the ITG growth rates, \textcolor{black}{or whether this stabilization can be ascribed to trapped or passing electrons. At these scales, $\nu_{ee} a/ v_{ti} \sim \textcolor{black}{0.8} \gg \gamma_{\mathrm{ITG} } a / v_{ti}$ and the modes with kinetic electron physics have a significant contribution of passing electrons due to the long electron ‘tails’ shown in \Cref{fig:lowereigenmodes} in \cref{sec:smallkrimodes}. Hence, at scales where there is ITG instability, the trapped electron response will be collisionally coupled to the large passing electron response.}

We now describe gyrokinetic simulations with the measured ion profiles. Compared with the equal profile case, `$T_{0i} = T_{0e}$,' once measured equilibrium profiles are included, the ITG growth rates decrease substantially. In \Cref{fig:ITGscan}(a), `Measured $T_{0i}$' is a simulation with the measured ion temperature profiles; the fastest growing modes at ITG scales are electron-driven modes with large electron tails \cite{Hallatschek2005} (see \cref{sec:smallkrimodes}), switching to a toroidal ETG mode once $k_y \rho_i \gtrsim 0.1$. In order to find the subdominant ITG instability, we must set $h_e = 0$ (otherwise electron-driven modes dominate), as shown in the `Measured $T_{0i}$, $h_e=0$' line. The ITG instability barely grows in the runs with adiabatic electrons, although there were well-resolved toroidal ITG eigenmodes. Using GS2's eigensolver function \cite{Dickinson2014}, we could not find any toroidal ITG instability for $k_y \rho_i \sim L_{Ti}/R_0$ when using the measured profiles and kinetic electrons, indicating that ITG is stable at $k_y \rho_i \ll 1$. However, at ETG scales ($k_y \rho_e \sim L_{Te}/R_0$), we did find weakly growing slab ITG modes by using adiabatic electrons, shown in \Cref{fig:ITGscan}(b) (`Measured $T_{0i}$, $h_e = 0$'), a result that was corroborated by very weakly growing slab ITG modes found using GS2's eigensolver. Therefore, for the measured profiles, ITG is extremely subdominant in JET shot 92174. Moreover, we will see in \Cref{sec:ExBeffects} that the slab ITG is easily quenched by $\mathbf{ E} \times \mathbf{ B}$ shear.

Heuristically, we can understand the stability of the toroidal ITG mode using a similar stability analysis performed for the toroidal ETG mode in \Cref{sec:critR0LTe}. In \Cref{fig:RLTeandShatThetaminScan}(b), we show the toroidal ETG mode being stabilized at $\eta_e \simeq \textcolor{black}{1.3}$ \textcolor{black}{(we checked the toroidal ETG growth rates went to zero for the low collisionality case; in the correct collisionality case shown in \Cref{fig:RLTeandShatThetaminScan}(b), a slab ITG mode appears before the toroidal ETG mode can be seen to be stabilized)}. Due to the isomorphism between toroidal ITG and toroidal ETG in the collisionless case where the other species is adiabatic, we can reasonably predict that toroidal ITG also has a similar critical $\eta_i \approx 1$. Examining the $\eta_i$ profile in \Cref{fig:1profiles}(c), we find that $\eta_i \simeq 0.8-1.2$ in the steep gradient region of the pedestal ($r/a \approx 0.97 - 0.99$). Hence, $\eta_i$ is very close to (and likely slightly below) its critical value in all regions of the pedestal for $\theta_0 = 0$, and it is unsurprising that the toroidal ITG mode is very weakly-driven. A broader question that merits examination is the physics that keeps $\eta_i$ close to its critical value, while $\eta_e$ is far above its critical value (although this is subject to uncertainties in the ion temperature profile, which could change $\eta_i$). Finally, the suppression of ITG instability in pedestals is not inconsistent with what has been observed in previous analyses; for example, \cite{Viezzer2017} found that the ion heat diffusion was close to neoclassical in ASDEX-U inter-ELM pedestal discharges.

\textcolor{black}{One might be concerned about the use of local simulations to analyze these large scale ITG modes. For JET shot 92174, at $r/a = 0.9743$ the local equations require $k_{\perp} \rho_i \gg \rho_i / L_{Te} = 0.12$ to be valid. Just as steep electron temperature gradients and FLR effects require toroidal ETG modes with wavenumber $k_y \rho_e \sim L_{Te}/R_0$ to satisfy $k_{\perp} \rho_e \sim 1$, steep ion temperature gradients and FLR effects require toroidal ITG modes with wavenumber $k_y \rho_i \sim L_{Ti}/R_0$ to satisfy $k_{\perp} \rho_i \sim 1$. Thus, even the long wavelength toroidal ITG does not violate $k_{\perp} \rho_i \gtrsim 0.12$. For example, we find that the toroidal ITG mode in \Cref{fig:ITGscan}(a) with $k_y \rho_i = 0.04$ ($`T_{0e} = T_{0i}, h_e = 0'$) has an eigenmode maximum at $k_{\perp} \rho_i = 0.9$, and so is far from violating the condition $k_{\perp} \rho_i \sim 0.12$. One might be concerned about the corresponding long wavelength slab ITG modes in \Cref{fig:ITGscan}(a), since $k_y \rho_i$ can be as small as $k_y \rho_i \simeq 0.05$ for the fastest growing slab ITG instability; however, similar to the slab ETG instability, these eigenmodes are still quite extended in $\theta$ for smaller values of $k_y \rho_i$. For a slab ITG mode with $k_y \rho_i = 0.05$, we find the eigenmode maximum occurring at a location where $k_{\perp} \rho_i = 0.2$, with many other peaks in the eigenmode with very similar amplitudes occurring at $\theta$ locations where $k_{\perp} \rho_i \gtrsim 1.0$. There may be subdominant ITG modes at $k_y \rho_i \sim L_{Ti}/ R_0$ that are very narrow in $\theta$, and hence have $k_{\perp} \rho_i \sim L_{Ti}/ R_0$; such modes would likely be poorly described by a local prescription, and therefore we would not dare to include such modes in our analysis. To summarize, the radial profile variation would be much more important for modes where $K_x \rho_i$ is sufficiently small, but we are examining modes that are typically much less extended in the radial direction than in the $y$ direction, and hence we do not expect a big difference between local and global simulations for these $K_x \gg k_y$ modes. Given the particular importance of FLR damping for these modes in the pedestal, having an accurate gyroaveraging scheme is also useful, which can be more challenging to implement for global simulations \cite{Guadagni2017}.}

To summarize, we find that with the measured ion temperature profiles, the ITG mode is stable for $k_y \rho_i \ll 1$, and there is very weakly-driven ITG at $k_y \rho_i \sim 1$. When the ion temperature profile is set equal to the electron profile and ITG modes become linearly unstable at very long wavelengths, the isomorphism between ITG with $h_e = 0$ and ETG with $h_i = 0$ holds. Electron collisions appear to decrease the ITG growth rate significantly. The detailed mechanism for this stabilizing impact of electron collisions requires further investigation.

\section{$\mathbf{ E} \times \mathbf{ B} $ Shear} \label{sec:ExBeffects}

In this paper we chose to perform most simulations without $\mathbf{ E} \times \mathbf{ B} $ shear, since in simulations with $\mathbf{ E} \times \mathbf{ B} $ shear, the electrostatic modes were barely modified compared to the simulations without $\mathbf{ E} \times \mathbf{ B} $ shear.

In this section, we present the results of gyrokinetic simulations with $\mathbf{ E} \times \mathbf{ B}$ shear. First, we discuss the validity of keeping $\mathbf{ E} \times \mathbf{ B}$ shear even though it is small in the low flow ordering. \textcolor{black}{After that, in} addition to the results we presented in \Cref{sec:jetprofiles} where KBMs were argued to be suppressed by $\mathbf{ E} \times \mathbf{ B}$ shear, we show the effect of $\mathbf{ E} \times \mathbf{ B}$ shear on KBMs, and ETG and ITG modes. We will see that while KBMs usually \textcolor{black}{are} easily suppressed by $\mathbf{ E} \times \mathbf{ B}$ shear, ETG modes are barely affected. ITG instability is easily stabilized when using the measured ion temperature profile, but is not fully-suppressed when the ion temperature profile is made equal to the electron temperature profile.

In our local linear simulations with $\mathbf{ E} \times \mathbf{ B}$ shear, we use a new $\mathbf{ E} \times \mathbf{ B}$ shear algorithm \cite{Christen2019}, and also tested that the results were qualitatively similar with the previous GS2 algorithm \cite{Hammett2006}. With the newer algorithm, a typical simulation with $\mathbf{ E} \times \mathbf{ B}$ shear contained a single poloidal mode, 150 radial wavenumbers with a spacing of $\Delta k_x \approx k_y$, and a $\mathbf{ E} \times \mathbf{ B}$ shear value of $\gamma_E a / v_{ti} = 0.56$. With the previous algorithm, the range of $k_x$ values was held fixed, but the $\Delta k_x$ spacing was reduced by a factor of 10.

In the low flow ordering, if one retains the $\mathbf{ E} \times \mathbf{ B} $ shear, one should also keep neoclassical corrections to the Maxwellian \cite{Parra2015,Lee2014two}, but for simplicity, we have neglected neoclassical corrections throughout this paper. When analyzing high $k_{\perp}$ modes for this equilibrium, it is inconsequential whether or not the $\mathbf{ E} \times \mathbf{ B}$ shear is kept, and we expect the neoclassical corrections to be similarly unimportant. However, for small $k_{\perp}$, we find the small $\mathbf{ E} \times \mathbf{ B}$ shear can suppress instabilities and hence one might expect that neoclassical corrections are also important.

The parallel flow is one of the main physical features of neoclassical corrections. Therefore to estimate the effect of these corrections, we will use previous studies on the parallel velocity gradient (PVG) instability \cite{Catto1973, Newton2010,Barnes2011b, Schekochihin2012,Lee2015,Ball2019}. The PVG growth rate is
\begin{equation}
\gamma_{\mathrm{PVG} } \sim \frac{d u_{\zeta i p}}{d r} k_y \rho_i. 
\end{equation}
In regions where we see ITG stabilization by $\mathbf{ E} \times \mathbf{ B}$ shear, $k_y \rho_i \sim 0.1$, and the PVG growth rate is much smaller than the $\mathbf{ E} \times \mathbf{ B}$ shear rate. From the measured ${}_6^{12}C^+$ rotation profiles at $r/a = 0.9743$, we find that $|du_{\zeta i p} / dr| a / v_{ti} \approx 1.4$, and thus $\gamma_{\mathrm{PVG} } a / v_{ti} \approx 0.14$. Therefore, given that $\gamma_E a / v_{ti} = 0.56 > \gamma_{\mathrm{PVG} } a / v_{ti}$, this PVG mode is likely stabilized by the $\mathbf{ E} \times \mathbf{ B}$ shear. Hence, we do not expect that the neoclassical flows will significantly modify a mode's growth rate, although the effect of neoclassical terms at these small scales merits further investigation.

\begin{figure}[t]
\begin{center}
 \includegraphics[width=\textwidth]{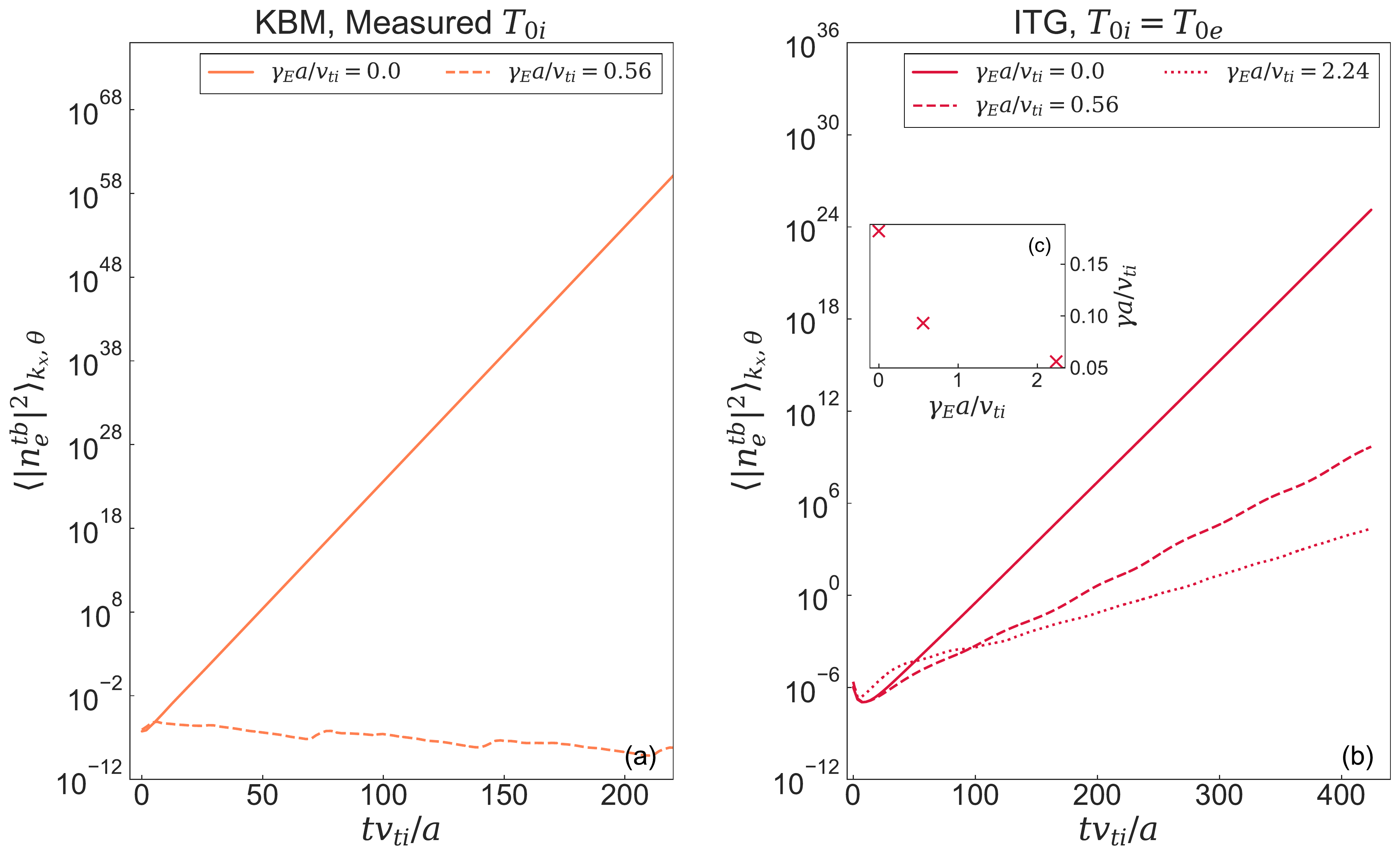}
\caption{Density time traces of KBM and ITG instabilities with and without $\mathbf{ E} \times \mathbf{ B}$ shear. (a): \textcolor{black}{T}he KBM is suppressed by the $\mathbf{ E} \times \mathbf{ B}$ shear consistent with the measured ion temperature profile. (b)\textcolor{black}{: T}he ITG is not fully suppressed by the $\mathbf{ E} \times \mathbf{ B}$ shear when the ion temperature and gradient are equal to the electron temperature and gradient. The two separate values of $\gamma_E a / v_{ti}$ correspond to its consistent value for the measured ion temperature profile ($\gamma_E a / v_{ti} = 0.56$) and when the ion temperature and gradients are equal to the electron temperature and gradient ($\gamma_E a / v_{ti} = 2.24$). (c): \textcolor{black}{T}he effective growth rates of the ITG instability for the three separate values of $\gamma_E a / v_{ti}$ in (b).}
\label{fig:flowshearresults}
\end{center}
\end{figure}

The $\mathbf{ E} \times \mathbf{ B}$ shear is usually more effective for low than for high $k_{\perp}$ modes, as shown in \Cref{fig:flowshearresults}. This is because the growth rate of the electrostatic instabilities that we are investigating typically scales with $\omega_{*s} \eta_s \sim k_y \rho_s v_{ts} / L_{Ts} $, and because of the differences in a mode's radial extent for different instabilities. If the typical timescale for an instability, $1/ \gamma$, is comparable to the $\mathbf{ E} \times \mathbf{ B} $ shearing time, $1/\gamma_E$, the $\mathbf{ E} \times \mathbf{ B}$ shear can be effective. However, when $1/\gamma_E \gg 1/\gamma \sim L_{Ts}/ k_y \rho_s v_{ts}$, the $\mathbf{ E} \times \mathbf{ B}$ shear is unable to shear the mode sufficiently quickly. Hence, $\mathbf{ E} \times \mathbf{ B} $ shear suppresses modes at smaller $k_y$, and barely modifies short wavelength modes.  Additionally, modes that are radially localized ($K_x \gg k_y$) are harder to shear than those with a wider radial width; this is apparent when examining the middle term in \Cref{eq:effectivewavenumberwithflowshear}. If the time independent piece of $|K_x|$ is already large, it will take a long time for flow shear to change $|K_x|$ substantially, by which time the linear mode will have likely already grown for multiple e-folding times. Hence, modes with $K_x \gg k_y$ are challenging to suppress with flow shear.

We now apply these two criteria (growth rate versus shearing rate, and radial extent of the mode) to explain our observations for which modes are suppressed by $\mathbf{ E} \times \mathbf{ B}$ shear. The KBM we discussed in \Cref{sec:jetprofiles} is easily suppressed by $\mathbf{ E} \times \mathbf{ B}$ shear because it is radially extended and is stable for a wide range of $\theta_0$ values (see \Cref{fig:electromagneticspectraearly}(d)). The KBM was shear suppressed even though $\gamma_{\mathrm{KBM} } > \gamma_E$. This suppression is demonstrated in \Cref{fig:flowshearresults}(a), where the mode's density is shown to decay in time.

Determining the effect of the $\mathbf{ E} \times \mathbf{ B}$ shear on toroidal and slab modes separately is challenging. To understand why this is the case, it will be useful to define an `effective' $\theta_0$ that now depends on time, 
\begin{equation}
\Theta_0(\gamma_E,t) = \theta_0 - k_y \frac{\gamma_E}{\hat{s}} t,
\label{eq:Theta0time}
\end{equation}
such that the time-dependent radial wavenumber is 
\begin{equation}
K_{x} = k_y \bigg{(} \hat{s}(\theta_0 - \theta) - \frac{r}{q} \frac{\partial \nu}{\partial r} \bigg{)} - k_y \gamma_E t = k_y \bigg{(} \hat{s} (\Theta_0-\theta) - \frac{r}{q} \frac{\partial \nu}{\partial r}.
\label{eq:effectivewavenumberwithflowshear}
\bigg{)}.
\end{equation}
The fact that the mode has different $\Theta_0$ values at different times considerably complicates understanding the effect of $\mathbf{ E} \times \mathbf{ B}$ shear on toroidal and slab ETG in the pedestal separately: for $k_y \rho_i \gtrsim 5$ in the absence of $\mathbf{ E} \times \mathbf{ B}$ shear, while for $\theta_0 = 0$ the fastest growing modes are slab ETG modes, for $\theta_0 \neq 0$ the fastest growing modes are almost always toroidal ETG modes. Since $\mathbf{ E} \times \mathbf{ B} $ shear changes $\Theta_0$ with time as described in \Cref{eq:Theta0time}, if at $t = 0$ a mode is a slab ETG mode (i.e. it has $\theta_0 = 0$), after a period of time it will become a toroidal ETG mode. Therefore, we can only determine if the $\mathbf{ E} \times \mathbf{ B}$ shear suppresses both slab and toroidal modes.

We now consider the effect of $\mathbf{ E} \times \mathbf{ B}$ shear on the ITG instability. Our simulations indicate that the effectiveness of $\mathbf{ E} \times \mathbf{ B}$ \textcolor{black}{shear} at suppressing ITG is sensitive to several parameters. We first test the effectiveness of $\mathbf{ E} \times \mathbf{ B} $ shear with the measured ion temperature profiles, which requires using adiabatic electrons, since electron temperature gradient-driven modes are the fastest growing at all scales (see \Cref{fig:ITGscan}). We test the $\mathbf{ E} \times \mathbf{ B} $ shear on an ITG mode with $k_y \rho_i = 0.7$, which has a modest growth rate of $\gamma a /v_{ti} \simeq 0.1$.  In simulations with $\mathbf{ E} \times \mathbf{ B}$ shear, the mode is easily suppressed. This is expected, since $1/\gamma_E \ll 1/\gamma$ for this ITG mode, and hence, both toroidal and slab ITG are suppressed by $\mathbf{ E} \times \mathbf{ B}$ shear at $k_y \rho_i = 0.7$ with the measured ion temperature profiles.

We also test the effectiveness of the $\mathbf{ E} \times \mathbf{ B} $ shear at suppressing the ITG instability when the ion temperature profiles are made equal to the electron temperature profiles (that is, $T_{0i} = T_{0e}$ and $L_{Ti} = L_{Te}$). To investigate this, we perform \textcolor{black}{ GS2 simulations} with $\mathbf{ E} \times \mathbf{ B}$ shear for a single toroidal ITG mode with $k_y \rho_i = 0.04$. Recall that we estimate the radial electric field by balancing it with the pressure gradient as in \Cref{eq:pressgradientradEfield}, which requires that $\gamma_E$ is roughly proportional to the second derivative of the pressure gradient, as in \Cref{eq:gammaEapproximation}. Therefore, when we quadruple $1/L_{Ti}$ for the case where the ion and electron temperature profiles are made equal, to be consistent with the temperature profile we must also roughly quadruple the value of $\gamma_E$. In \Cref{fig:flowshearresults}(b), we show the time trace of the density for three simulations of the ITG mode with $T_{0i} = T_{0e}, \; L_{Ti} = L_{Te}$, where the value of $\gamma_E$ varies in each simulation. We show the ITG mode in the absence of $\mathbf{ E} \times \mathbf{ B}$ shear, the mode with $\gamma_E a/v_{ti} = 0.56$, which is consistent with the measured ion temperature gradients, and the mode with $\gamma_E a/v_{ti} = 2.24$, which is consistent with the steepened ion temperature gradients. To calculate the effective growth rate, we used a similar technique to that in \cite{Roach2009}, which involves fitting the mode amplification in time. As shown in \Cref{fig:flowshearresults}(c), while the consistent value of $\mathbf{ E} \times \mathbf{ B}$ shear, $\gamma_E a/v_{ti} = 2.24$, reduces the growth rate by 70 \%, it does not fully suppress the ITG instability. We also found a range of additional parameters that determined how successfully the $\mathbf{ E} \times \mathbf{ B}$ shear suppressed the high gradient ITG mode such as $T_{0i}/T_{0e}$; more work is required to understand the resilience of strongly-driven pedestal ITG to $\mathbf{ E} \times \mathbf{ B}$ shear.

We now discuss the ETG instability. We found that $\mathbf{ E} \times \mathbf{ B} $ shear was insufficient to quench the ETG modes. Even tripling the value of $\gamma_E$ at $k_y \rho_i = 2.8$ barely changed the growth rates of the toroidal and slab ETG modes. The ineffectiveness of the $\mathbf{ E} \times \mathbf{B} $ shear for ETG modes is due to $\gamma \gg \gamma_E$ for these modes. There is likely no experimentally-realizable value of $\gamma_E$ that would suppress these ETG modes in the pedestal.

Thus, to summarize, we establish the following hierarchy for the efficiency of $\mathbf{ E} \times \mathbf{ B} $ shear at reducing the growth rates of linear modes. KBMs are completely suppressed by $\mathbf{ E} \times \mathbf{ B} $ shear, and ITG is also fully suppressed when using the measured ion temperature profiles. Using profiles with ion gradients as steep as the electron gradients, while the toroidal ITG growth rate is significantly reduced by $\mathbf{ E} \times \mathbf{ B}$ shear, it is not necessarily stabilized. ETG is very resistant to $\mathbf{ E} \times \mathbf{ B}$ shear.

\section{Discussion} \label{sec:discussion}

In the steep gradient region of the fully developed pedestal of a JET H-mode discharge (92174) where measurements indicate that $T_{0i} > T_{0e}$ and $R_0/L_{Te} > R_0/L_{Ti}$, local gyrokinetic simulations demonstrate that electron-driven modes are the fastest growing modes at all length scales perpendicular to $\mathbf{B} $. Linearly, KBMs are quenched by $\mathbf{ E} \times \mathbf{ B}$ shear, as is ITG when the measured ion temperature profiles are used. This leaves ETG at $0.1 \lesssim k_y \rho_i \lesssim 400$.

Using $R_0/L_{Te} \gg 1$, we predicted that a novel type of toroidal ETG would be driven at $k_y \rho_i \sim 1$ and $K_x \rho_e \sim 1$, which we have confirmed in gyrokinetic simulations. This toroidal ETG at $k_y \rho_i \sim 1$ in the linear growth rate spectrum seems to be a robust feature of steep temperature gradient regions, having been seen in all three other pedestals we examined (see \Cref{fig:allspectra}, and \cref{app:dischargeparams} for experimental information), as well as in other works: DIII-D \cite{Fulton2014, Kotschenreuther2019}, NCSX \cite{Baumgaertel2011}, and ASDEX-U \cite{Told2008,Jenko2009,Told2012,Kotschenreuther2019}. It is also likely that a toroidal ITG mode of a similar nature has been observed at $k_y \rho_i \sim L_{Ti}/ R_0$ in \cite{Bowman2018}.

A notable success of this work is that a simple theoretical model predicted the linear growth rates of the toroidal and slab ETG and the poloidal location of the toroidal ETG mode fairly well. If the ion temperature profile is set equal to the electron temperature profile, ITG modes grow fastest for $k_y \rho_i \lesssim 0.5$, and ETG modes grow fastest for $0.5 \lesssim k_y \rho_i \lesssim 400$. With equal ion and electron temperature profiles, one might be concerned about significant transport caused by the toroidal ITG at scales as small as $k_y \rho_i \sim L_{Ti}/ R \ll 1$, since nonlinearly these instabilities might produce large eddies that cause substantial heat transport. However, our simple estimate of $\gamma_E$ by balancing the radial electric field with the pressure gradient found that $\mathbf{ E} \times \mathbf{ B}$ shear could fully suppresses the ITG instability for certain temperature ratios $T_{0i}/T_{0e}$ when the ion temperature gradients are as steep as the electron temperature gradients. While the $\mathbf{ E} \times \mathbf{ B}$ shear frequency is too small to damp the ETG, impurities are known to damp ETG \cite{Jenko2001,Reshko2008}. Therefore, further investigation might explore the effect of impurities on toroidal ETG instability in pedestals. Work has already shown that impurities can produce non-negligible ion-scale pedestal transport \cite{Kotschenreuther2017,Hatch2019}.

With the measured ion temperature profiles, it is likely that the nonlinear state of JET shot 92174's pedestal is dominated by electron-driven transport. Indeed, the novel toroidal ETG modes we have described in this work could be important for transport, as evidenced by the heuristic estimate of $\gamma/k_{\perp}^2$ in \Cref{fig:theta0growthrateandmodeclassification}. Careful work will be required to resolve these modes in nonlinear simulations. We have not included results from nonlinear simulations in this paper because the linear results of this work demonstrate how challenging these simulations are to correctly resolve. For example, to resolve the fastest growing linear modes --- toroidal ETG modes --- from $1 \lesssim k_y \rho_i \lesssim 100$ in a nonlinear simulation requires significant $k_x$ resolution, as well as a sufficiently large number of independent $\theta_0$ modes. In addition, the slab ETG modes require increasingly fine $\theta$ grids to resolve at higher values of $k_y \rho_i$, which significantly increases computational cost. Caution is required in attempting to infer transport properties from these linear results: the modes we observe span a wide range of perpendicular scales, and complex multiscale interactions could be important \cite{Maeyama2015,Howard2016b,Maeyama2017,Hardman2019,Hardman2019b}.

While in this paper we have focused on a single radial location for a single discharge, we have also investigated the growth rates at various radial locations using gyrokinetic simulations. These simulations have demonstrated a significant sensitivity of the growth rates to the radial location because of the sensitivity of the instabilities to local gradients. Nevertheless, certain features such as (i) the dominance of ETG at all scales, and (ii) the toroidal ETG at $k_y \rho_e \sim L_{Te}/R_0$ were robust features. Due to the sensitivity of microstability to the radial location, we caution against using the local growth rates at any given flux surface to infer global properties about the pedestal, such as its width or height. We have observed that some pedestals have consistently lower growth rates than others, but more work, particularly nonlinear simulations, is required to connect gyrokinetic analysis with predictions of pedestal structure.

\section*{Data and Code Accessibility}

The data and code versions used for the material in this paper are available at the following dataset archive \cite{Parisi2020b}.

\section*{Acknowledgements}

We gratefully acknowledge useful conversations with I. G. Abel, T. Adkins, O. Beeke, B. Chapman, N. Christen, D. Dickinson, M. R. Hardman, W. Guttenfelder, Y. Kawazura, A. Mauriya, J. Ruiz Ruiz, S. Saarelma, A. A. Schekochihin, and D. A. St-Onge. JFP is supported by EPSRC Scholarship No 3000207032. FIP and MB are supported in part by the Engineering and Physical Sciences Research Council (EPSRC) [Grant Number EP/R034737/1]. CMR is supported in part by the RCUK Energy-Programme (grant number EP/EPI501045). PGI is supported by an EU H2020 grant, agreement No 3000207035. Computational time provided by Plasma HEC Consortium EPSRC (grant number EP/L000237/1). This work has been carried out within the framework of the EUROfusion Consortium and has received funding from the Euratom research and training programme 2014-2018 and 2019-2020 under grant agreement No 633053 and from the RCUK Energy Programme (grant number EP/T012250/1 and EP/P012450/1). Withal, this work has been carried out within the framework of the Contract for the Operation of the JET Facilities and has received funding from the European Union’s Horizon 2020 research and innovation programme. The views and opinions expressed herein do not necessarily reflect those of the European Commission. The authors acknowledge EUROfusion, the EUROfusion High Performance Computer (Marconi-Fusion), the use of ARCHER through the Plasma HEC Consortium EPSRC grant numbers EP/L000237/1 and EP/R029148/1 under the projects e281-gs2, and software support from Joseph Parker through the Plasma-CCP Network under EPSRC grant number EP/M022463/1. This work also received computational support from Joseph Parker of CoSeC (the Computational Science Centre for Research Communities), funded through CCP-Plasma/Plasma-HEC EPSRC grants EP/R029148/1 and EP/M022463/1. \textcolor{black}{This work was supported by the US Department of Energy through grant DE-FG02-93ER-54197.} JFP acknowledges travel support from Merton College, Oxford. We are grateful for the hospitality of the Wolfgang Pauli Institute, University of Vienna.

\clearpage

\appendix

\section{Other Discharges} \label{app:dischargeparams}

\begin{table}[h]
\begin{center}
    \begin{tabular}{ccccc}
    \hline
    Discharge & 82550 &  92167 & 92168 & 92174  \\ \hline \hline
    Experimental Parameters \\
    $I_p$ [MA] & 2.5 & 1.4 & 1.4 & 1.4 \\ 
    $B_{T0}$ [T] & 2.7 & 1.9 & 1.9 & 1.9 \\ 
    $H_{98(y,2)}$ & 0.7 & 0.9 & 1.0 & 1.0 \\ 
    $n_G$ & 0.8 & 0.6 & 0.7 & 0.7 \\ 
    $R_D$ [electrons/s $\times 10^{22}$] & 2.3 & 0.8 & 0.4 & 0.9 \\ 
    $q_{95}$ & 3.3 & 4.3 & 4.4 & 4.2 \\ 
    $\mathrm{Z}_{\mathrm{eff} } $ & 1.2 & 1.8 & 1.8 & 1.8 \\ 
    $P_{\mathrm{NBI}}$ [MW] & 14.4 & 17.4 & 17.6 & 17.4 \\ 
    $\beta_N$ & 1.1 & 2.2 & 2.6 & 2.5 \\ \hline \hline \hline
    Simulation Parameters \\
     $r/a$ & 0.9660 & 0.9784 & 0.9713 & 0.9743 \\ 
     $q$ & 3.65 & 5.14 & 5.07 & 5.08 \\ 
     $\hat{s}$ & 4.92 & 3.93 & 4.62 & 3.36 \\ 
     $a/L_{Te}$  & 57  & 41 & 29 & 42  \\ 
     $a/L_{Ti}$  & 12 & 19 & 16  & 11 \\ 
     $a/L_{n}$  & 23 & 8 & 10 & 10 \\ 
     $\kappa$  & 1.61 & 1.54 & 1.54 & 1.55 \\ 
     $\delta$  & 0.30 & 0.26 & 0.26 & 0.26 \\ 
     $a \beta'$  & -0.09 & -0.06 & -0.07 & -0.08 \\ 
     $\textcolor{black}{dR_c/dr}$  & -0.17 & -0.34 & -0.36 & -0.35 \\ 
     $a(d \kappa / dr)$ & 1.11 & 1.15 & 0.81 & 0.95 \\ 
     $a(d \delta / dr)$ & 0.97 & 0.85 & 0.67 & 0.74 \\ 
    \end{tabular}
\end{center}
        \caption{Experimental and simulation parameters for the discharges in this work.}
    \label{tab:expdetails}
\end{table}

Here we present the results of gyrokinetic analysis for three other JET-ILW H-mode pedestal discharges. The basic experimental and simulation parameters for these JET-ILW discharges in addition to the discharge discussed in the main text (shot 92174) are shown in \Cref{tab:expdetails}. Discharge 82550 is a very highly-fueled deuterium discharge with high triangularity and low ion temperature, 92167 is a highly-fueled deuterium discharge, 92168 is a weakly-fueled deuterium discharge, and 92174 is a highly-fueled deuterium discharge with deuterated ethylene ($\mathrm{C}_2 \mathrm{D}_4$) injection. In \Cref{tab:expdetails}, the quantity $q_{95}$ is the safety factor measured at the location where the normalized poloidal flux is equal to $0.95$. For more information on these data types, refer to the JET data handbook.

\begin{figure}[h]
\centering
    \includegraphics[width=\textwidth]{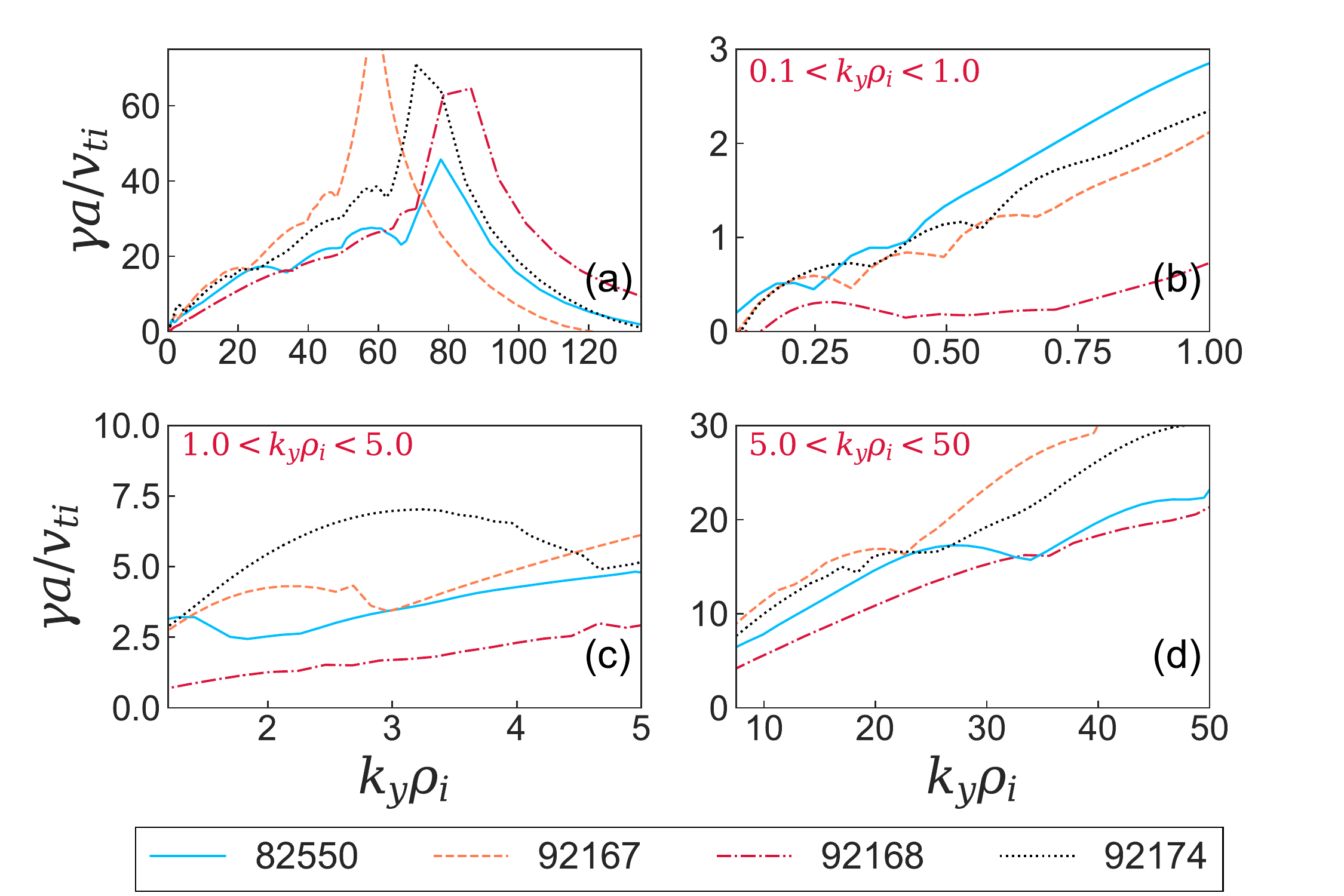}
 \caption{GS2 gyrokinetic pedestal electrostatic growth rates for 4 JET equilibria with $\theta_0 = 0$ for different ranges of $k_y \rho_i $. (a) $1 \lesssim k_y \rho_i \lesssim 135$. (b) $0.1 \lesssim k_y \rho_i \lesssim 1.0$. (c) $1 \lesssim k_y \rho_i \lesssim 5$. (d) $5 \lesssim k_y \rho_i \lesssim 50$.}
  \label{fig:allspectra}
\end{figure}

\Cref{fig:allspectra} shows results from local gyrokinetic microinstability analysis at the radial location with the maximum pressure gradient (and therefore close to the maximum $\gamma_E$) in the four JET-ILW H-mode pedestals described in \Cref{fig:allspectra}. These are all electrostatic, linear GS2 simulations performed without $\mathbf{ E} \times \mathbf{ B}$ shear and with $\theta_0 = 0$. While JET shot 92168 does not appear to have the characteristic toroidal ETG bump at $k_y \rho_i \sim 1$, an analysis of the eigenmodes demonstrates that toroidal ETG modes are indeed the fastest growing modes for $1 \lesssim k_y \rho_i \lesssim 7$ with $\theta_0 = 0$.

\begin{figure}
\centering
    \includegraphics[width=\textwidth]{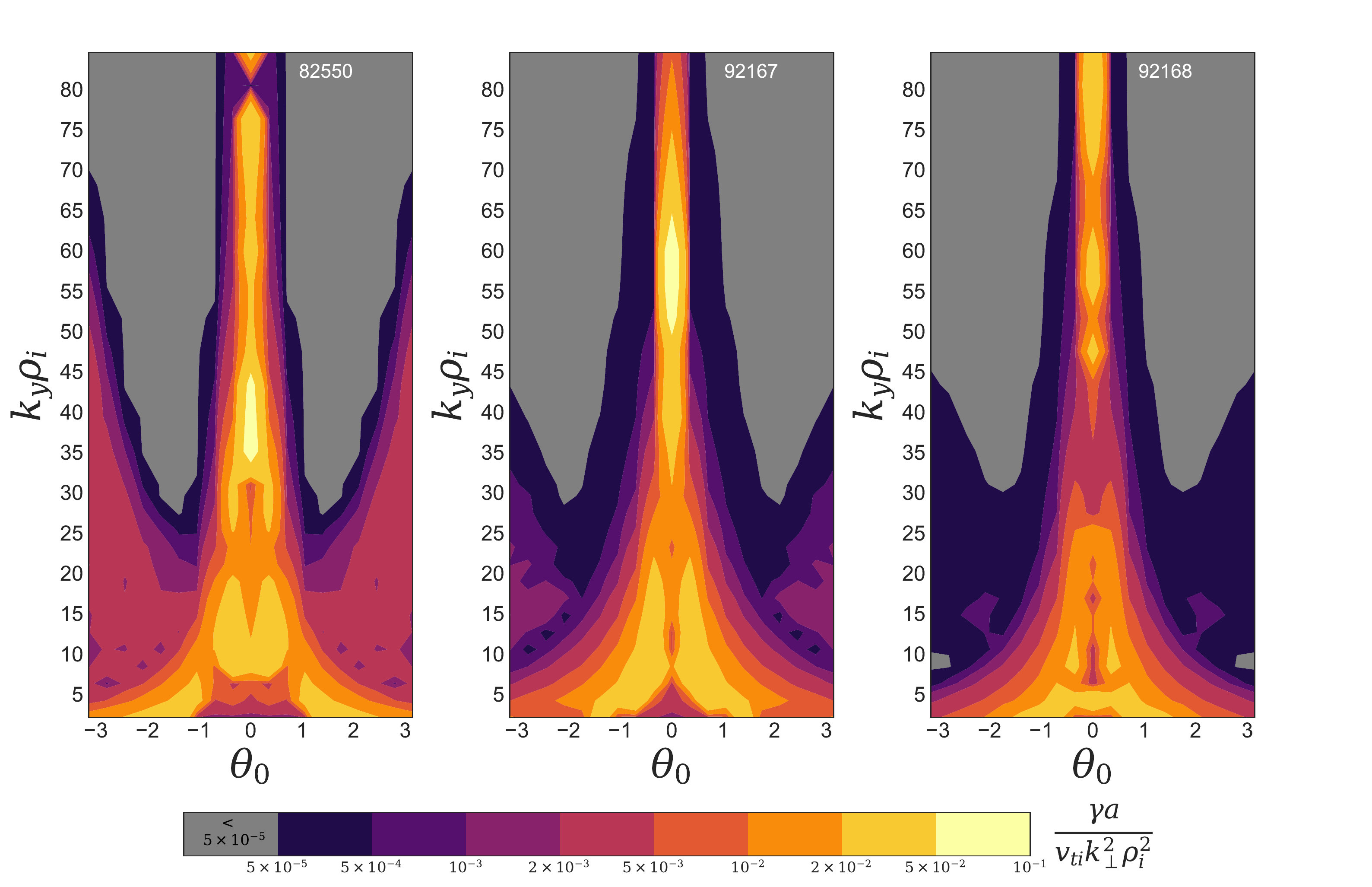}
 \caption{\textcolor{black}{Quasilinear estimates of $\gamma a / v_{ti} k_{\perp}^2$ for JET shots 82550, 92167, and 92168.}}
  \label{fig:gammoverkperp2}
\end{figure}

\textcolor{black}{In \Cref{fig:gammoverkperp2}, we also plot quasilinear transport estimates for JET shots 82550, 92167, and 92168 using $\gamma / k_{\perp}^2$. These estimates demonstrate that $\gamma / k_{\perp}^2$ depends non-trivially on $\theta_0$, similar to JET shot 92174, which is shown in \Cref{fig:theta0growthrateandmodeclassification}(b).}

\section{Electrostatic modes at $k_y \rho_i \lesssim 1.0$} \label{sec:smallkrimodes}

\begin{figure}[t]
\begin{center}
 \includegraphics[width=\textwidth]{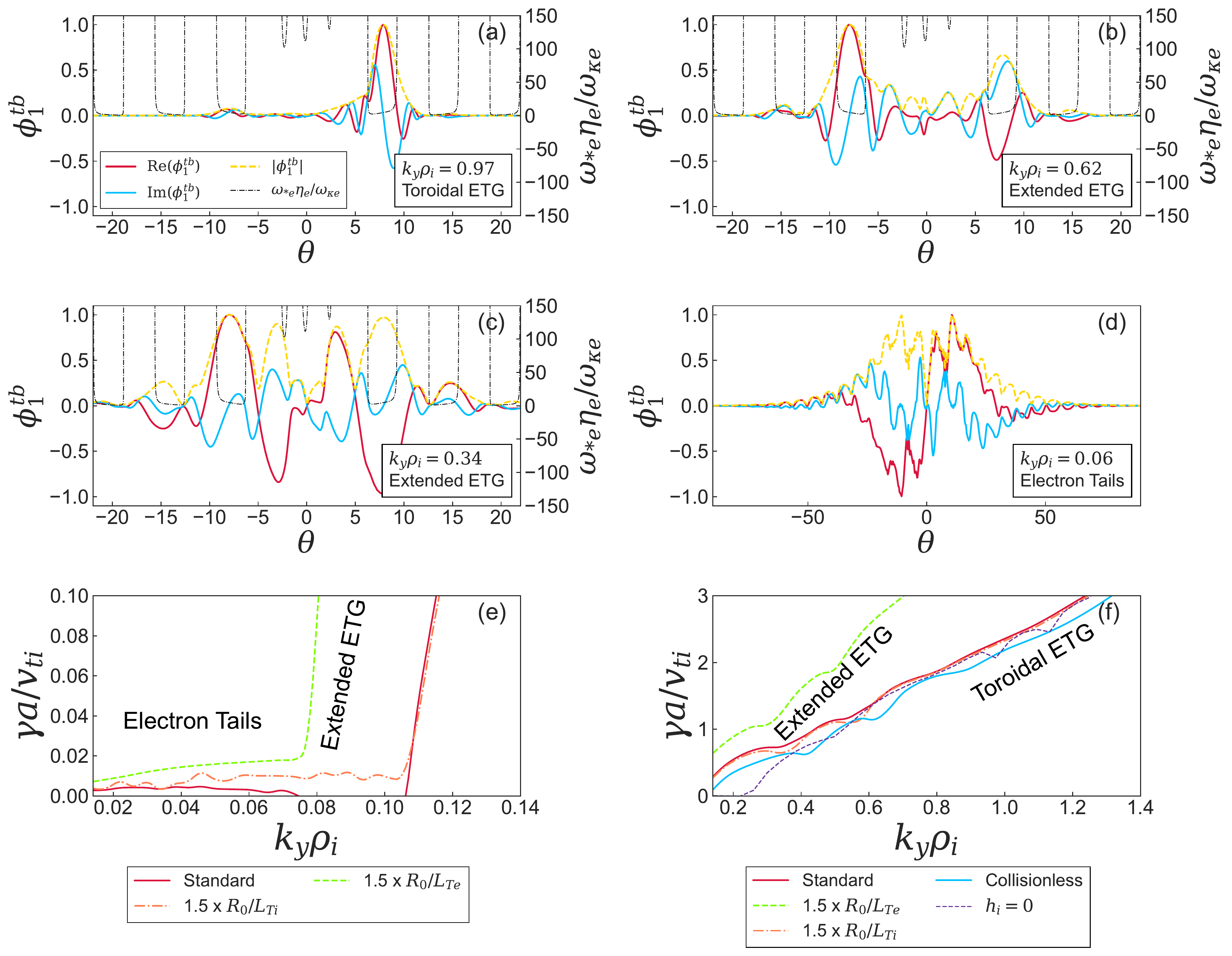}
\end{center}
 \caption{Eigenmodes for $k_y \rho_i \lesssim 1$ and $\theta_0 = 0.05$ for JET shot 92174. In (a), (b), and (c), the quantity $\omega_{*e} \eta_e/ \omega_{\kappa e}$ is plotted only when it is positive. In (a)-(d), the crimson lines are $\mathrm{Re}(\phi ^{tb} _1)$, the blue lines are $\mathrm{Im}(\phi ^{tb} _1)$, the gold dashed lines are $|\phi ^{tb} _1|$, and the black dashed lines are $\omega_{*e} \eta_e/ \omega_{\kappa e}$. (a): $k_y \rho_i = 0.97$, toroidal ETG with large amplitude far down the field line. (b): $k_y \rho_i = 0.62$, extended ETG, (c): $k_y \rho_i = 0.34$, extended ETG, and (d): $k_y \rho_i = \textcolor{black}{0.06}$: modes with electron tails. Growth rates for $k_y \rho_i \lesssim 1.0$ modes with scans in temperature gradients, collisions, and kinetic/adiabatic ions: (e): $k_y \rho_i < 0.14$ modes, and (f): $0.14 < k_y \rho_i < 1.4$ modes.}
\label{fig:lowereigenmodes}
\end{figure}

For completeness, we briefly detail the electrostatic modes at $k_y \rho_i \lesssim 1.0$. We describe their eigenmode structure as well as growth rate sensitivity scans in temperature gradients and collisionality.

All of these simulations are performed with $\theta_0 = 0.05$. For $0.1 \lesssim k_y \rho_i \lesssim 1.0$, we observe modes that become increasingly extended in $\theta$ with decreasing values of $k_y \rho_i$. For $k_y \rho_i \approx 1$, the fastest growing mode is still the toroidal ETG mode described throughout this paper, shown in \Cref{fig:lowereigenmodes}(a). Once $k_y \rho_i$ decreases, the eigenmodes become more complicated and more extended in $\theta$, as shown by \Cref{fig:lowereigenmodes}(b) and (c); we refer to these modes as `extended ETG.' We also plot the quantity $\omega_{*e} \eta_e / \omega_{\kappa e}$ when it is positive in \Cref{fig:lowereigenmodes}(a), (b), and (c) --- we observe that the extended ETG tends to have maxima of $|\phi ^{tb} _1|$ in bad curvature regions. This leads us to speculate that the extended ETG modes are a more complicated version of the toroidal ETG modes described throughout this paper. The extended ETG modes in \Cref{fig:lowereigenmodes}(b) and (c) have tearing parity for both $\mathrm{Re}(\phi ^{tb} _1)$ and $\mathrm{Im}(\phi ^{tb} _1)$. We normalize the eigenmodes in \Cref{fig:lowereigenmodes}(a), (b), (c), and (d) such that the maximum of $|\phi ^{tb} _{1}|$ is 1, and such that the value of $\phi ^{tb} _1 $ is purely real at that location.  In \Cref{fig:lowereigenmodes}(f), we perform a growth rate sensitivity scan for these modes; the growth rate of these extended modes is very sensitive to $R_0/L_{Te}$ and only slightly sensitive to $R_0/L_{Ti}$ and collisions for smaller values of $k_y \rho_i$. The extended ETG modes are stable when run with adiabatic ions for $k_y \rho_i \lesssim 0.2$. 

For $k_y \rho_i \lesssim 0.1$, we observe extremely extended eigenmodes, shown in \Cref{fig:lowereigenmodes}(d) --- the mode extends as far as $\theta \approx \textcolor{black}{50}$ before the typical $|\phi_1 ^{tb}|$ value is less than 10 \% of the eigenmode maximum value. The modes are reminiscent of modes with extended electron tails \cite{Hallatschek2005}. There is no apparent relationship between the maxima of $|\phi ^{tb} _1|$ and bad curvature regions, unlike for the extended toroidal ETG modes. The mode shown in \Cref{fig:lowereigenmodes}(d) has tearing parity for both $\mathrm{Re}(\phi ^{tb} _1)$ and $\mathrm{Im}(\phi ^{tb} _1)$. Sensitivity scans in \Cref{fig:lowereigenmodes}(e) show that these modes are very sensitive to $R_0/L_{Te}$, but insensitive to $R_0/L_{Ti}$. The modes with electron tails were stable for collisionless simulations.

\section{Full Dispersion Relation} \label{app:disprlnfull}

Using \Cref{eq:hsomegakappanormalized} in the quasineutrality \Cref{eq:turnQNexplicit}, we find \Cref{eq:localdispreln} with
\begin{equation}
\fl
\eqalign{
 D_s & \equiv \bigg{(} \frac{e  \phi ^{tb} _1 n_{0e}}{Z_s T_{0e}} \bigg{)}^{-1} \int h_s d^3 v
= \frac{2 i Z_s^2}{\pi^{1/2} v_{ts}^3} \frac{T_{0e}}{T_{0s}} \frac{n_{0s}}{n_{0e}} \int_0^{\infty} d \lambda \int_0^{\infty} dv_{\perp} v_{\perp} \int_{-\infty}^{\infty} d v_{\parallel } \\
& \times \exp\bigg{(}i \lambda \left( \arc{\omega} - \sigma \hat{v}_{\parallel }^2 - \arc{\omega}_{\nabla B s} \frac{\hat{v}_{\perp}^2}{2} - \arc { k}_{\parallel} \hat{v}_{\parallel } \right) - \hat{v}_{\parallel }^2 - \hat{v}_{\perp}^2\bigg{)} \\ &
\times \bigg{[} - \arc {\omega } + \arc {\omega}_{*s} \bigg{(} 1 + \eta_s \left(\hat{v}_{\parallel }^2 + \hat{v}_{\perp }^2 - \frac{3}{2}\right)  \bigg{)} \bigg{]} J_0^2 \left( \sqrt{2 b_s} \hat{v}_{\perp} \right),
}
\end{equation}
where we have used \cite{Biglari1989}
\begin{equation}
\fl
i \int_0^{\infty} d\lambda \exp\bigg{(}i \lambda \left(\arc{\omega } - \sigma \hat{v}_{\parallel }^2 - \arc{\omega}_{\nabla B s} \hat{v}_{\perp}^2/2 - \arc {k}_{\parallel} \hat{v}_{\parallel } \right)\bigg{)} = \frac{1}{-\arc{\omega  }+ \arc {k}_{\parallel} \hat{v}_{\parallel } +\sigma \hat{v}_{\parallel }^2 + \arc{\omega}_{\nabla B s} \hat{v}_{\perp}^2/2 }.
\end{equation}
To find growing solutions and obtain a converged integral, we require that $ \mathrm{Im}(\arc {\omega }) > 0$. Evaluating the integral in $\hat{v}_{\parallel }$ gives
\begin{equation}
\fl
\eqalign{
D_s = & 2 i \frac{Z_s^2}{ v_{ts}^2} \frac{T_{0e} n_{0s}}{ T_{0s} n_{0e}} \int_0^{\infty} d \lambda \int_0^{\infty} dv_{\perp} v_{\perp} \frac{1 }{(1 + i \sigma \lambda)^{1/2}}  \exp\bigg{(}  i \lambda \arc {\omega } - \hat{v}_{\perp }^2 (1 + i \arc{\omega}_{\nabla B s} \lambda / 2) \bigg{)}  \\
& \times \exp\bigg{(}- \frac{(\lambda \arc { k}_{\parallel})^2}{ 4(1 + i \sigma \lambda)}\bigg{)} \\ 
& \times \bigg{[} - \arc {\omega } + \arc {\omega}_{*s} \bigg{(} 1 + \eta_s \left( \frac{2(1 + i \sigma \lambda) - (\arc {k}_{\parallel}\lambda)^2}{4 (1+i \sigma \lambda)^2}  + \hat{v}_{\perp }^2 - \frac{3}{2} \right) \bigg{)}   \bigg{]} J_0^2 \left( \sqrt{2 b_s} \hat{v}_{\perp} \right) .
}
\end{equation}
The integral in $\hat{v}_{\perp}$ gives \Cref{eq:newDR}, where we used the integrals
\begin{equation}
\int_0^{\infty} x J_0^2 (a x) \exp(-b x^2) d x = \frac{1}{2b} I_0 \bigg{(} \frac{a^2}{2 b} \bigg{)} \exp(-a^2/2b) = \frac{1}{2b} \Gamma_0 \left( \frac{a^2}{2b} \right),
\label{eq:BessFunIntegral}
\end{equation}
and
\begin{equation}
\eqalign{
\int_0^{\infty} x^3 J_0^2 (a x) \exp(-b x^2) d x & = \frac{-(a^2 - 2b) \Gamma_0 \bigg{(} a^2/2b \bigg{)} + a^2 \Gamma_1 \bigg{(} a^2/2b \bigg{)}}{4 b^3},
}
\end{equation}
which is found by differentiating \Cref{eq:BessFunIntegral} with respect to $b$.

We proceed to explain the numerical technique used to calculate the $\lambda$ integral in \Cref{eq:newDR}. The $\lambda$ integral in \Cref{eq:newDR} along the real $\lambda$ axis is highly oscillatory when $\gamma \to 0$, and standard numerical integration methods can make substantial errors in the low growth rate limit. Similarly, a straightforward change of variables such as $\lambda \to i \overline{ \lambda}$ will fail for nonzero $k_{\parallel}$ and $b_s$ due to exponential singularities caused by $k_{\parallel}$ and $b_s$ (at $\lambda = \sigma i$ and $2i / \arc{\omega}_{\nabla B s} $, respectively). To avoid these problems, we introduce a numerically robust path of integration that avoids singularities and significantly reduces the number of oscillations. 

In the limit $\lambda \to \infty$, the exponential in \Cref{eq:newDR} reduces to,
\begin{equation}
\exp \left[   i  \left( \arc{\omega} + \frac{\arc{k}_{\parallel}^2}{4 \sigma} \right) \lambda \right].
\end{equation}
Thus, if we wish to minimize oscillations, we should choose our path such that the imaginary component of the exponential is constant. This is achieved with the integral path
\begin{equation}
\lambda = i \left( \arc{\omega}^* + \frac{\arc{k}_{\parallel}^2}{4 \sigma} \right) \overline{\lambda} + a,
\label{eq:integrationpath}
\end{equation}
where $a$ is a constant that we need to choose to improve integral convergence.
\begin{figure}[t]
\centering
    \includegraphics[width=\textwidth]{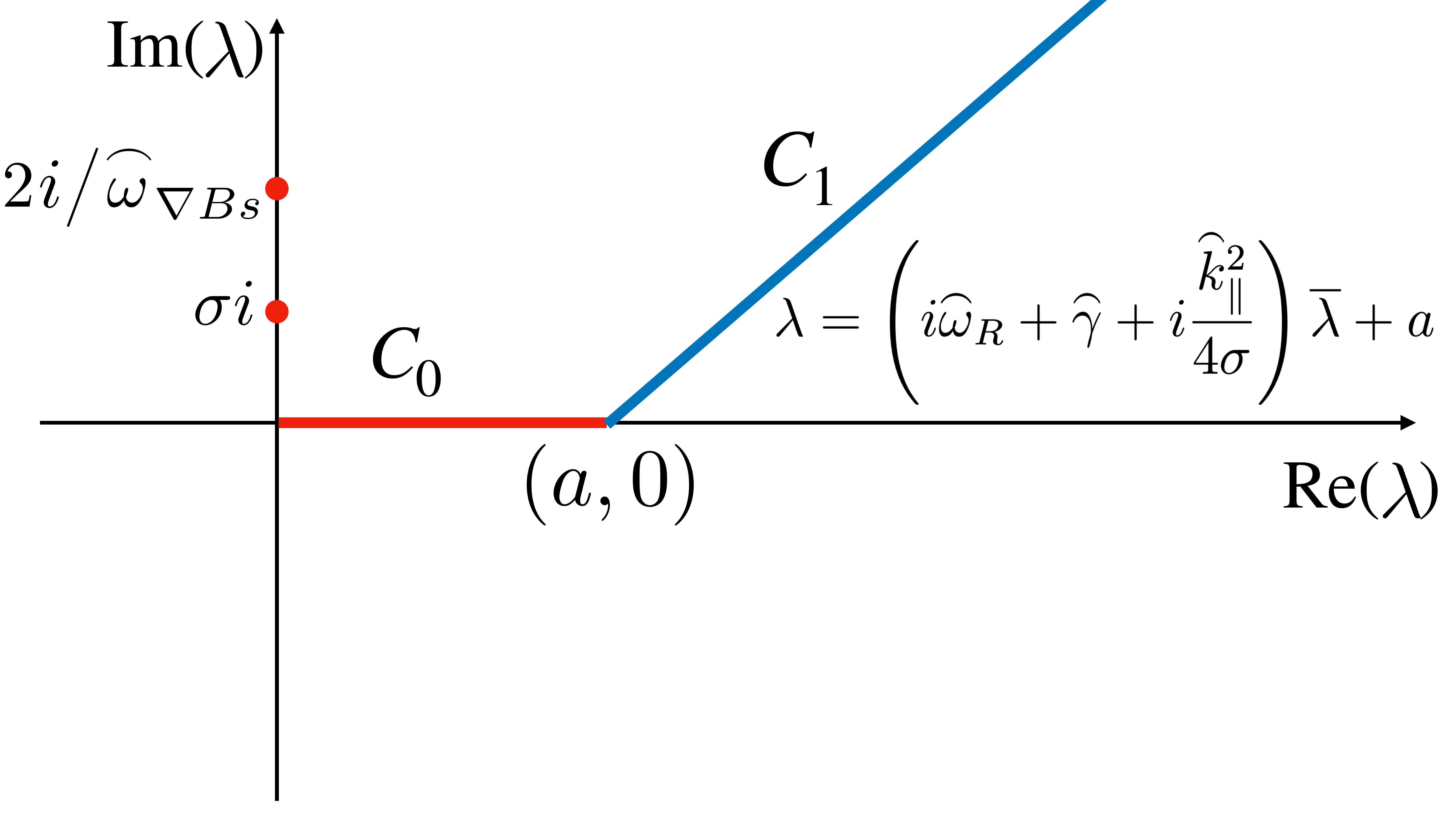}
 \caption{Contour paths $C_0$ and $C_1$, constructed to avoid the poles along the imaginary $\lambda$ axis at $\sigma i$ and $2 \sigma i$, as well as minimizing exponential oscillations.}
 \label{fig:newstabcontour}
\end{figure}
Therefore, we choose an integration path composed of two different paths, $C_0$ and $C_1$. The first path, $C_0$, goes a short distance $a$ along the real $\lambda$ axis. The second path, $C_1$, is the one given in \Cref{eq:integrationpath}. The total integration path is shown in \Cref{fig:newstabcontour}. The integration path in \Cref{fig:newstabcontour} gives the same result as the original path because the integrand in Equation (51) decays as $|\lambda| \rightarrow \infty$. The constant $a$ needs to be sufficiently large to avoid the singularities at $\lambda = \sigma i$ and $2 i/ \arc{\omega}_{\nabla B s}$. A value $a = 0.5$ is usually sufficiently large.

\clearpage

\printbibliography

\end{document}